\documentclass[12pt]{amsart}

\usepackage{amsfonts,amssymb,amsmath,amscd}
\usepackage{bbm}
\usepackage[ps,dvips,matrix,arrow,frame,import,curve,color]{xy}
\newcommand{\raa}{\rightarrow}
\newcommand{\bul}{\bullet}
\newcommand{\Tr}{{\rm Tr}}

\newcommand{\ZZ}{{\mathbb Z}}
\newcommand{\RR}{{\mathbb R}}
\newcommand{\CC}{{\mathbb C}}

\newcommand{\bi}{{\bar i}}
\newcommand{\bj}{{\bar j}}
\newcommand{\cS}{{\mathcal S}}

\newcommand{\cO}{{\mathcal O}}
\newcommand{\bpartial}{{\bar\partial}}
\newcommand{\eps}{\epsilon}

\newcommand{\U}{{\mathcal U}}

\newcommand{\bz}{{\bar z}}

\newcommand{\HK}{{hyper-K\"ahler}}
\newcommand{\cA}{{\mathcal A}}
\newcommand{\bA}{{\bar A}}
\newcommand{\bB}{{\bar B}}
\newcommand{\Hom}{{\rm Hom}}
\newcommand{\cU}{U}
\newcommand{\cB}{{\mathcal B}}
\newcommand{\bb}{{\bar b}}
\newcommand{\obul}{{\Omega^{0,\bullet}}}
\newcommand{\dt}{{d_t}}
\newcommand{\bk}{{\bar k}}
\newcommand{\bp}{{\bar p}}

\newcommand{\tq}{{\tilde q}}
\newcommand{\cV}{{\mathcal V}}
\newcommand{\cR}{{\mathcal R}}
\newcommand{\smpf}{\Omega}
\newcommand{\Z}{\mathcal Y}
\newcommand{\Ext}{{\rm Ext}}

\newcommand{\tF}{{\tilde F}}
\newcommand{\ba}{{\bar a}}
\newcommand{\cM}{{\mathcal M}}

\newcommand{\ver}{v}
\newcommand{\hor}{h}
\newcommand{\cK}{{\mathcal K}}
\newcommand{\cN}{{\mathcal N}}
\newcommand{\End}{{\rm End}}

\newcommand{\STr}{{\rm STr}}
\newcommand{\Hol}{{\rm Hol}}
\newcommand{\pt}{{pt}}
\newcommand{\tp}{{\tilde p}}
\newcommand{\tsmpf}{{\tilde \smpf}}
\newcommand{\Sym}{{\rm Sym}}
\newcommand{\bm}{{\bar m}}
\newcommand{\bI}{{\bar I}}
\newcommand{\bJ}{{\bar J}}
\newcommand{\by}{{\bar y}}
\newcommand{\bx}{{\bar x}}
\newcommand{\tA}{{\tilde A}}
\newcommand{\bfx}{{\bf x}}
\newcommand{\bfy}{{\bf y}}
\newcommand{\brho}{{\bar\rho}}
\newcommand{\bl}{{\bar l}}
\newcommand{\codim}{{\rm codim}}
\newcommand{\tx}{{\tilde x}}
\newcommand{\tz}{{\tilde z}}
\newcommand{\hY}{{\hat Y}}

\newcommand{\hnabla}{{\hat\nabla}}
\newcommand{\cF}{{\mathcal F}}
\newcommand{\tdelta}{{\tilde\delta}}
\newcommand{\mA}{{\mathsf A}}
\newcommand{\Coh}{{\rm Coh}}
\newcommand{\corn}{{\,\begin{picture}(5,5)(0,0)\put(0,0){\line(1,0){5}}\put(0,0){\line(0,1){5}}\end{picture}\,}}

\def\be{\begin{equation}}
\def\ee{\end{equation}}
\def\bear{\begin{eqnarray}}
\def\eear{\end{eqnarray}}

\def\half{{ \frac{1}{2} }}

\def\a{{\alpha}}
\def\b{\beta}

\def\eps{{\epsilon}}

\def\cT{{{\mathcal T}}}
\def\cL{{{\mathcal L}}}

\def\tq{{{\tilde q}}}

\def\c{{\gamma}}

\def\p{{\partial}}

\def\vert{{|}}

\def\xId{ \mathbbm{1} }

\title[3d TFT and symplectic algebraic geometry I]{Three-dimensional topological field theory and symplectic algebraic geometry I}

\author{Anton Kapustin}\address{California Institute of Technology}\email{kapustin@theory.caltech.edu}
\author{Lev Rozansky}\address{University of North
Carolina}\email{rozansky@math.unc.edu}
\author{Natalia Saulina}\address{California Institute of Technology}\email{saulina@theory.caltech.edu}

\begin{document}

\begin{titlepage}
\maketitle

\begin{abstract}
We study boundary conditions and defects in a three-dimensional
topological sigma-model with a complex symplectic target space $X$
(the Rozansky-Witten model). We show that boundary conditions
correspond to complex Lagrangian submanifolds in $X$ equipped with
complex fibrations. The set of boundary conditions has the structure
of a 2-category; morphisms in this 2-category are interpreted
physically as one-dimensional defect lines separating parts of the
boundary with different boundary conditions. This 2-category is a
categorification of the $\ZZ_2$-graded derived category of $X$; it
is also related to categories of matrix factorizations and a
categorification of deformation quantization (quantization of
symmetric monoidal categories). In the appendix we describe a
deformation of the B-model and the associated category of branes by
forms of arbitrary even degree.

\end{abstract}

%\vspace{-7in}
%\parbox{\linewidth}
%{\small\hfill \shortstack{CALT-xx-xxxx}} \vspace{6in}

\end{titlepage}

\tableofcontents

\section{Introduction}

Topological field theory (TFT) provides a bridge between quantum
field theory with its nonrigorous path-integrals and rigorous
mathematics. Two-dimensional topological field theories have been
studied in great detail, mostly in connection with mirror symmetry.
Three-dimensional topological field theories are much less
understood, with the exception of the Chern-Simons gauge theory
\cite{Witten:CS}. But Chern-Simons theory is a rather degenerate
example of a 3d TFT. It has no nontrivial local observables or
boundary conditions; the only topological observables are Wilson
loops which are localized on closed 1-dimensional submanifolds. Our
experience with 2d TFT teaches us that a theory is much more
interesting and nontrivial when one can define it on a manifold with
boundaries.

In this paper we study a 3d TFT of a geometric nature which admits
nontrivial boundary conditions. This theory is a topological
sigma-model which was introduced by E.~Witten and the second author
in \cite{RW}. We will call it the RW model. The target manifold for
such a model must be a \HK\ manifold $X$, or more generally a
holomorphic symplectic manifold. This theory is closely related to
the B-model with target $X$ and reduces to it upon compactification
on a circle. We will see that it admits a variety of boundary
conditions which are analogous to topological branes in the B-model.

There are several reasons to study this model. Firstly, it is rich
enough to elucidate the general structure of 3d TFT. The original
definition of a TFT in arbitrary dimension given by M.~Atiyah
\cite{Atiyah} did not allow for boundaries; neither did it allow for
defects of  higher codimension (such as Wilson loops in Chern-Simons
theory). It was realized in mid 90's that such a generalization is
rather intricate and is most conveniently formulated in the language
of category theory. In two dimensions examples of the resulting
structure are provided by A and B-models associated to Calabi-Yau
manifolds \cite{Wittenmir,Douglas}, and more abstractly by
Calabi-Yau $A_\infty$ categories
\cite{KontsevichSoibelman,Costello}. The structure of 3d TFT is even
more complex and requires higher category theory (see e.g.
\cite{Freed,BaezDolan,CraneYetter}), but concrete examples have been
sparse. The path-integral for the RW model can be regarded as a
heuristic tool for constructing such an example, or rather, a class
of examples.

Another mathematical reason to study the RW model with boundaries is
its relationship with the problem of ``deformation quantization'' of
the derived category of coherent sheaves on a complex manifold,
regarded as a symmetric monoidal category. This relationship will be
explained below. We also find a close connection between RW theory
and categorified algebraic geometry \cite{To-Ve,BFN}. Namely, in the
case when the target space of the RW model has the form $T^*_Y$,
where $Y$ is a complex manifold, the 2-category of boundary
conditions is very similar to the 2-category of derived categorical
sheaves on $Y$ as defined in \cite{To-Ve}. This will discussed in
detail in the follow-up paper \cite{KR2}.

Finally, a physical motivation to study the RW model is its
connection with the 3d mirror symmetry of K.~Intriligator and
N.~Seiberg \cite{IS}. This is a conjectural duality between 3d gauge
theories with $N=4$ supersymmetry. An $N=4$ $d=3$ theory can be
twisted into a 3d topological field theory, and in some special
cases the twisted theory turns out to be equivalent to the RW model
(with a noncompact target space $X$). Understanding the RW model
thus represents a step towards understanding 3d mirror symmetry.

Here is a brief summary of the results of the paper. We show that
the simplest topological boundary conditions in the RW model with
target $X$ correspond to complex Lagrangian submanifolds of $X$.
More generally, we show how to associate a boundary condition to a
fibration over a complex Lagrangian submanifold $Y$ whose fiber is a
Calabi-Yau manifold $Z$. One should think of this fibration as a
B-model with target $Z$ fibered over $Y$. This already points to a
connection with derived algebraic geometry, which we discuss in
section \ref{sec:concl}. Even more generally, one might consider
fibering over $Y$ a deformation of the B-model by forms of even
degrees. For example, if the form is a holomorphic function on $Z$,
one gets a fibered Landau-Ginzburg model. We argue that such a
generalization is necessary if we want the set of boundary
conditions to be closed with respect to some natural operations.

The set of boundary conditions for the RW model with target $X$
should be thought of as the set of objects of a 2-category. The set
of morphisms from one boundary condition to another is interpreted
physically as the set of topological defects of real dimension one
separating one boundary condition from the other. We call such
defects boundary line operators; they are similar to Wilson line
operators in Chern-Simons theory and in the twisted $N=4$ $d=4$
gauge theory considered in \cite{GL}. Boundary line operators
corresponding to a pair of boundary conditions form a category; we
show that it is closely related to a $\ZZ_2$-graded version of the
derived category of coherent sheaves. In certain cases, including
the case when $X=T^*_Y$ for some complex manifold $Y$, one can
promote $\ZZ_2$-grading to $\ZZ$-grading.

If the two boundary conditions coincide, the corresponding category
of boundary line operators has a monoidal structure, i.e. an
associative tensor product. Physically, the composition of line
operators arises from fusing them together. This composition is
noncommutative in general. It is a deformation of the usual tensor
product of holomorphic vector bundles on $Y$.

Let us say a few words about the organization of the paper. In
section \ref{sec:bdries} we study boundary conditions in the RW
model. In section \ref{sec:defects} we study surface operators, line
operators, and point operators. In section \ref{sec:concl} we
discuss the algebraic interpretation of the 2-category of boundary
conditions in the RW model. We argue that it is best expressed in
the language of categorified algebraic geometry as discussed
recently in \cite{To-Ve,BFN}. In the appendix we review the B-model
and introduce a generalization of the B-model which we call the
curved B-model. The category of branes for the curved B-model plays
a central role in this paper and is a deformation the $\ZZ_2$-graded
derived category of coherent sheaves.

In the second part of this paper \cite{KR2} we will reformulate our
results in the language of the Atiyah-Segal axiomatic approach to
Topological Field Theory. We will also discuss the relationship
between the RW model and categorified algebraic geometry.

A. K. would like to thank Alexei Bondal, Dennis Gaitsgory, David
Ben-Zvi, David Nadler, Dan Freed, Andrei Mikhailov, and Dima Orlov
for discussions. L. R. would like to thank Dima Arinkin, David
Ben-Zvi and David Nadler for discussions. N. S. would like to thank
Andrei Mikhailov for discussions. Part of this work was done during
the BIRS workshop ``Matrix Factorizations in Physics and
Mathematics'', May 2008. A. K. and L. R. are grateful to the
organizers for the invitation and to the Banff International
Research Station for hospitality. The work of A. K. and N. S. was
supported in part by the DOE grant DE-FG03-92-ER40701. The work of
L. R. was supported by NSF grant DMS-0808974.

\section{Topological boundary conditions in the RW
model}\label{sec:bdries}

\subsection{Review of the RW model}

Let $M$ be a Riemannian 3-manifold with local coordinates $x^\mu$,
$\mu=1,2,3,$ and $X$ be a \HK\ manifold $X$ of complex dimension
$dim_{\CC}X=2n.$ The fields of the theory are
\begin{equation} \label{e2.1} \text{bosonic:}\quad\phi^{I},\phi^{\bar
I},\qquad\text{fermionic:}\quad \eta^{\bar I},\chi_{\mu}^I.
\end{equation}
where $I,\, {\bar I}=1,\ldots,2n,\quad \mu=1,2,3.$ Bosonic fields
$\phi^I(x),\phi^{\bar I}(x)$ describe a map $\phi$ from $M$ to $X$
(in complex coordinates). The fermionic fields $\chi_\mu^I$ are
components of a 1-form $\chi^I$ on $M$ with values in $\phi^*(T_X)$,
where $T_X$ is the holomorphic tangent bundle of $X$. The fermionic
field $\eta^{\bar I}$ is a 0-form on $M$ with values in the
complex-conjugate bundle $\phi^*({\overline T}_X)$. BRST
transformations of the fields are
\begin{equation} \label{e2.2} \delta_Q\phi^{I} = 0,\quad\delta_Q\phi^{\bar I} =
\eta^{\bar I},\quad \delta_Q\eta^{\bar I}=0, \quad\delta_Q\chi^I =
d\phi^I.
\end{equation}

This transformation is nilpotent, $\delta_Q^2=0$, and is a
derivation of the algebra generated by fields and their derivatives.
The cohomology of $\delta_Q$ is called the algebra of topological
observables.

The BRST invariant action consists of two parts:
\begin{equation} \label{e2.3} S=\int_M\, \cL = \int_M\, \cL_1 +
\int_M\, \cL_2, \end{equation}
where
\begin{align}\label{e2.4}
\cL_1 &= \delta_Q \left(g_{I\bar J}\chi^I \wedge *d\phi^{\bar J}
\right)=g_{I\bar J}\, d\phi^I\wedge *d\phi^{\bar J}-g_{I\bar
J}\chi^I \wedge *\nabla \eta^{\bar J},\\ \label{e2.6} \cL_2
&=\frac12 \smpf_{IJ}\left(\chi^I\wedge
\nabla\chi^J+\frac{1}{3}\cR^J_{KL\bar
M}\chi^I\wedge\chi^K\wedge\chi^L\wedge\eta^{\bar M}\right).
\end{align}
Here $\smpf_{IJ}$ is the holomorphic symplectic form on $X$, star
denotes the Hodge star operator on forms on $M$ with respect to a
Riemannian metric $h_{\mu\nu}$, and the covariant derivatives are
defined using the pull-back of the Levi-Civita connection:
\begin{align*}
\nabla\eta^{\bar J} &=d\eta^{\bar J}+ \Gamma^{\bar J}_{\bar I \bar
K}d\phi^{\bar
I}\wedge \eta^{\bar K},\\
\nabla\chi^{J} &=d\chi^{J}+ \Gamma^{J}_{I K}d\phi^{I}\wedge\chi^{K}.
\end{align*}
(From now on, we will omit the sign $\wedge$ when writing the
exterior product of forms on $M$.) Finally, $\cR^J_{KL\bar M}$
denotes the curvature tensor of the Levi-Civita connection on $X$:
$$\cR^{J}_{K L \bar M}={\p \Gamma^{J}_{K L}\over \p \phi^{\bar M}},\quad
\Gamma^I_{JK}=\left(\p_J g_{K \bar M}\right)g^{I \bar M}.$$

Note that only the BRST exact piece $\cL^{bulk}_1$ depends on the
choice of the metric $h_{\mu \nu}$ on $M$. Therefore the partition
function and correlators of BRST-invariant observables are
independent of the metric. In other words, the theory is a
topological field theory.

Correlators of BRST-invariant observables are called topological
correlators. As usual, any topological correlator involving a
BRST-exact observable vanishes. Thus one may regard topological
correlators as multilinear functions on the BRST cohomology, i.e. on
the algebra of topological observables. The algebra of local
topological observables for the RW model is isomorphic to
\begin{equation}\label{bulkobs}
\oplus_p H^p(\cO_X).
\end{equation}
Indeed, given a $(0,p)$-form $\omega_{{\bar I_1}\ldots {\bar I_p}}$
we let
$$
\cO_\omega=\frac{1}{p!}\omega_{{\bar I_1}\ldots {\bar I_p}}\,
\eta^{\bar I_1}\ldots\eta^{\bar I_p}.
$$
Then the BRST-operator acts by
$$
\delta_Q\left(\cO_\omega\right)=\cO_{\bpartial\omega},
$$
and therefore the BRST cohomology is isomorphic to the Dolbeault
cohomology. The algebra structure is given by the ordinary exterior
product of forms.\footnote{This analysis is classical, but we will
show below that no quantum corrections are possible.}

The term $\cL_2$ in the Lagrangian depends only on the holomorphic
symplectic form $\smpf$ on $X$. Thus topological correlators do not
depend on the K\"ahler form of $X$ and are holomorphic functions of
$\smpf$. In fact, it is not necessary to require the $(1,1)$ form
$g_{I\bar J}$ to be closed, and the theory makes sense when $X$ is a
not-necessarily-K\"ahler complex symplectic manifold. The action is
a bit more complicated since the natural connection in the
non-K\"ahler case is not the Levi-Civita connection. To simplify
formulas, in the rest of the paper we will assume that $X$ is \HK.
However, all the results hold for arbitrary complex symplectic
manifolds.

As a consequence of BRST symmetry, the path-integral computing
topological correlators localizes on BRST-invariant field
configurations. These can be determined by requiring the BRST
variations of fermionic fields to vanish. From
$\delta_Q\chi^I=d\phi^I$ it follows that the path-integral localizes
on constant maps to $M$. The contribution of a particular map is
given by the ratio of certain determinants (the Ray-Singer torsion)
times the sum of Feynman diagrams; the Feynman rules are explained
in detail in \cite{RW}. The correlator is obtained by integrating
over the moduli space of constant maps to $X$, i.e. over $X$.

In addition to BRST symmetry, the theory also has a $\ZZ_2$ symmetry
which multiplies all fermionic fields by $-1$. If the Lagrangian
contained only the $\cL_1$ term, this $\ZZ_2$ could be extended to a
$U(1)$ symmetry, so that the field $\chi$ has $U(1)$ charge $-1$ and
$\eta$ has charge $+1$. This symmetry is broken by $\cL_2$ down to
$\ZZ_2$. One can formally restore it by assigning the holomorphic
symplectic form $\smpf$ charge $+2$. We will call this symmetry the
ghost number symmetry.

The ghost number symmetry has interesting consequences for the
dependence of the partition function on the symplectic form $\smpf$.
If the measure in the path-integral were invariant under ghost
number symmetry, one could conclude that the partition function is
invariant under a rescaling of $\smpf$ and therefore, by holomorphy,
independent of $\smpf$. As a matter of fact, because of fermionic
zero modes the measure has charge $2n(b_1(M)+1)$ \cite{RW}, so under
the rescaling
$$
\smpf\mapsto\lambda^2\smpf
$$
the partition function scales as
$$
Z(M)\mapsto \lambda^{2n(b_1(M)+1)} Z(M).
$$
As explained in \cite{RW}, this scaling implies that for a
particular $M$ only a finite number of Feynman diagrams may
contribute to the partition function.

From the physical viewpoint, it is convenient to redefine
$$
\smpf\mapsto \hbar^{-1}\smpf,\quad g_{I\bar J}\mapsto
\hbar^{-1}g_{I\bar J},
$$
so that the action (\ref{e2.3}) is proportional to $\hbar^{-1}$ and
one can interpret $\hbar$ as the Planck constant. The above
discussion of ghost number symmetry can be restated by saying that
topological correlators of observables with a fixed ghost number
receive contributions only from a particular order in the
$\hbar$-expansion. In what follows, we will often deal with
observables which do not have a definite ghost number. Correlators
of such observables are formal power series in $\hbar$. However, for
target $X$ of a fixed dimension no problems with convergence arise,
since higher powers of $\hbar$ are accompanied by higher powers of
fermionic fields, and there is only a finite number of the latter.

\subsection{Reduction of the RW model on a circle}

If $M=S^1\times\Sigma$, where $\Sigma$ is a two-dimensional oriented
manifold, the RW model reduces to the B-model on $\Sigma$ with
target $X$ \cite{Thompson}. Indeed, since the model is topological,
we may assume that the size of the circle is very small,
Fourier-expand all the fields on the circle and retain only the
constant modes. The field $\eta^\bI$ becomes a 0-form on $\Sigma$.
If we denote by $x^3$ the coordinate on $S^1$, the 1-form field
$\chi^I$ can be expanded as
$$
\chi^I=\chi^I_\Sigma+\chi_3^I dx^3,
$$
where $\chi^I_\Sigma$ is a pull-back of a 1-form $\rho^I$ on
$\Sigma$. As for the field $\chi_3^I$, it becomes a 0-form on
$\Sigma$. We can define
$$
\theta_I=\smpf_{IJ}\chi_3^J.
$$
Note that since $\smpf_{IJ}$ has ghost number $2$, the fields
$\theta_I$ and $\eta^\bI$ have ghost number $1$, while $\rho^I$ has
ghost number $-1$. Thus the ghost number symmetry of the RW model
becomes the usual ghost number symmetry of the B-model.

The BRST transformation restricted to constant Fourier modes read
\begin{align}\label{BmodelBRSTreduced}
\delta\phi^I & =0, & \delta\phi^\bI &=\eta^\bI,\\
\delta\eta^\bI & =0, & \delta\theta_I &= 0,\\
\delta\rho^I & = d\phi^I. & &
\end{align}
These are the BRST transformations of the B-model with target $X$.
It is easy to check that the action of the RW model also reduces to
the action of the B-model. In what follows we will often use the
known properties of the B-model as a guide to understanding the RW
model.

The above analysis of the reduction was classical. In the quantum
theory, instead of setting nonconstant Fourier modes to zero, one
has to integrate over them in the path-integral. This could induce
corrections to the action of the B-model. However, using the ghost
number symmetry, it is easy to show that no such corrections are
possible, and the above classical result is exact. Indeed,
infinitesimal deformations of the B-model with target $X$ are
parameterized by even elements of the $\ZZ_2$-graded vector space
$$
\oplus_{p,q} H^p(\Lambda^q T_X).
$$
Since the RW model on $S^1\times\Sigma$ has ghost number symmetry,
the action of the reduced model also should preserve this symmetry.
This implies that possible deformations of the B-model must have
$p+q=2$. On the other hand, the class in question must be
constructed from the curvature of $X$, which is an element of
$H^1(\Omega^1\otimes \End\, T_X)$, and the symplectic form $\smpf$.
The curvature has ghost number $0$, $\smpf$ has ghost number $2$.
The deformation of type $(q,p)$ must be a polynomial of degree $p$
in the curvature (to match the cohomological degree) and of degree
$0$ in $\smpf$ (to have zero ghost number zero). But then it is
impossible to contract the holomorphic indices to get a section of
$\Lambda^q T_X$ with $q\geq 0$. Hence no quantum deformation of the
classical result is possible.

The reduction of the RW model to the B-model has a further subtlety,
which becomes apparent when one compares the partition function of
the RW theory on $S^1\times\Sigma$ and the partition function of the
B-model on $\Sigma$ \cite{Thompson}. When $\Sigma$ is a torus, they
are both equal to the Euler characteristic of $X$. For $g(\Sigma)>1$
both partition functions vanish because of zero modes for the
fermionic field $\chi$. But while the partition function of the
B-model vanishes for $g(\Sigma)=0$, the partition function of the RW
theory is equal to
$$
\int_X f(R_X),
$$
where $f(R_X)$ is a certain characteristic class of $X$. To explain
this disagreement, note that the action of the B-model can be
modified by a term
$$
\int_\Sigma R^{(2)} \cO,
$$
where $R^{(2)}$ is the curvature 2-form representing $2\pi$ times
the Euler class of $\Sigma$ and $\cO$ is any BRST-invariant local
operator built from fields $\eta$ and $\theta$. We can choose
$R^{(2)}$ so that it is concentrated at a single point on $\Sigma$.
Then, when $\Sigma$ is a sphere, the partition function of the
deformed B-model is the same as the expectation value of the
operator $e^{4\pi\cO}$ in the undeformed B-model. Note further that
observables in the B-model can be identified with forms on $X$ using
the isomorphism $\smpf: T_X\raa T^*_X$. Thus if we set $4\pi\cO={\rm
log} f(R_X)$, we can reproduce the partition function of the RW
theory on $S^1\times\Sigma$. This curvature-dependent correction to
the action of the B-model is again a quantum effect.

\subsection{Complex lagrangian submanifolds}\label{sec:complagrbdry}

First we will look for BRST-invariant boundary conditions for the
sigma-model which do not involve additional degrees of freedom on
the boundary.

Suppose the boundary is locally given by the equation $x^3=0$, and
the interior of $M$ is $x^3>0$. Let us temporarily regard $x^3$ as
the time coordinate. Then from the point of view of the canonical
formalism, the space of fields and their conjugate momenta at
$x^3=0$ is an infinite-dimensional symplectic manifold. A boundary
condition then corresponds to a choice of a Lagrangian submanifold
in this symplectic manifold. The equations defining the submanifold
must be local in the $x^1,x^2$ directions, to ensure that the
boundary condition is local. Upon quantization, this Lagrangian
submanifold will become what is known as the boundary state. We also
require the Lagrangian submanifold to be BRST-invariant.

On the other hand, one could also consider $x^1$ as the time
coordinate.\footnote{In this paper we will only consider the
boundary conditions which do not break rotational invariance on the
boundary, so one could equally well take any other direction along
the boundary as the time.} Then the space of fields and their
conjugate momenta on the hypersurface $x^1=0$ must have a
well-defined symplectic structure. This provides an additional
constraint on allowed boundary conditions.

To apply the canonical formalism, it is convenient to choose local
coordinates in the collar neighborhood of $\partial M$ so that the
metric on $T_M\vert_{\partial M}$ has the form
$$
h_{\a\b} dx^\a \otimes dx^\b+dx^3\otimes dx^3,
$$
where the indices $\a,\b$ take values in the set $\{1,2\}$ and label
the coordinates on the boundary. If we regard $x^3$ as the time
coordinate, the momenta conjugate to $\phi^I$ and $\phi^\bJ$ are
\begin{align}
\label{e2.7} \pi_{\phi^I} & =
\frac{\partial\cL^{bulk}}{\partial(\p_3\phi^I)} =vol_{\partial M}
g_{I\bar J}\p_3 \phi^{\bar J}
-\frac12\smpf_{KJ}\Gamma^{J}_{IL}\chi^{K}\chi^{L}\vert_{\partial M}
\\
\label{e2.8} \pi_{\phi^{\bar J}} & =
\frac{\delta{\cL^{bulk}}}{\delta(\p_3 \phi^{\bar J})} =vol_{\partial
M} \bigl(g_{I \bar J}\p_3 \phi^I - g_{I \bar K}\Gamma^{\bar K}_{\bar
J \bar L}\,\chi_{3}^{I}\,\eta^{\bar L}\Bigr)
\end{align}
Here $vol_{\partial M}$ is the volume form of the induced metric:
$$
vol_{\partial M}=\sqrt h dx^1 dx^2,\quad h=\det h_{\a\b}.
$$
The symplectic form on the space of fields can be read off the
boundary part of the variation of the bulk action \begin{equation}
\label{genvar} -\int_{\partial M} \Biggl[ \pi_{\phi^I}\delta \phi^I
+\pi_{\phi^{\bar K}}\delta \phi^{\bar K} - vol_{\partial M}\, g_{I
\bar J}\chi_3^I \,\delta \eta^{\bar J}-
\frac12\smpf_{IJ}\chi^I\delta\chi^J\Biggr].
\end{equation}
This expression defines a 1-form on the space of fields whose
exterior differential is the symplectic form.

The simplest boundary condition restricts the map $\phi: M\raa X$ so
that $\partial M$ is mapped to some submanifold $Y\subset X$. Such
boundary conditions are called D-branes in the B-model. Every
BRST-invariant boundary condition in the RW model gives, upon
reduction on a circle, a BRST-invariant boundary condition in the
B-model with the same target. It is well-known that $Y$ corresponds
to a BRST-invariant D-brane in the B-model only if it is a complex
submanifold. Thus we may assume that $Y$ is complex.

We will now show that in the RW model $Y$ has to be a Lagrangian
submanifold with respect to the holomorphic symplectic structure
$\smpf_{IJ}$. We may assume without loss of generality that locally
$Y$ is given by the equations
$$
\phi^I = 0,\quad I=2n-m+1,\ldots ,  2n.
$$
Here $m$ is the complex codimension of $Y$. Let us adopt the
convention that lower-case unprimed indices range from $1$ to
$2n-m$, while lower-case primed indices range from $2n-m+1$ to $2n$.
Then the equation of the submanifold $Y$ can be concisely written as
\begin{equation}\label{phiboundcond}
\phi^{i'}=0,
\end{equation}
and BRST-invariance requires \be\label{e2.11} \eta^{\bar
i'}=0,\quad\chi_{\a}^{i'}=0. \end{equation}But since the Poisson
bracket of the fields $\chi_1^I$ and $\chi_2^I$ is given by the
inverse of the holomorphic symplectic form $\smpf$ (if we regard
$x^3$ as the time coordinate), the last equation implies that $Y$
must be Lagrangian with respect to $\smpf$. One can express the
boundary conditions on the fields $\eta^{\bar J}$ and $\chi_{\a}^J$
more invariantly by saying that the vector fields $\eta$ and
$\chi_{\a}, \a=1,2,$ are tangent to $Y$.

It is convenient from now on to assume that local coordinates
$\phi^i,\phi^{i'}$ have been chosen so that the symplectic form
$\smpf$ is off-diagonal:
$$
\smpf=\smpf_{j k'} d\phi^j d\phi^{k'}.
$$
The matrix inverse to $\smpf_{j k'}$ will be denoted $\smpf^{k' j}$.

To complete the determination of boundary conditions, let us now
regard $x^1$ as the time coordinate. With respect to $x^1$, the
momentum conjugate to $\eta^{\bar j}$ is $g_{\bar j K}\sqrt h
h^{1\b}\chi_\b^K$. Since the fields $\chi_\b^j,$ $\b=1,2,$ are
unconstrained on the boundary, the same should be true about
$\eta^{\bar j}$. On the other hand, with respect to $x^3$ the field
$\eta^{\bar j}$ is canonically conjugate to $g_{\bar i J}\chi_3^J$,
so we need to require
\begin{equation}
g_{\bar i J}\chi_3^J=0.
\end{equation}
In other words, the vector field $\chi_3$ must be orthogonal to
$T_Y$ with respect to the Hermitian metric. Finally, since the
bosonic fields $\phi^i$ and $\phi^{\bar i}$ are unconstrained, we
must require their conjugate momenta (\ref{e2.7},\ref{e2.8}) to
vanish on the boundary: \be\label{piboundcond}
 \pi_{\phi^i}=0,\quad \pi_{\phi^{\bar i}}=0.
\end{equation}One can verify that these boundary conditions are
BRST-invariant.\footnote{When checking the BRST-invariance of the
first condition in (\ref{piboundcond}) one has to use the equation
of motion for $\chi_3^i$
\begin{equation} \label{e2.11'} -g_{i \bar
J}vol_{\partial M}\nabla_3 \eta^{\bar J}+\smpf_{i j'}
\Biggl(\nabla\chi^{j'}+\frac12\cR^{j'}_{kl\bar m}\chi^k\chi^l
\eta^{\bar m}\Biggr)=0, \end{equation}where all fields and the
covariant derivative are restricted to the boundary.}

We would like to stress that boundary conditions involving
non-Lagrangian holomorphic submanifolds of $X$ are not allowed in
the RW model, even though they correspond to valid topological
branes in the B-model. The only way to avoid this restriction is to
break explicitly the diffeomorphism invariance on the boundary.

So far the discussion was purely classical. We will see later that
on the quantum level one has to require $Y$ to be a Calabi-Yau
manifold to make sure that the measure in the path-integral is
BRST-invariant.

\subsection{Deformations of boundary conditions}\label{sec:defs}

We can find more general boundary conditions for the RW model by
deforming the ones we constructed above. To deform a boundary
condition, one adds a boundary term to the action and simultaneously
modifies the BRST transformations of the fields by terms localized
on the boundary, so that the whole action is BRST-invariant.

A systematic way to construct such a deformation is the descent
procedure \cite{Wittentop}. Let $\cO$ be an even (i.e. bosonic)
local topological observable on the boundary. Its descendants
$\cO^{(1)}\in \Omega^1(M)$ and $\cO^{(2)}\in \Omega^2(M)$ are
defined by
$$
d\cO=\delta_Q \cO^{(1)}+\ldots,\quad d\cO^{(1)}=\delta_Q
\cO^{(2)}+\ldots,
$$
where dots denote terms proportional to equations of motion. Now
consider adding to the action a boundary term
$$
S^{bulk}\mapsto S^{bulk} + \eps \int_{\partial M}\cO^{(2)}.
$$
where $\eps$ is a formal parameter. It follows from the definition
of $\cO^{(2)}$ that the modified action is BRST-invariant up to
terms proportional to equations of motion. The latter can be removed
by modifying the BRST transformations of bulk fields by terms
proportional to the delta-function localized on the boundary. In
general, $\cO^{(2)}$ is not invariant with respect to modified BRST
transformations so one has to correct the boundary action further by
terms of order $\eps^2$. Luckily, in all cases of interest to us
this is not necessary, and the procedure stops here.

It is easy to see that local topological observables on the boundary
are of the same form as in the bulk, i.e. they have the form
$$
\frac{1}{p!}\omega_{{\bar i}_1\ldots {\bar i}_p}\,\eta^{{\bar
i}_1}\ldots\eta^{{\bar i}_p},
$$
for some $(0,p)$-form $\omega$ on $Y$. On the classical level the
BRST operator again acts as the Dolbeault operator and therefore the
space of boundary topological observables is isomorphic to
\begin{equation}\label{bdryobs}
\oplus_{p=0}^n H^p(\cO_Y).
\end{equation}
The $\ZZ_2$-grading is given by $p\,{\rm mod}\, 2$. In what follows
we will denote $p\, {\rm mod}\, 2$ by $\hat p$. Thus the even
subspace of $H^\bul(\cO_Y)$ will be denoted $H^{\hat 0}(\cO_Y)$, and
the odd subspace will be denoted $H^{\hat 1}(\cO_Y)$. On the
classical level, the algebra structure is given by the usual
exterior product. We will show later that the algebra of boundary
observables is not modified by quantum corrections.

Infinitesimal deformations of the  boundary condition are
parameterized by $H^{\hat 0}(\cO_Y)$. An element of this space can
be represented by a $\bpartial$-closed inhomogeneous form $W$ of
even degree. The corresponding observable can be thought of as an
even function $W(\phi,\eta)$ of bosonic variables $\phi^i,\phi^\bi$
and fermionic variables $\eta^\bi$ satisfying
\begin{equation}\label{Wclosed}
\eta^\bi\ \frac{\partial W(\phi,\eta)}{\partial\phi^\bi} =0.
\end{equation}
More invariantly, one can say that $W$ is a function on the odd
tangent bundle $\Pi{\bar T}_Y$ annihilated by the odd vector field
$$
\bpartial=\eta^\bi\frac{\partial}{\partial\phi^\bi}.
$$
We will call this function a curving on $Y$. The origin of this
terminology will become clear later.

Next we construct the descendants of $W$. Let $\partial_i$ and
$\partial_\bi$ denote partial derivatives with respect to $\phi^i$
and $\phi^\bi$ respectively. We first note that the equation
(\ref{Wclosed}) implies
$$
\partial_\bj W=\eta^\bi \partial_\bi \frac{\partial
W}{\partial\eta^\bj}=\delta_Q \frac{\partial
W}{\partial\eta^\bj},\quad \partial_\bi \frac{\partial W}{\partial
\eta^\bj}-\partial_\bj \frac{\partial W}{\partial
\eta^\bi}=\delta_Q\frac{\partial^2
W}{\partial\eta^\bi\partial\eta^\bj}.
$$
Using these relations, we find
\begin{align}
W^{(1)}&=\chi^i \partial_i W +d\phi^\bi\frac{\partial
W}{\partial\eta^\bi},\\
W^{(2)}&=\frac12 \chi^i\chi^j \hnabla_i \partial_j W+\chi^i
d\phi^\bj
\partial_i \frac{\partial W}{\partial\eta^\bj}+\frac12 d\phi^\bi
d\phi^\bj \frac{\partial^2 W}{\partial\eta^\bi\partial\eta^\bj}-
vol_{\partial M} \smpf^{k'i} A^K_{k'}\partial_i W g_{K\bar
J}\partial_3\phi^{\bar J}.
\end{align}
Here $\hnabla$ is the covariant differential with respect to the
induced metric on $Y$ and $A^K_{k'}$ is defined by
\begin{equation}\label{Adef}
A^{i'}_{k'}=\delta^{i'}_{k'},\quad A^i_{k'}=-g_{k' \bj} {\hat
g}^{\bj i},
\end{equation}
where ${\hat g}^{\bj i}$ is the matrix inverse to $g_{i\bj}$. The
matrix $A^K_{k'}$ represents the unique bundle map $A:N_Y\raa
T_X\vert_Y$ which splits the exact sequence
\begin{equation}\label{exactsequenceY}
0\raa T_Y\raa T_X\vert_Y\raa N_Y\raa 0.
\end{equation}
and identifies $N_Y$ with the orthogonal complement of $T_Y$. The
splitting $A$ is nonholomorphic, and in general no holomorphic
splitting exists; some consequences of this are described in the
next subsection.

The descendants satisfy
$$
\delta_Q W^{(1)}=dW,\quad \delta_Q W^{(2)}= d W^{(1)}+
\smpf^{k'i}\partial_i W A_{k'}^K \frac{\delta S^{bulk}}{\delta
\chi^K_3}.
$$
We define the boundary action as
$$
S^{bry}=\int_{\partial M} W^{(2)},
$$
and modify the BRST transformation for $\chi_3$ by a boundary term:
\begin{equation}\label{modifiedBRSTchi}
\delta_{Q,W} \chi_3^K=\p_3 \phi^K-\delta(x_3)A^K_{k'} \smpf^{k'
i}\p_i W.
\end{equation}
Then the total action is BRST-invariant up to a total derivative on
the boundary:
$$
\delta_{Q,W} (S^{bulk}+S^{bry})=\int_{\partial M} d W^{(1)}.
$$

It is instructive to look at a couple of special cases. If $W$ is a
degree-0 form, then it is simply a holomorphic function on $Y$. The
boundary action simplifies to
\begin{equation} \label{e2.14}
S^{bry}=\int_{\partial M} \Bigl(\frac12\hnabla_i\p_jW \chi^i
\chi^j-vol_{\partial M}\smpf^{k' i}\p_iW A^K_{k'} g_{K \bar J} \p_3
\phi^{\bar J}\Bigr)
\end{equation}
The first term in this action is reminiscent of the Landau-Ginzburg
deformation of the B-model (see Appendix), with $W$ playing the role
of the superpotential. We will see later on that this is more than a
mere similarity.

If $W$ has degree $2$, then it corresponds to a $\bpartial$-closed
$(0,2)$ form $B_{\bi\bj}$ on $Y$:
$$
W=\frac12 B_{\bi\bj}\eta^\bi\eta^\bj.
$$
The corresponding boundary action is similar to the deformation of
the B-model by a $(0,2)$ B-field. To make the similarity more
obvious, let us recall that $X$ and therefore $Y$ are K\"ahler
manifolds. If we assume in addition that $Y$ is compact, then the
Dolbeault cohomology class of $B$ contains a closed 2-form, and
therefore we may assume without loss of generality that $B$ is
closed. Then the boundary action simplifies to
$$
S^{bry}=\int_{\partial M} \phi^* B
$$
In the context of 2d sigma-models such a term in the bulk action is
known as the B-field. In 3d sigma-model it deforms the boundary
action instead. Note that in this case no modification of the BRST
transformation is necessary.

Finally, we need to consider deformations of the boundary action
corresponding to geometric deformations of the Lagrangian
submanifold. An infinitesimal deformation of $Y$ corresponds to a
holomorphic section $\xi$ of the normal bundle $N_Y$. The symplectic
form identifies $N_Y$ with $T_Y^*$; the requirement that the
deformed submanifold be Lagrangian is equivalent to $\partial(\smpf
\xi)=0$. Thus infinitesimal deformations are in one-to-one
correspondence with closed holomorphic 1-forms on $Y$. Locally, such
a form can be integrated to a holomorphic function $W$ on $Y$. This
suggests that an infinitesimal deformation of the Lagrangian
submanifold corresponding to a 1-form $\alpha$ can be described by
the boundary action (\ref{e2.14}) where $\partial W$ is replaced by
$\alpha$ times a constant factor.

To prove this statement, note that the variation of the boundary
action contains a peculiar term
$$
-\int_{\partial M} vol_{\partial M} \smpf^{k'i}\partial_i W A^K_{k'}
g_{K\bar J} \delta(\partial_3\phi^{\bar J}).
$$
To cancel such a term by a boundary term in the variation of the
bulk action, we must assume that the field $\phi^K$ has a step-like
discontinuity on the boundary:
\begin{equation}\label{phijump}
\lim_{x_3\raa
0+}\phi^{K}(x_1,x_2,x_3)=\phi^{K}(x_1,x_2,0)+A^K_{k'}\smpf^{k'i}\partial_i
W.
\end{equation}
Therefore deforming the boundary action by $W$ is equivalent to
deforming $Y$ along a normal vector field
$$
\xi^{k'}=\smpf^{k' i}\partial_i W.
$$

We can describe this result a bit differently by introducing local
Darboux coordinates $q^i, p_i$, $i=1,\ldots,n,$ so that the
symplectic form is $\smpf=dq^i dp_i$ and the undeformed Lagrangian
is given by the equations $p_i=0$. Given a holomorphic function $W$
on $Y$, we define a deformed Lagrangian submanifold $\tilde Y$ by
$$
p_i=\frac{\partial W}{\partial q^i},\quad i=1,\ldots,n.
$$
In symplectic geometry, the function $W$ is known as the generating
function of the Lagrangian $\tilde Y$. We have shown above that on
the infinitesimal level the boundary condition corresponding to
$\tilde Y$ is equivalent to the deformation of $Y$ by the generating
function $W$.

It is tempting to extend this result from infinitesimal to finite
deformations and identify the superpotential $W$ with the generating
function of the deformed Lagrangian submanifold $\tilde Y$. But such
an identification appears problematic because the generating
function depends on the choice of Darboux coordinates, while $W$
does not. The likely resolution is that the superpotential
deformation is not well-defined beyond leading order in $W$, and the
additional choices one has to make to define it amount to a choice
of local Darboux coordinates. Such ambiguities in perturbation
theory typically arise from short-distance divergences which require
renormalization.

\subsubsection{Obstructions}\label{sec:obstructions}

While classically adding a descendant 2-form to the boundary action
preserves BRST-invariance, this is not necessarily true on the
quantum level beyond first order in the deformation. Consequently,
at second order in $W$ one may have an obstruction to deformation.
An analogous phenomenon is well-known in the context of B-branes:
deformations of complex submanifolds, holomorphic vector bundles or
more general objects of the derived category of coherent sheaves may
be obstructed. A less-known case where such obstructions occur is
the B-model itself, in the case when the target manifold is not
compact and K\"ahler. (One can show that the deformation theory of
compact K\"ahler Calabi-Yau manifolds is unobstructed
\cite{Bogomolov,Todorov,Tian}). We will see below that the situation
for boundary deformations in the RW model is similar to that in the
B-model: if $Y$ is compact and K\"ahler, the deformation theory is
unobstructed, while in general there are obstructions.

As shown above, geometric deformations of $Y$ are essentially a
special case of deformations by means of a curving $W$. A geometric
counterpart of our result is a theorem proved by C.~Voisin
\cite{Voisin} which implies that deformations of complex Lagrangian
submanifolds in a complex symplectic manifold are unobstructed in
the compact K\"ahler case.

One difference between deformations of the B-model and boundary
deformations of the RW model is that in the latter case obstructions
arise from the external geometry of $Y$ in $X$. It is well-known
that in the real case the formal neighborhood of a submanifold is
isomorphic to the total space of its normal bundle. In the complex
case this is no longer the case. To describe the difference,
consider again a short exact sequence of sheaves
(\ref{exactsequenceY}). It defines a class
$[\beta]\in\Ext^1(N_Y,T_Y)=H^1(T_Y\otimes N_Y^*)$ which measures the
failure of the exact sequence to split holomorphically. This
cohomology class also describes the leading deviation of the complex
geometry of the neighborhood of $Y$ in $X$ from the geometry of the
total space of the holomorphic vector bundle $N_Y$. To be more
concrete, let $x^i, i=1,\ldots,\dim_\CC Y$ denote local complex
coordinates on $Y$ and $z^\alpha,\alpha=1,\ldots,\codim_\CC Y,$
denote complex linear coordinates on the fibers of $N_Y$ with
respect to some local trivialization $e_\alpha$. Then to leading
order in $z$ the difference between the $\bpartial$ operator on the
neighborhood of $Y$ and the $\bpartial$-operator on the total space
of $N_Y$ is
\begin{equation}\label{deviationbeta}
z^\alpha \beta_{\alpha \bk}^i dx^\bk \partial_i ,
\end{equation}
where
$$
\beta_{\alpha\bk}^i e^\alpha dx^\bk \partial_i
$$
is a Dolbeault representative of $[\beta]$. From the
differential-geometric viewpoint, it can be represented by the
second fundamental form of $Y$.

An equivalent description of this class using Cech cohomology goes
as follows. Let us choose local coordinates $x^i,z^\alpha$ on a
neighborhood $\hat Y$ of $Y$ so that $Y$ is given by $z^\alpha=0$,
$\alpha=1,\codim_\CC Y$. The normal bundle of $Y$ is spanned by
partial derivatives along $z^\alpha$ at $z^\alpha=0$. On an overlap
of two charts we have two sets of coordinates $x^i,z^\alpha$ and
$\tx^i,\tz^\alpha$. We consider a section of $\Hom(N_Y,T_Y)$ given
by
$$
\frac{\partial \tx^i}{\partial z^\alpha}\,\left(dz^\alpha \otimes
\frac{\partial}{\partial \tx^i}\right).
$$
It defines a cocycle $\beta$ with values in $\Hom(N_Y,T_Y)$. If
$\hY$ happens to be isomorphic to $N_Y$, one can choose the
coordinates $x^i$ on all charts so that on double overlaps $\tx^i$
depends only on $x^i$ and not on $z^\alpha$; then the cocycle
$\beta$ is trivial.

In the case when $Y$ is a complex Lagrangian submanifold of a
complex symplectic manifold $X$, $N_Y$ is isomorphic to $T_Y^*$, and
the class $[\beta]$ can be regarded as an element of $H^1(T_Y\otimes
T_Y)$. Moreover, an easy computation in holomorphic Darboux
coordinates shows that $[\beta]$ belongs to $H^1({\rm Sym}^2\,
T_Y)$, i.e. one can assume that $\beta^{ij}_\bk=\beta^{ji}_\bk$. A
particular Dolbeault representative of $[\beta]$ is given by
\begin{equation}\label{betaA}
\beta^{ij}_\bk=\smpf^{k' i}\partial_\bk A^j_{k'}.
\end{equation}
Note that $A^j_{k'}$ does not transform as a tensor, but its
$\bpartial$-differential does, so the form $\beta$ is not
necessarily exact.

The description of the formal neighborhood of $Y$ can be extended to
all orders in the normal coordinates. Let us choose local Darboux
coordinates $x^i,p_i$ such that $Y$ is given by $p_i=0$. In such
coordinates the transition functions must arise from a generating
function on an overlap of two coordinate charts, which upon Taylor
expansion in powers of $p_i$ can be regarded as a holomorphic
section of
$$
\oplus_\ell\, \Sym^\ell T_Y .
$$
In addition, the Taylor expansion must start with quadratic terms,
so that in new coordinates $Y$ is still given by the equation
$p_i=0$. Thus the transition functions can be encoded in an element
of
$$
\oplus_{\ell\geq 2} H^1(\Sym^\ell T_Y).
$$
We will denote by $[\beta^{(\ell)}]$ the $\ell^{\rm th}$ homogeneous
component of this class. The class $[\beta]$ we considered before is
the same as $[\beta^{(2)}]$. Note that multiplying the symplectic
form $\smpf$ by a factor $\lambda^2$ requires multiplying the
coordinates $p_i$ and the generating function by the same factor.
Hence the class $\beta^{(\ell)}$ has ghost number $2-2\ell$. This
will be useful later on.

Let us now explain how the obstruction arises at second order in
$W$. The key observation is that adding $W$ to the action modifies
the BRST transformation not only of $\chi_3^K$, but also of
$\chi^i_{1,2}$ and $\phi^i$. Indeed, one can easily see that unless
we modify the BRST transformation of $\phi^i$, the BRST
transformation (\ref{modifiedBRSTchi}) does not satisfy
$\delta_{Q,W}^2\chi_3=0$. We can rectify this by letting
$$
\delta_{Q,W}\phi^i(x^3)= -\theta(-x^3)\beta^{ij}_\bk\partial_j W
\eta^\bk,
$$
where $\theta(x)$ is a unit step-function, i.e. $\theta(x)=1$ if
$x\geq 0$ and $\theta(x)=0$ if $x<0$. Then
$$
\partial_3\delta_{Q,W}\phi^i=\delta(x^3)\beta^{ij}_\bk\partial_j W
\eta^\bk,
$$
and we see that $\delta_{Q,W}^2\chi_3=0$. Alternatively, we can
derive this formula from (\ref{phijump}), (\ref{betaA}) and
$$
\lim_{x^3\raa 0+} \delta_{Q,W}\phi^i(x^1,x^2,x^3)=0.
$$

Since $\Omega^{k' i}$ is implicitly proportional to the Planck
constant, we should regard this as the leading quantum correction to
the BRST transformation. Similarly, one can determine quantum
corrections to the BRST variations of the fields $\chi^i_{1,2}$.

Note that the variation of the bulk action is unaffected by the
modifications in the BRST transformations of the fields
$\phi^i,\chi^i_{1,2}$, because these modifications are concentrated
on the boundary and the momenta conjugate to these fields vanish on
the boundary. But the curving $W$ is no longer BRST-invariant:
$$
\delta_{Q,W} W=-\beta^{ij}_\bk \eta^\bk\partial_i W\partial_j W .
$$
Similarly, $W^{(2)}$ is no longer BRST-invariant up to total
derivatives and the equation of motion for $\chi_3^K$. To rectify
this, let us replace $W$ with a power series in the Planck constant
$W=W_0+W_1+\ldots$, where the term $W_n$ is of order $\hbar^n$. Then
BRST invariance requires $W_0$ to satisfy (\ref{Wclosed}), while
$W_1$ must satisfy
\begin{equation}\label{deltaW}
\bpartial W_1=\beta(dW_0,dW_0).
\end{equation}
This equation has solutions if and only if the cohomology class
\begin{equation}\label{Wobstruction}
[\beta(dW_0,dW_0)]\in H^{\hat 1}(\cO_Y)
\end{equation}
is trivial. This is the leading obstruction to the boundary
deformation corresponding to $W_0$.

An alternative derivation of the same result goes as follows.
Consider the path-integral with a boundary insertion
$$
W(0) \int_{\partial M} W^{(2)}.
$$
Naively, since the BRST-variation of $W^{(2)}$ is a total
derivative, and $W$ is BRST-invariant, this expression is also
BRST-invariant. But because of possible short-distance singularities
when the insertion point of $W^{(2)}$ approaches $0$, one should be
more careful. Let $D_\eps\subset \partial M$ be a small disk
centered at $0$. One can define the above product as the limit
$$
\lim_{\eps\raa 0} W(0)\int_{\partial M\backslash D_\eps} W^{(2)}.
$$
The BRST-variation $\delta_Q$ of this expression is
\begin{equation}\label{opeobstruction}
\oint W^{(1)} W(0),
\end{equation}
where integration is over a small circle on $\partial M$ centered at
$0$. If the operator product of $W^{(1)}$ and $W$ is singular, this
expression may be nonzero. In fact, as shown in appendix
\ref{app:fusion}, there is a singularity in the boundary operator
product of the fields $\chi^i\vert_{\partial M}$ and $\delta\phi^j$,
where $\delta\phi^j$ is a deviation of $\phi^j$ from its classical
background value. If $z$ is a local complex coordinate on $\partial
M$, the singularity in the OPE is
\begin{align}
\chi_z^i(z,\bz)\delta\phi^j(0) &\sim \frac{C}{z}\beta^{ij}_{\bk}\eta^\bk +\ldots,\\
\chi_\bz^i(z,\bz)\delta\phi^j(0) &\sim
-\frac{C}{\bz}\beta^{ij}_{\bk}\eta^\bk +\ldots ,
\end{align}
where $C$ is a constant. Therefore the operator product
(\ref{opeobstruction}) is proportional to
\begin{equation}\label{obstruction}
\beta^{ij}_\bk \eta^\bk \partial_i W\partial_j W.
\end{equation}
If this form vanishes identically, there is no obstruction at second
order in $W$. More generally, if the form does not vanish but is a
trivial class in $\bpartial$-cohomology, we can restore
BRST-invariance at second order by writing $W=W_0+W_1+\ldots$ as
above and requiring $W_1$ to satisfy (\ref{deltaW}).

If $Y$ is compact and K\"ahler, we can choose a representative of
the Dolbeault cohomology class $[W_0]$ which is annihilated by
$\partial$ (namely, the harmonic representative). This implies that
the Dolbeault cohomology class of $\partial W_0$ is trivial, and
therefore the obstruction (\ref{obstruction}) is trivial.

Let us compare this physical obstruction to the one arising from
deformation theory of complex Lagrangian submanifolds. As shown by
Kodaira \cite{Kodaira}, all obstructions to deforming a complex
submanifold $Y$ of a complex manifold $X$ take values in $H^1(N_Y)$.
In the case when $Y$ is Lagrangian, we have $N_Y\simeq T_Y^*$, and
obstructions take values in $H^1(\Omega^1_Y)$. If we want the
deformed manifold to be Lagrangian, we must require that the
holomorphic 1-form $\alpha$ on $Y$ corresponding to $\xi$ is
$\partial$-closed. Then a somewhat lengthy computation shows that
the first obstruction is the Dolbeault cohomology class of
$$
\partial (\beta(\alpha\otimes\alpha))\in \Omega^{1,1}_Y.
$$
In the situation considered above, $\alpha=\partial W$, where $W$ is
a holomorphic function, and we see that the obstruction to deforming
the submanifold is $\partial$ of (\ref{Wobstruction}). In the
compact K\"ahler case, the obstruction vanishes, in agreement with
\cite{Voisin}. Note that the geometric problem of deforming a
complex Lagrangian submanifold is a special case of the physical
problem of deforming the boundary condition in the RW theory: in the
former case $\partial W$ is a closed holomorphic 1-form, while in
the latter case it is a $\partial$ and $\bpartial$-closed form of
type $(1,\hat 0)$.

Deforming $Y$ as a submanifold of $X$ may deform the complex
structure of $Y$. In general, an infinitesimal deformation of the
complex structure of $Y$ is described by a class $[\mu]\in
H^1(T_Y)$. Its Dolbeault representative $\mu$ is called a Beltrami
differential. Given a Beltrami differential, we can deform the
$\bpartial$ operator on $Y$ by letting
$$
\bpartial^{new}=\bpartial - \mu(\partial).
$$

It is easy to see that the Beltrami differential $\mu$ corresponding
to a normal vector field $\xi$ on $Y$ is
$$
\mu=\beta(\xi),
$$
where we regard $\beta$ a section of
$Hom(N_Y,T_Y)\otimes\Omega^{0,1}_Y$. Therefore the deformation of
the $\bpartial$ operator on $Y$ corresponding to a curving $W$ is
$$
\bpartial^{new}=\bpartial - \beta^{ij}_\bk dx^\bk\partial_j
W\partial_i.
$$
The corresponding modification of the BRST transformation for the
fields $\phi^i$ and $\phi^\bi$ is
$$
\delta_{Q,W}\phi^\bi=\eta^\bi,\quad \delta_{Q,W}\phi^i=-
\beta^{ij}_\bk\partial_j W \eta^\bk.
$$
This is precisely what we found above for the BRST transformation of
the fields $\phi^\bi,\phi^i$ evaluated at $x^3=0$.

\subsection{Calabi-Yau fibrations}\label{sec:fibrations}

\subsubsection{Boundary degrees of freedom}

In the case of a two-dimensional sigma-model, an important class of
boundary conditions arises from vector bundles on the target space
or its submanifolds. These boundary conditions involve extra degrees
of freedom on the boundary and therefore are not really conditions
but boundary terms in the action which describe interactions between
bulk and boundary degrees of freedom.

A systematic way to deduce the existence of such boundary conditions
in the 2d sigma-model is to consider first boundary degrees of
freedom completely decoupled from the bulk. Such degrees of freedom
are described by ordinary quantum mechanics; in the topological
case, this means that states are described by elements of a
finite-dimensional vector space, or a graded vector space, while
observables are arbitrary linear operators on this space. If the
vector space has dimension $n$, this corresponds to a trivial
rank-$n$ vector bundle on the brane. Next one considers deformations
which introduce interactions between bulk and boundary degrees of
freedom. It turns out that the most general such deformation
corresponds to a complex of vector bundles over the brane; in the
B-model BRST-invariance requires the vector bundles to be
holomorphic and the differential to commute with the $\bpartial$
operator. This line of thought eventually leads one to identify
D-branes in the B-model with objects of the bounded derived category
of coherent sheaves.

Let us apply the same strategy to the RW model. If the boundary
degrees of freedom are decoupled from the bulk, they must be
described by a two-dimensional topological field theory. Since the
bulk theory resembles the B-model, it is very natural to consider a
B-model with a Calabi-Yau target $Z$ living on the boundary of $M$.
Recall that in general a deformation of a B-model is described not
in terms of classical geometry but by an $A_\infty$ Calabi-Yau
category \cite{KontsevichSoibelman,Costello}. To keep the discussion
concrete, we will limit ourselves to a particular class of
deformations, namely the ones involving even $\bpartial$-closed
forms of type $(0,p)$. As discussed in the appendix, the case $p=0$
corresponds to a superpotential deformation, while the case $p=2$
corresponds to a B-field. The advantage of this class of topological
field theories is that they can be described by a relatively simple
explicit action. More general deformations involving forms with
values in polyvector fields are possible, but the corresponding
deformations are difficult to write down in a closed form
\cite{Wittenmir}.

Calabi-Yau sigma-model with target $Z$ deformed by a holomorphic
function $W$ on $Z$ is known as a Landau-Ginzburg model. Of course,
in order for a nonconstant superpotential $W$ to exist, $Z$ must be
noncompact. This does not cause any problems provided the critical
set of $W$ is compact, and the resulting topological field theory is
very similar to a B-model with a compact target space. There is one
important distinction though: while the ordinary B-model is usually
regarded as $\ZZ$-graded (the grading is provided by the vector
R-charge), the Landau-Ginzburg model is only
$\ZZ_2$-graded.\footnote{Note that if one sets $W=0$ in the
Landau-Ginzburg model, one recovers not the usual B-model with a
noncompact target $Z$, but its $\ZZ_2$-graded version. In
particular, the resulting category of B-branes is not the bounded
derived category of $Z$, but the derived category of 2-periodic
complexes of $\cO_Z$-modules with coherent cohomology.} For this
reason, it is natural to consider, along with the superpotential
deformation, deformations by arbitrary even $(0,p)$ forms, $p>0$. It
is convenient to combine all these deformations into an
inhomogeneous even form on $Z$; BRST-invariance requires it to be
$\bpartial$-closed. We will call such a form the curving, and the
corresponding 2d TFT the curved B-model; its properties are
discussed in section \ref{app:curved}. Since the RW model is
$\ZZ_2$-graded, it is very natural to consider coupling it to a
curved B-model on the boundary.

We are thus led to consider boundary degrees of freedom
corresponding to a curved B-model with a Calabi-Yau target $Z$.
Allowing interactions between bulk and boundary degrees of freedom
amounts to deforming the boundary action by descendants of local
operators of the form
\begin{equation}\label{WYWZ}
W_Y\cdot W_Z,
\end{equation}
where $W_Y\in \oplus_r H^r(\cO_Y)$ is a boundary topological
observable constructed from the restrictions of bulk degrees of
freedom, and $W_Z$ is a topological observable constructed from
boundary degrees of freedom. For example, if the boundary degrees of
freedom are described by the ordinary B-model with a Calabi-Yau
target $Z$, $W_Z$ is an element of $\oplus_{p,q} H^p(\Lambda^q
T_Z).$ In general, the only constraint on the local operator
(\ref{WYWZ}) is that it must be even. One can think of such a
deformation as an even element of the vector space
\begin{equation}\label{bdrydefstotal}
\oplus_{s,r} H^s(\Lambda^r T^\ver_\Z),
\end{equation}
where $\Z=Y\times Z$ is regarded as a trivial fibration with base
$Y$ and fiber $Z$, and $T^\ver_\Z$ is the vertical tangent bundle of
$\Z$, i.e. the bundle of vector fields tangent to the fibers.

In particular, elements of the subspace
$$
H^1(T^\ver_\Z)
$$
correspond to infinitesimal deformations of the product $Y\times Z$
into a general complex fibration with base $Y$. We conclude that in
general boundary degrees of freedom can be thought of as taking
values in the fiber of such a fibration.

We will begin by constructing a BRST-invariant boundary action for
such degrees of freedom. Then we will consider a class of
deformations of the boundary action parameterized by the subspace
$$
\oplus_s H^s(\cO_\Z).
$$
More general deformations corresponding to graded components of the
vector space (\ref{bdrydefstotal}) with $r>0$ also exist, but are
more difficult to describe in a closed form.

\subsubsection{The Ehresmann connection}

To write down a boundary action which is invariant under a change of
local coordinates on $\Z$, we need to discuss connections on complex
fibrations. Let the boundary $\p M$ be mapped to the total space of
the fibration $p:\Z \rightarrow Y.$ The map $p$ is assumed to be
holomorphic, and a fiber of $p$ is a Calabi-Yau m-fold. As before,
$Y$ is a complex Lagrangian submanifold of a \HK\ target space $X.$
Let \begin{equation} \label{e1.7} \phi^{A},\quad A=1,\ldots,n+m,
\end{equation} be local complex coordinates on the fibration $\Z
\rightarrow Y$ so that \begin{equation} \label{e1.8} \phi^{i},\quad
i=1,\ldots,n, \end{equation} are local complex coordinates on the
base $Y$, while \be \label{e1.9} \phi^{a},\quad a=n+1,\dots,n+m
\end{equation}are local complex coordinates on the fiber.

A vector field on $\Z$ is called vertical if it is tangent to the
fibers of $p$, i.e. it belongs to the kernel of $dp$. In local
complex coordinates, a vertical $(1,0)$ vector field $\xi$ has the
form
$$
\xi^a\frac{\partial}{\partial \phi^a},
$$
where the functions $\xi^a$ may depend on both base and fiber
coordinates. Vertical vector fields form a smooth complex subbundle
of $T_\Z$ which we denote $T^\ver_\Z$. The quotient vector bundle
$T_\Z/T^{\ver}_\Z$ will be denoted $T^\hor_\Z$; it is naturally
isomorphic to $p^*T_Y$.

A 1-form on $\Z$ is called horizontal if it annihilates vertical
vector fields. In local coordinates a horizontal (1,0)-form has the
form
$$
\alpha_i\, d\phi^i.
$$
The bundle of horizontal (1,0)-forms is a subbundle of $T^*_\Z$
which we denote $T^{*\hor}_\Z$.

An Ehresmann connection on a smooth fibration $\Z$ is a splitting of
the tangent bundle $T_\Z$ into a sum \begin{equation}
\label{splitting} T^{\ver}_\Z\oplus T^\hor_\Z.\end{equation} We will
assume that this splitting is complex, i.e. respects the complex
structure of the bundles. Such a splitting always exists. We do not
assume that the splitting is holomorphic, since a holomorphic
splitting rarely exists. In local coordinates the splitting
$\xi\mapsto (\xi_\ver,\xi_\hor)$ is described by
\begin{equation} \label{e1.18a} \xi_{\ver} = \left(\xi^a - V_i^a
\xi^i\right)\frac{\partial}{\partial\phi^a}, \quad \xi_{\hor} =
\xi^i\frac{\partial}{\partial\phi^i}, \end{equation}
and similarly for vector fields of type $(0,1)$. Given an Ehresmann
connection $(V^a_i,V^\ba_\bi)$, we can define a horizontal vector
field on $\Z$ as a vector field which belongs to $T^\hor_\Z$. Any
vector field $\psi$ on the base $Y$ is lifted uniquely to a
horizontal vector field $\psi^\hor$ on $\Z$.

We define the covariant derivative of a section
$$
s:Y\raa \Z,\quad s:(\phi^i,\phi^\bi)\mapsto s^a(\phi^i,\phi^\bi)
$$
along a vector field $\xi=(\xi^i,\xi^\bi)$ on $Y$ by local formulas
\begin{equation}
(\nabla_\xi s)^a =\xi^i \nabla_i s^{a} +\xi^\bi \nabla_\bi s^a=
\xi^i\left(\p_{i}s^a - V_i^a\right)+\xi^\bi \partial_\bi s^a.
\end{equation}
Note that the covariant derivative in the $(0,1)$ direction
coincides with the ordinary derivative. This happens because we
assumed that the Ehresmann connection is compatible with the complex
structure of $T_\Z$ and as a consequence the mixed component of the
connection $V^a_\bi$ is zero.

In the theory of Ehresmann connections, the Lie algebra of vertical
vector fields plays the same role as the Lie algebra of the gauge
group in Yang-Mills theory. Thus we should think of $V^a_i d\phi^i$
as components of a horizontal 1-form on $\Z$ with values in vertical
vector fields. Similarly, the curvature of the Ehresmann connection
is a horizontal 2-form with values in vertical vector fields. In our
case, it only has $(1,1)$ and $(2,0)$ parts:
\begin{align}
\cR^a_{i \bar j} & = \p_{\bar j}V_i^a + V_{\bar j}^{\bar b} \p_{\bar
b} V_i^a\\
\cR^a_{ij} & = \p_j V_i^a -\p_i V_j^a+ V_j^b\p_b V_i^a-V_i^b\p_b
V_j^a.
\end{align}

\subsubsection{An affine connection on the fibration}

To construct a boundary action, we will need a K\"ahler metric on
the fibers of $\Z$ (actually, a Hermitian one would suffice) and an
affine torsion-free connection on the vector bundle $T_\Z$
compatible with the complex structure. In principle, any such
connection would do, and the topological correlators can be shown
not to depend on the choice. But to minimize the geometric input, it
is convenient to construct the affine connection starting from the
fiberwise K\"ahler metric, the K\"ahler metric on the base $Y$, and
an Ehresmann connection on $\Z$. Note that we do not require the
total space of the fibration to be K\"ahler: that would be an
unnecessary and rather strong restriction.

The vector bundles $T^{\ver}_\Z$ and $T^{\hor}_\Z$ have natural
connections. The connection on $T^{\hor}_\Z$ is a pull-back of the
Levi-Civita connection of the induced metric on $Y$. The
corresponding Christoffel symbols are
$$
\hat\Gamma^i_{jk}=\Gamma^i_{jk}-A^i_{i'}\Gamma^{i'}_{jk},
$$
where $A^i_{i'}$ is defined by (\ref{Adef}). Thus if $\xi$ is a
horizontal vector field and $\zeta$ is an arbitrary vector field, we
define
\begin{equation} \label{e1.19a} \nabla_{\zeta}\xi^i = \zeta^{A}\p_A \xi^i +
\hat\Gamma^i_{jk} \zeta^j \xi^k. \end{equation}
In order to describe the affine connection on $T^{\ver}_{\Z}$, it is
convenient to split $\zeta$ according to (\ref{e1.18a}) and define
covariant derivatives $\nabla^{\ver}_a$ for $\zeta_{\ver}^a \in
T^{\ver}_{\Z}$ and $\nabla^{\hor}_i$ for $\zeta_{\hor}^i \in
T^{\hor}_{\Z}$ separately:
\begin{equation} \label{e1.20a} \nabla^{\ver}_b \xi_{\ver}^a = \p_b \xi_{\ver}^a
+ \c^a_{bc}\xi_{\ver}^c,\quad \nabla^{\hor}_i \xi_{\ver}^a = \p_i
\xi_{\ver}^a + \left(L_V \xi_{\ver}\right)^{a}_{i}, \end{equation}
where the Lie derivative $L_V$ along the vector field $V_i^a$ is
defined as
\begin{equation} \label{e1.21a} \left(L_V \xi_{\ver}\right)^{a}_{i} =V_i^b \p_b
\xi_{\ver}^a -\xi_{\ver}^b \p_b V_i^a. \end{equation}
Hence if $\xi_{\ver}$ is a vertical vector field, then its total
covariant derivative is
\begin{equation} \label{e1.22a} \nabla_{\b} \xi_{\ver}^a =\p_{\b}
\xi_{\ver}^a+\c^a_{bc}\nabla_{\b}\phi^b \xi_{\ver}^c -\left(\p_b
V_i^a\right)\left(\p_{\b}\phi^i\right)\xi_{\ver}^b. \end{equation}

Since both components (\ref{splitting}) of $T_\Z$ have canonical
affine connections, they define a canonical affine connection on
$T_\Z$. It is given by
\begin{equation} \label{e1.23a} \tilde\nabla_{B} \xi^A = \p_B \xi^A+{\tilde
\cA}^A_{BC}\xi^C, \end{equation}
where
\begin{eqnarray}
\label{e1.24a} &{\tilde \cA}^i_{jk}= \hat\Gamma^i_{jk},\qquad
{\tilde \cA}^i_{bk}={\tilde \cA}^i_{jc}={\tilde \cA}^i_{bc}=0,\qquad
{\tilde \cA}^a_{bc} = \c^a_{bc},
\\
\label{e1.25a} &{\tilde \cA}^a_{kb} ={\tilde \cA}^a_{bk} = -
\c^a_{bc} V_k^c-\p_b V_k^a ,
\\
\label{e1.26a} & {\tilde \cA}^a_{jk} =  -\p_j V_k^a + V_k^c\p_c
V_j^a+\c^a_{bc}V_j^bV_k^c+V_i^a\hat\Gamma^i_{jk}.
\end{eqnarray}
This connection has torsion:
\begin{equation} \label{e1.27a} \cT^A_{BC} = {\tilde \cA}^A_{BC} - {\tilde
\cA}^A_{CB}, \end{equation}
whose only non-zero components are
\begin{equation} \label{e1.28a} \cT^a_{jk} = \cR^a_{jk}. \end{equation}
Hence a modified connection
\begin{equation} \label{e1.29a} \cA^A_{BC} = {\tilde \cA}^A_{BC} -
\frac{1}{2}\cT^A_{BC}, \end{equation}
is torsion-free. Its components are
\begin{eqnarray}
\label{e1.30a} &\cA^i_{jk} = \hat\Gamma^i_{jk},\qquad
\cA^i_{bk}=\cA^i_{jc}=\cA^i_{bc}=0,\qquad \cA^a_{bc} = \c^a_{bc},
\\
\label{e1.31a} &\cA^a_{kb}= \cA^a_{bk} = - \c^a_{bc}V_k^c - \p_b
V_k^a ,
\\
\label{e1.32a} & \cA^a_{jk}= \half \cR^a_{jk}
+\c^a_{bc}V_j^bV_k^c+V_i^a\hat\Gamma^i_{jk}.
\end{eqnarray}

\subsubsection{BRST transformations of boundary degrees of freedom}

We consider the following fields on $\partial M$:
\begin{align} \label{e3.1} & \text{bosonic:}\ \phi^A,\phi^{\bar
A}\in {\rm Map}(\Sigma,\Z), F^A\in\Gamma\left(\phi^*T_\Z\otimes
\Lambda^2
T^*_\Sigma\right),\\
& \text{fermionic:}\ H^{\bar A}\in\Gamma\left(\phi^*{\bar
T}_\Z\right), P^A\in\Gamma\left(\phi^*T_\Z\otimes
T^*_\Sigma\right),\theta_a\in\Gamma\left(\phi^*T^{\ver\,
*}_\Z\right).
\end{align}
The fields are essentially those of the B-model with target $\Z$.
Note though that instead of a fermionic 0-form taking value in the
pull-back of $T^*_\Z$, we have a field $\theta_a$ taking value in
the pull-back of the quotient bundle $T_\Z^{\ver \,
*}=T_\Z^*/T_\Z^{*\hor}$. The bosonic field $F^A$ is an auxiliary
field. The BRST transformations are taken to be
\begin{equation} \label{e3.2} \delta_Q\phi^{A} = 0,\quad \delta_Q\phi^{\bar A}
=H^{\bar A},\quad \delta_Q H^{\bar A}=0,\quad \delta_Q
P^A=d\phi^A,\quad \delta_Q\theta_a=0, \end{equation}
and for the auxiliary field
\begin{equation} \label{e3.3} \delta_Q F^A = \nabla P^A +
\frac{1}{2}\cR^A_{BC\bar D}P^B P^C H^{\bar D}. \end{equation}
In the latter formula the covariant derivative and its curvature are
those of (\ref{e1.30a}-\ref{e1.32a}): since the connection has zero
torsion, the BRST transformation satisfies $\delta_Q^2 F^A=0$. In
fact, $\delta_Q$ is nilpotent when acting on any field, not just
$F^A$.

Next we need to restrict the fields on the boundary so that for
fixed values of the bulk fields they describe a B-model whose target
is a fiber of $\Z$. We impose the following projection conditions:
\begin{equation} \label{e3.4} H^{\bar i} = \eta^{\bar i}\vert_{\partial M},\quad
P^i=\chi^i\vert_{\partial M},\quad F^i = vol_{\partial M}\smpf^{ij'}
A_{j'}^K g_{K \bar J} \p_3 \phi^{\bar J}. \end{equation}

The unrestricted parts of the fields can be parameterized by their
projections to $T^{\ver}\Z$. Hence we introduce the fields
\begin{equation} \label{3.5} \eta^{\bar a} = H^{\bar a} - V_{\bar i}^{\bar
a}\eta^{\bar i},\quad \rho^a=P^a-V_i^a\chi^i,\quad f^a = F^a -
V_i^aF^i. \end{equation}
Their BRST transformations can be computed to be \begin{equation}
\label{e3.6} \delta_Q \phi^a=0,\quad \delta_Q \phi^{\bar
a}=\eta^{\bar a}+V_{\bar i}^{\bar a}\eta^{\bar i},\quad \delta_Q
\theta_a=0\end{equation}
\begin{equation} \label{e3.7}  \delta_Q \eta^{\bar a}=-\left(\p_{\bar b} V_{\bar
i}^{\bar a}\right)\eta^{\bar b} \eta^{\bar i} +\half \cR^{\bar
a}_{\bar i \bar j} \eta^{\bar i}\eta^{\bar j},\quad \delta_Q
\rho^a=\nabla\phi^a-\cR^a_{i \bar j}\eta^{\bar j}\chi^i-
\left(\p_{\bar b} V_i^a \right)\eta^{\bar b}\chi^i,\end{equation}
\begin{multline} \label{e3.8}  \delta_Q f^a= \nabla\rho^a+\cR^a_{i \bar
j}d\phi^{\bar j}\chi^i+\left(\p_{\bar b}V_i^a\right)\nabla\phi^{\bar
b}\chi^i+\half \cR^a_{bc \bar d}\rho^b \rho^c\eta^{\bar
d}\\
-\nabla_c\left(\p_{\bar d} V_j^a\right)\chi^j \rho^c\eta^{\bar d}
-\half \nabla_j\left(\p_{\bar d} V_k^a\right)\chi^j\chi^k\eta^{\bar
d}+ \Bigl(-\nabla_c \cR^a_{j \bar l}+\left(\p_c V_{\bar l}^{\bar
b}\right)\left(\p_{\bar b}V_j^a\right)\Bigr)\chi^j\rho^c\eta^{\bar
l}\\
+\half \Bigl(-\nabla_j \cR^a_{k \bar l}+\left(\p_{\bar b}
V_{k}^{a}\right) \cR^{\bar b}_{\bar l j}\Bigr)\chi^j\chi^k\eta^{\bar
l} - vol_{\partial M}\Bigl( \left(\p_{\bar b}V_i^a\right)\eta^{\bar
b}+R^a_{i \bar j}\eta^{\bar j}\Bigr) \smpf^{ij'} A^K_{j'} g_{K\bar
J}\p_3\phi^{\bar J} .
\end{multline}
To compute the BRST transformation of $f^a$ we first used
(\ref{e3.3}) for BRST transformations of $F^a$ and $F^i$ and then
substituted the projected expressions (\ref{e3.4}). In
(\ref{e3.6}-\ref{e3.8}) we used
$$\nabla\phi^a=d\phi^a-V^a_i d\phi^i,$$
while the other covariant derivatives are defined as follows:
\begin{align} \label{3.9} \nabla\rho^a &= d\rho^a+
\c^a_{bc}\left(\nabla\phi^b\right)\rho^c -\left(\p_c
V_j^a\right)\left(d\phi^j\right)\rho^c+\half \cR^a_{kj}
\left(d\phi^k\right)\chi^j,
\\
\label{3.10} \nabla_c \left(\p_{\bar d}V_j^a\right) &= -\p_c\p_{\bar
d}V_j^a + \c^a_{bc} \p_{\bar d} V_j^a,
\\
\label{3.11} \nabla_j\left(\p_{\bar d} V_k^a\right) &= \p_j\p_{\bar
d}V_k^a - \hat\Gamma^i_{jk}\p_{\bar d}V_k^a+ \left( L_V\p_{\bar
d}V_k\right)^a_j,
\\
\label{3.12} \left( L_V\p_{\bar d}V_k\right)^a_j &=V_j^b\p_b
\p_{\bar d} V_k^a-\left(\p_{\bar d} V_k^b\right) \p_b V_j^a,
\\
\label{3.13} \nabla_{c}\cR^{a}_{j \bar l} &= \p_c\cR^a_{j \bar l} +
\c^a_{bc}\cR^b_{j \bar l},
\\
\label{3.14} \nabla_j\cR^a_{k \bar l} &= \p_j\cR^a_{k \bar l} -
\hat\Gamma^i_{jk} \cR^a_{i \bar l} + \left(L_V \cR_{k \bar
l}\right)^a_j,
\\
\label{3.15} \left(L_V \cR_{k \bar l}\right)^a_j&= V_j^b\p_b
\cR^a_{k \bar l}-\cR^b_{k \bar l}\p_b V_j^a.
\end{align}
Note that if the Ehresmann connection is zero,  $V_i^a=0,$ then the
equations (\ref{e3.6}-\ref{e3.8}) become BRST transformations of the
B-model with target space $Z.$

\subsubsection{The boundary action}

Now we can construct a covariant and BRST-invariant boundary action
$S^{bry}=\int_{\partial M} \cL^{bry}$ which for trivial Ehresmann
connection becomes the usual action of B-model: \begin{equation}
\label{e3.16} \cL^{bry}=\cL^{bry}_1+\cL^{bry}_2, \end{equation}
\be\label{e3.17} \cL^{bry}_1=\delta_Q\left(\cV^{bry}_1\right),\quad
\cV^{bry}_1=g_{a \bar b} \rho^a\wedge *\nabla\phi^{\bar
b},\end{equation}
\begin{multline} \label{e3.18} \cL^{bry}_1=vol_{\partial M} h^{\a \b}\Biggl[
g_{a \bar b}\left(\nabla_{\a}\phi^a\right)
\left(\nabla_{\b}\phi^{\bar b}\right)-g_{a \bar b} \rho_{\a}^a
\nabla_{\b}\eta^{\bar b}+ \left(\nabla_{\bar i}g_{a \bar
b}\right)\left(\nabla_{\a}\phi^{\bar b}\right)\eta^{\bar
i}\rho_{\b}^a\\
+\Bigl(\cR^{\bar b}_{\bar i j}\p_{\a}\phi^j+\cR^{\bar b}_{\bar i
\bar j}\p_{\a}\phi^{\bar j}+ \left(\p_c V_{\bar i}^{\bar
b}\right)\nabla_{\a}\phi^{c}\Bigr)g_{a \bar b}\eta^{\bar
i}\rho_{\b}^a- g_{a \bar b} \cR^a_{i \bar
j}\left(\nabla_{\a}\phi^{\bar b}\right)\eta^{\bar j}\chi_{\b}^i\\
-g_{a \bar b} \left(\p_{\bar c} V_{i}^a\right)
\left(\nabla_{\a}\phi^{\bar b}\right) \eta^{\bar
c}\chi_{\b}^i\Biggr]
\end{multline}
\begin{equation} \label{e3.20} \cL^{bry}_2=-\theta_a\delta_Q f^a.\end{equation}

BRST variation of the boundary action (\ref{e3.16}-\ref{e3.20}) is
given by
\begin{equation}\label{e3.21}
\delta_Q \cL^{bry}=\smpf^{i j'} \Bigl( \left(\p_{\bar
b}V_i^a\right)\eta^{\bar b}+\cR^a_{i \bar j}\eta^{\bar j}\Bigr)
\theta_a A^K_{j'} {\delta S^{bulk}\over \delta \chi_3^K}.
\end{equation}
Let us modify BRST transformation of the bulk field $\chi_3^K$ in
the following way:
\begin{equation} \label{e3.22}
\tdelta_Q\chi_3^K=\p_3 \phi^K- \delta(x_3) A^K_{j'} \smpf^{j'i}
\theta_a\Bigl( \left(\p_{\bar b}V_i^a\right)\eta^{\bar b}+\cR^a_{i
\bar j}\eta^{\bar j}\Bigr).
\end{equation}
Then the total action $S^{tot}=\int_M \cL^{bulk}+\int_{\partial M}
\cL^{bry}$ is BRST invariant, and moreover $\cL^{bulk}_1$ combines
with the last term in $\cL^{bry}_2$ into
$$\tdelta_Q\left( \cV^{bulk}_1\right),$$
so that the dependence of $S^{tot}$ on the metric on $M$ enters only
through BRST-exact terms.

The boundary term in the BRST variation of $\chi_3^K$ is
proportional to
$$
\left(\p_{\bar b}V_i^a\right)\eta^{\bar b}+\cR^a_{i \bar
j}\eta^{\bar j}=\partial_{\bar A} V^a_i H^{\bar A}
$$
The tensor
$$
\cF^a_{i\bar A}=\partial_{\bar A} V^a_i
$$
is a $\bpartial$-closed section of $\Omega^{0,1}(T^{\ver}_\Z\otimes
T^{\hor *}_\Z)$, and its cohomology class has a simple geometric
meaning: it is the Atiyah class of the fibration $\Z\raa Y$, i.e. an
obstruction to the existence of a holomorphic Ehresmann connection.

The Atiyah class also corrects the boundary BRST variations of other
fields. Let us consider the field $\phi^I$. The boundary action
(\ref{e3.20}) contains a term proportional to $\partial_3\phi^{\bar
J}$, therefore $\partial_3\phi^I$ has a delta-function singularity
on the boundary:
$$
\partial_3\phi^I=\delta(x^3)A^I_{k'}\smpf^{k'j}\theta_a H^{\bar A}\cF^a_{j\bar
A}.
$$
This means that $\phi^I$ itself has a step-like discontinuity on the
boundary. Using (\ref{betaA}) we find
\begin{equation}\label{tdeltaphii}
\tdelta_Q\phi^i=\theta(-x^3)\beta^{ij}_\bk \cF^a_{j\bar
A}\theta_a\eta^\bk  H^{\bar A}.
\end{equation}
This should be regarded as the leading quantum correction to the
BRST variation of $\phi^i$.

Once the BRST transformations for $\phi^i$ are modified, covariance
with respect to holomorphic changes of coordinates on $\Z$ requires
one to modify BRST transformations for $\phi^a$. In mathematical
terms, one needs to lift the cohomology class $[\beta\cF]\in
H^2(T^{\hor}_\Z\otimes T^{\ver}_\Z)$ to $H^2(T_\Z\otimes
T^{\ver}_\Z)$. That is, we would like to find a $\bpartial$-closed
section of $T_\Z\otimes T^{\ver}_\Z\otimes\Omega^{0,2}_\Z$ whose
image under the projection $T_\Z\raa T^{\hor}_\Z$ is the section
$$
\beta^{ij}_\bk\cF^a_{j\bar A} \partial_i\otimes \partial_a\otimes
d\phi^\bk\wedge d\phi^{\bar A}\in \Omega^{0,2}(T^{\hor}_\Z \otimes
T^{\ver}_\Z).
$$
From the short exact sequence
$$
0\raa T^{\ver}_\Z\raa T_\Z\raa T^{\hor}_\Z\raa 0
$$
we see that there is an obstruction for doing this taking values in
$H^3(T^{\ver}_\Z\otimes T^{\ver}_\Z)$.

To write the obstruction and the modified transformation law for
$\phi^a$ more explicitly, we will adopt the following notation. We
will regard inhomogeneous forms of type $(0,p)$ as functions on the
odd tangent bundle of $\Z$, i.e. as functions of odd fields $H^{\bar
A}$, and will keep this dependence implicit. Thus $\beta^{ij}$ will
mean $\beta^{ij}_\bk \eta^\bk$, and $\cF^a_j$ will mean
$\cF^a_{j\bar A} H^{\bar A}$. Then the most general covariant
transformation law for $\phi^a$ can be written as
$$
\tdelta_Q\phi^a=V^a_i\beta^{ij}\cF^b_j\theta_b +\rho^a,
$$
where $\rho^a$ is a section of $T^{\ver}_\Z$ depending on other
fields. Requiring $\tdelta_Q^2\phi^a=0$ we find that $\rho^a$ has
the form
$$
\rho^a=\rho^{ab}\theta_b,
$$
where $\rho^{ab}$ satisfies
\begin{equation}\label{rhoeq}
\bpartial\rho^{ab}=\beta^{ij}\cF^a_i\cF^b_j.
\end{equation}
This means that the cohomology class of
\begin{equation}\label{obstructioncF}
\beta^{ij}\cF^a_i\cF^b_j\in\Omega^{0,3}(\Lambda^2 T^{\ver}_\Z),
\end{equation}
must be trivial. This cohomology class is an obstruction for
coupling the B-model with target $\Z$ and the RW model with target
$X$.

If the obstruction vanishes, then one picks a particular solution
$\rho^{ab}$ of (\ref{rhoeq}) and defines the modified BRST
transformation laws for $\phi^a$ by
\begin{equation}\label{tdeltaphia}
\tdelta_Q\phi^a=\left(\rho^{ab}+V^a_i\beta^{ij}\cF^b_j\right)\theta_b
.
\end{equation}
One may always choose $\rho^{ab}$ to be antisymmetric.

BRST transformations of the fields $\theta_a$ and $P^A$ also need to
be corrected to ensure that the boundary action is BRST-invariant.
In effect, this amounts to a further deformation of the boundary
B-model. We will not pursue this issue here.

\subsubsection{Boundary local observables}
Local observables on the boundary are constructed in complete
analogy with the B-model:
$$
\cO_\omega=\omega_{{\bar A}_1\ldots {\bar A}^s}^{a_1\ldots a_r}
H^{{\bar A}_1}\ldots H^{{\bar
A}_s}\theta_{a_1}\ldots\theta_{a_r},\quad \omega\in
\Omega^{0,s}\left(\Lambda^r T^{\ver}_\Z\right).
$$
On the classical level, the BRST operator acts by
$$
\delta_Q \cO_\omega=\cO_{\bpartial\omega}.
$$
Thus we may identify the classical algebra of boundary topological
observables with
\begin{equation}\label{classicalobs}
\oplus_{s,r} H^s(\Lambda^r T^\ver_\Z)
\end{equation}

As discussed above, the BRST differential receives quantum
corrections. Eqs.~(\ref{tdeltaphii}) and (\ref{tdeltaphia}) tell us
how the BRST variation of $\phi^A$ is modified to leading order in
the Planck constant. Since we have not computed how the BRST
variation of $\theta_a$ is modified, we will only consider the
observables which do not depend on $\theta_a$. More precisely, we
expand an observable $W$ into a Taylor series in the Planck
constant, $W=W_0+W_1+\ldots ,$ where $W_k$ is of order $\hbar^k$,
and require $W_0$ to depend only on $\phi^A,\phi^{\bar A}$ and
$H^{\bar A}$. Then BRST-invariance requires $W_0$ to satisfy
$$
\bpartial W_0=0,
$$
while $W_1$ satisfies
$$
\bpartial W_1+\beta^{ij}\cF^b_j \theta_b\nabla_i W_0 +
\rho^{ab}\partial_a W_0\theta_b =0.
$$
From this equation we see that $W_1$ must have the form
$W_1=\psi^a\theta_a$, where $\psi^a$ is an element of
$\Omega^{0,{\hat 1}}(T^{\ver}_\Z)$ satisfying
\begin{equation}\label{psieq}
\bpartial\psi^a=-\beta^{ij}\cF^a_j \nabla_i W_0 +\rho^{ab}\partial_b
W_0.
\end{equation}
It follows from (\ref{rhoeq}) that the right-hand-side of
(\ref{psieq}) is $\bpartial$-closed and therefore defines a class in
$H^{\hat 0}(T^{\ver}_\Z)$. For the equation (\ref{psieq}) to admit
solutions, this cohomology class must be trivial. One can interpret
this mathematically by saying that there is a spectral sequence
which converges to the quantum algebra of boundary observables whose
first term is given by (\ref{classicalobs}) and the action of the
first differential on the subspace with $r=1$ is given by the
right-hand-side of (\ref{psieq}).

\subsubsection{Deformations of the boundary action}
Infinitesimal deformations of the boundary action correspond to even
elements of the algebra of boundary observables. Let us again
consider an observable $W=W_0+W_1+\ldots$ such that $W_0$ does not
depend on $\theta_a$. On the classical level, we may drop all the
terms except $W_0$, and the BRST invariance requires $W_0$ to
satisfy $\bpartial W_0=0$. The corresponding deformation of the
boundary action is obtained by applying the descent procedure. The
results are very much like in subsection \ref{sec:defs}, with $Y$
replaced with the total space of $\Z$:
\begin{align}\label{W1}
W^{(1)}& =P^A \partial_A W +d\phi^{\bA}\frac{\partial
W}{\partial H^\bA},\\
W^{(2)}&=\frac12 P^A P^B \nabla_A \partial_B W+P^A d\phi^\bB
\partial_A \frac{\partial W}{\partial H^\bB}+\frac12 d\phi^\bA
d\phi^\bB \frac{\partial^2 W}{\partial H^\bA\partial
H^\bB}\\
 & - vol_{\partial M} \smpf^{k'i}\left(\partial_i
W+V_i^a\partial_a W\right) A^K_{k'} g_{K\bar J}\partial_3\phi^{\bar
J}.
\end{align}
The descendants satisfy
$$
\delta_Q W^{(1)}=dW,\quad \delta_Q W^{(2)}=dW^{(1)}+\partial_a
W\frac{\delta S^{bry}}{\delta\theta_a}+ \smpf^{k' i}
\left(\partial_i W+V_i^a\partial_a W\right)A^K_{k'}\frac{\delta
S^{bulk}}{\delta \chi_3^K}.
$$
We define the deformed action to be
$$
S^{tot\, def}=\int_M \cL^{bulk}+\int_{\partial M}
\cL^{bry}+\int_{\partial M} W^{(2)}.
$$
and modify the BRST transformation for the fields $\chi_3^K$ and
$\theta_a$:
\begin{align}
\tdelta_{Q,W}\chi_3^K&=\partial_3\phi^K - \delta(x_3)A^K_{k'}
\smpf^{k'i} \theta_a H^{\bar A}\cF^a_{i\bar A}-\delta(x^3) A^K_{k'}
\smpf^{k'
i} \left(\partial_i W+V_i^a\partial_a W\right),\\
\tdelta_{Q,W}\theta_a &=-\partial_a W .
\end{align}
Then the total action is BRST-invariant up to a total derivative:
$$
\tdelta_{Q,W} S^{tot\, def}=\int_{\partial M} dW^{(1)}.
$$
In particular, if we take $W$ to be a holomorphic function on $\Z$,
the deformed action describes a family of Landau-Ginzburg models
fibered over $Y$, with the superpotential depending both on the
fiber and base coordinates. In general, we will call $W$ the
curving, since in section \ref{app:curved} the B-model deformed by
an even inhomogeneous $\bpartial$-closed form is called the curved
B-model.

As discussed above, for a given $W_0$ there are obstructions for
finding $W_1,W_2,...,$ satisfying the requirements of BRST
invariance. However even if all these obstructions vanish, there may
be an obstruction to extending an infinitesimal deformation
$W=W_0+W_1+\ldots$ to a finite deformation already at second order
in $W$. This is a quantum effect which arises because adding
$W^{(2)}$ to the action necessitates a change in the BRST
transformations of fields on the boundary. As in section
\ref{sec:obstructions}, we can determine the modification of the
BRST transformation of $\phi^i$ by noting that $\phi^I$ has a
discontinuity at $x^3=0$:
$$
\partial_3\phi^K=\delta(x^3)
A^K_{k'}\smpf^{k'j}\left(-\cF^a_j\theta_a+\nabla_jW_0\right)
$$
Then the BRST variation of $\phi^i$ is given by
$$
\tdelta_{Q,W}\phi^i=\theta(-x^3)\beta^{ij}\left(\cF^a_j\theta_a-\nabla_jW_0\right).
$$
(We replaced $W$ with $W_0$ because we are only interested in
leading quantum corrections to BRST transformations). It is easy to
see that this ensures $\tdelta_{Q,W}^2\chi_3^K=0$. One can also
check that $\tdelta_{Q,W}^2\phi^i=0$.

Once the BRST transformation of $\phi^i$ has been modified,
covariance with respect to changes of coordinates requires us to
modify the BRST transformation of $\phi^a$. The most general
covariant BRST variation is
$$
\tdelta_{Q,W}\phi^a=V^a_i \tdelta_{Q,W}\phi^i+\rho_W^a,
$$
where $\rho_W^a$ is a field-dependent and possibly $W$-dependent
section of $T^{\ver}_{\Z}$. Requiring $\tdelta_{Q,W}^2\phi^a=0$, we
find that $\rho_W^a$ must have the form
$$
\rho_W^a=\rho_W^{ab}\theta_b+\psi_W^a,
$$
where $\rho^{ab}_W$ and $\psi_W^a$ are independent of $\theta_a$ and
satisfy
\begin{align}\label{rhoWeq}
\bpartial\rho^{ab}_W &=\beta^{ij}\cF^a_i\cF^b_j,\\
\bpartial\psi^a_W &=\rho^{ab}\partial_bW_0-\beta^{ij}\cF^a_i\nabla_j
W_0.
\end{align}
Comparing these equations with (\ref{rhoeq}) and (\ref{psieq}) we
see that particular solutions are obtained by letting
$$
\rho^{ab}_W=\rho^{ab},\quad \psi_W^a=\psi^a.
$$
We now found the leading quantum corrections to BRST transformations
of $\phi^A$ including terms linear in $W_0$. The obstruction
quadratic in $W_0$ arise from the requirement that $W$ be
BRST-invariant with respect to these corrected transformations. This
requirement implies that $W_1$ must have the form
$$
W_1=\psi^a\theta_a+w,
$$
where $w$ is independent of $\theta_a$ and satisfies
$$
\bpartial w=-2\psi^a\partial_a W_0+\beta^{ij}\nabla_i W_0\nabla_j
W_0.
$$
The right-hand-side of this equation is $\bpartial$-closed, thanks
to (\ref{psieq}) and the antisymmetry of $\rho^{ab}$. Thus it
defines a cohomology class in $H^{\hat 1}(\cO_\Z)$ which is an
obstruction for the existence of $w$. This is the first obstruction
for extending an infinitesimal deformation $W$ to a finite
deformation; as expected, it is quadratic in $W$.

\subsection{Reduction on a circle in the presence of
boundaries}\label{sec:redcircle}

Let $M=S^1\times\Sigma$, where the Riemann surface $\Sigma$ has a
nonempty boundary. Since the dimensional reduction of the RW model
is the B-model with the same target, every boundary condition in the
RW model gives rise to a boundary condition for the B-model. In this
subsection we will determine which B-branes can be obtained in this
way.

First let us consider the case without boundary degrees of freedom,
i.e. when $\Z=Y$. Let $S^1$ be parameterized by $x^2\in [0,2\pi)$.
The reduction to a 2d field theory occurs in the limit when the
circumference of $S^1$ goes to zero, i.e. $h_{22}\raa 0$. We
decompose the 1-form $\chi$ into components along $\Sigma$ and
$S^1$:
$$
\chi^I=\chi^I_\Sigma+\chi^I_2 dx^2,
$$
Reduction along $S^1$ amounts to requiring all fields to be
independent of $x^2$, so $\chi^I_\Sigma$ is a pull-back of a 1-form
$\rho^I$ on $\Sigma$. We also define
$$
\theta_I=\int \smpf_{IJ} \chi_2^{J} dx^2.
$$
The reduced model also contains fields $\eta^{\bar I},\phi^I$ and
$\phi^{\bar I}$. Their BRST transformations are
$$
\delta\phi^I=0,\quad \delta\phi^{\bar I}=\eta^{\bar I},\quad
d\rho^I=d\phi^I,\quad \delta\theta_I=0.
$$
These are the BRST transformations of the B-model. On the boundary
$\partial\Sigma$ the bosonic fields $\phi^I,\phi^{\bar I}$ take
values in $Y\subset X$, while the conjugate momenta satisfy
$$
\pi_{\phi^{i}}=\frac{\partial\cL^{bulk}}{\partial
\left(\partial_3\phi^{i}\right)}=0,\quad
\pi_{\phi^{\bi}}=\frac{\partial\cL^{bulk}}{\partial
\left(\partial_3\phi^{\bi}\right)}=0.
$$
The fermionic fields satisfy the boundary conditions
$$
\eta^{\bi'}=0,\quad \theta_i=0,\quad g_{\bi J}\rho_3^J=0,\quad
\rho_1^{i'}=0.
$$
These are precisely the boundary conditions for a B-brane
corresponding to the submanifold $Y\subset X$.

In general, when boundary degrees of freedom are present, we have to
perform the reduction both for bulk and boundary degrees of freedom.
For simplicity, let us consider the case when the fiber of $\Z$ is
compact. If the bulk fields are fixed, the space of states of
boundary degrees of freedom is the space of states of the curved
B-model whose target is the fiber of $\Z$. If the fiber is K\"ahler,
this space is the de Rham cohomology of the fiber, with the natural
$\ZZ_2$ grading (see section \ref{app:curved}). As one varies the
boundary values of the bulk fields, these $\ZZ_2$-graded vector
spaces fit into a vector bundle which can be described as fiberwise
de Rham cohomology. This vector bundle is flat, the flat connection
being the Gauss-Manin connection. We conclude that a Lagrangian
submanifold $Y$ equipped with a fibration $\Z$ whose fibers are
compact and K\"ahler reduces to a B-brane $Y$ equipped with a flat
vector bundle (the fiberwise de Rham cohomology of $\Z$).

\subsection{Reduction on an interval}\label{sec:intervalred}

We have explained above that the RW theory with target $X$ on
$S^1\times\Sigma$ is equivalent to the B-model with the same target
on $\Sigma$. We can also consider the RW model on a manifold of the
form $I\times \Sigma$, where $I$ is an interval. The boundary of
this 3-manifold is a disjoint union of two copies of $\Sigma$, so we
can choose two different boundary conditions corresponding to two
complex Lagrangian submanifolds $Y_1$, $Y_2$ equipped with
fibrations $\Z_1,\Z_2$ over them. In this situation the RW theory
must reduce to some 2d topological field theory on $\Sigma$. In this
subsection we will determine this theory. This will enable us in the
next section to identify the categories of line defects separating
different boundary conditions.

\subsubsection{Identical submanifolds at the two ends}

Since the general case is rather complicated, we will first consider
some special cases. The simplest case is $Y_1=Y_2=Y$, and the
boundary degrees of freedom are absent: $\Z_1=\Z_2=Y$. The boundary
conditions restrict the bosonic fields to the submanifold $Y$. If we
expand all bosonic fields into Fourier series on the interval $I$,
then all the modes for $\phi^{i'}$ will be massive, while the
constant mode for $\phi^i$ will be massless. The masses of the
massive modes are of order $1/\sqrt {h_{33}}$, so in the limit
$h_{33}\raa 0$ one can set $\phi^{i'}=0$ and assume that the fields
$\phi^i$ are constant on $I$. Similarly, the massless fermionic
fields are constant modes of $\eta^\bi,\chi_{1,2}^i,$ and
\be\label{theta} \theta_i=\int \smpf_{ij'}\chi_3^{j'} dx^3.
\end{equation} These are the fields of the B-model with target $Y$.
It is easy to check that the BRST transformations and the action are
also those of the B-model with target $Y$.

The above derivation of the effective 2d theory was classical; now
let us analyze possible quantum corrections. As in the case of the
circle reduction, on the quantum level it is not correct simply to
set nonconstant modes to zero: one must integrate them out, and this
could induce corrections to the effective 2d theory. We will now
show that ghost number symmetry does not allow any such corrections.
Quantum corrections would deform the B-model while preserving the
ghost number symmetry and therefore would take values in
\begin{equation}\label{defsBmodelY}
\oplus_{p+q=2} H^p(\Lambda^q T_Y).
\end{equation}
The corresponding class must be built from the curvature tensor of
$Y$, which represents a class in $H^1(\Omega^1\otimes\End\, T_Y)$,
and the classes $\beta^{(\ell)}\in H^1(\Sym^\ell T_Y)$, $\ell\geq
2$, describing the formal neighborhood of $Y$. Consider a monomial
which contains $m$ curvature tensors and the classes
$\beta^{(\ell_i)}$, $i=1,\ldots,N$. The integers $p$ and $q$ in
(\ref{defsBmodelY}) are given by
$$
p=m+N,\quad q=\sum_i\ell_i - m.
$$
Since $p+q=2$, we must have
$$
\sum_i \ell_i +N=2.
$$
But since $\ell_i\geq 2$ $\forall i$ and $N>0$, this equality cannot
be satisfied.

The quantum B-model with target $Y$ is anomalous unless $Y$ is a
half-Calabi-Yau manifold, i.e. unless the square of the canonical
line bundle $K_Y$ is trivial. Since a consistent 3d topological
field field theory must reduce to a consistent 2d topological field
theory, we conclude that the RW model on a 3-manifold with boundary
is anomalous unless $Y$ is a half-Calabi-Yau manifold. In fact, it
is easy to see that the RW model is anomalous unless $Y$ is a
Calabi-Yau manifold. Consider the path-integral on a ball $D^3$ on
whose boundary we specify a boundary condition associated to $Y$.
The bosonic zero modes are $\phi^i,\phi^{\bar i}$, $i=1,\ldots,n$,
the fermionic zero modes are $\eta^{\bar i}$, $i=1,\ldots,n$, and a
BRST-invariant measure
$$
\omega_{i_1\ldots i_n}d\phi^{i_1}\ldots d\phi^{i_n} d\phi^{\bar
1}\ldots d\phi^{\bar n} d\eta^{\bar 1}\ldots d\eta^{\bar n}
$$
requires a choice of a holomorphic volume form $\omega_{i_1\ldots
i_n}$ on $Y$. Thus $K_Y$ must be trivial.

\subsubsection{Including the curving}

Now let us keep $\Z_1=\Z_2=Y_1=Y_2=Y$, but allow for nontrivial
curvings $W_1$ and $W_2$ on $Y_1$ and $Y_2$. An important difference
compared to the previous case is that the field $\phi^K$ has
step-like discontinuities at $x^3=0$ and $x^3=1$:
\begin{align}\label{discW12}
\lim_{x^3\raa 0+}\phi^K(x^1,x^2,x^3)-\phi^K(x^1,x^2,0)& =A^K_{k'}\smpf^{k' i}\partial_i W_1,\\
\lim_{x^3\raa 1-}\phi^K(x^1,x^2,x^3)-\phi^K(x^1,x^2,0)&
=A^K_{k'}\smpf^{k' i}\partial_i W_2.
\end{align}
As a consequence, the field $\phi^I$ on the interval $(0,1)$ will
not be constant either, even in the limit when its length goes to
zero. To determine the reduced model, a convenient short-cut is to
focus on BRST transformations of the fields. The action of the
B-model will be determined by these BRST transformations uniquely,
up to BRST-exact terms.

The bosonic fields of the reduced model are
$$
\phi_0^i=\int_0^1 \phi^i dx^3,\quad \phi_0^\bi=\int_0^1 \phi^\bi
dx^3.
$$
The fermionic 0-form fields are
$$
\eta_0^\bi=\int_0^1 \eta^\bi dx^3,\quad \theta_i=\int_0^1
\smpf_{ik'}\chi_3^{k'} dx^3.
$$
The fermionic 1-form fields are
$$
\rho^i=\int_0^1 \chi^i_\Sigma dx^3.
$$
The BRST variations of the fields are
\begin{align}\label{BRSTreducedW}
\delta_{Q,W}\phi_0^i & =0, & \delta_{Q,W}\phi_0^\bi &=\eta_0^\bi,\\
\delta_{Q,W}\eta_0^\bi &= 0, &
\delta_{Q,W}\theta_i &=-\partial_i(W_1-W_2),\\
\delta_{Q,W}\rho^i &=d\phi_0^i, &
\end{align}
where $W_{1,2}$ are regarded as functions of $\phi_0^i,\phi_0^\bi,$
and $\eta_0^\bi$.

The field content and BRST transformations of the reduced model
appear to be the same as in the curved B-model with target $Y$ and
curving $W_1-W_2$. But on the quantum level there are corrections to
this statement which originate from the fact that in general the
formal neighborhood of $Y$ differs from $T^*_Y$. Indeed, the fields
$\phi^{i'}$ have discontinuities at $x^3=0,1$, and since the
transition functions for $\phi^i$ depend on the values of
$\phi^{i'}$, this affects the transition functions for $\phi_0^i$,
i.e. the complex structure of the target of the B-model. To leading
order in $\phi^{i'}$, this modification of the complex structure of
$Y$ is represented by a Cech 1-cocycle with values in $T_Y$ which
represents the cohomology class $[\beta(\xi)]$, where $\beta\in
H^1(T_Y\otimes N_Y^*)$ is defined by the exact sequence
(\ref{exactsequenceY}), and $\xi$ is the normal vector field on $Y$
specifying the discontinuity of $\phi^{i'}$.

In our case $\xi$ is given by
$$
\xi^{k'}=\smpf^{k'i}\partial_i W,
$$
where $W$ is either $W_1$ or $W_2$ depending on whether we take
$x^3=0$ or $x^3=1$. In the limit $h_{33}\raa 0$ we may assume that
$\phi^{i'}$ depends linearly on $x^3$, so the deformation of the
transition function for $\phi_0^i$ is the average of the
deformations at $x^3=0$ and $x^3=1$.\footnote{For $W_1\neq W_2$, we
cannot assume that $\phi^{i'}$ is independent of $x^3$ because such
an ansatz would not satisfy the boundary conditions.} The conclusion
is that the reduced theory is a B-model whose target is a
deformation of $Y$ corresponding to the cohomology class
$$
\left[\frac12\beta(\partial(W_1+W_2)\right].
$$

Note that since $W_{1,2}$ are not simply functions, but
$\bpartial$-closed $(0,\hat 0)$-forms, $\xi$ is not really a normal
vector field but a $\bpartial$-closed section of $\Omega^{0,\hat
0}(N_Y)$. Consequently, the class above takes values not in
$H^1(T_Y)$, but in $H^{\hat 1}(T_Y)$, and the modified ``transition
functions'' are not holomorphic functions but invertible
$\bpartial$-closed inhomogeneous forms of type $(0,p)$ with even
$p$. If desired, one can make a change of field variables
$\phi_0^i,\phi_0^\bi$ and $\eta_0^\bi$ which eliminates such strange
``transition functions'', but at the expense of modifying the BRST
transformations for $\phi^i_0$. If $\beta$ is a Dolbeault
representative of $[\beta]$, the corresponding modification of the
BRST transformation for $\phi_0$ will read
$$
\delta\phi_0^i=-\frac12\beta^{ij} \partial_j(W_1+W_2)
$$
Since the curvings $W_1,W_2$ depend on the fields $\eta_0^\bj$, the
right-hand-side of this formula may contain arbitrary odd powers of
$\eta_0^\bj$.

\subsubsection{Including the fibrations}

The next case is $Y_1=Y_2=Y$, but the fibrations $\Z_1, \Z_2$ are
otherwise arbitrary. We also allow for curvings $W_1$ and $W_2$ on
$\Z_1$ and $\Z_2$. For simplicity we will neglect the effects of the
class $\beta$, i.e. the discussion will be classical. As before, in
the 2d limit the bosonic fields $\phi^{i'}$ vanish, while the
boundary degrees of freedom at $x^3=0$ and $x^3=1$ are not affected
by the reduction. Thus the reduced theory will be a sigma-model with
target $\Z_1\times_Y \Z_2$, where $\times_Y$ denotes the fibered
product. The map from $\Sigma$ to $\Z_1\times_Y \Z_2$ can be
described by the bosonic fields
$\phi^A=(\phi^i,\phi_1^{a_1},\phi_2^{a_2})$.

Let $\pi_1$ and $\pi_2$ be the projections from $\Z_1\times_Y \Z_2$
to $\Z_1$ and $\Z_2$, respectively, and $p_{12}$ be the projection
from $\Z_1\times_Y \Z_2$ to $Y$. The fermionic fields of the reduced
model come from the boundary fermionic fields and the constant
Fourier modes of the bulk fields. The former give rise to
\begin{align}
\eta_1^{\ba_1}&\in\Gamma(\phi^*\pi_1^*{\bar T}^\ver_{\Z_1}), &
\rho_1^{a_1}&\in \Gamma(\phi^*\pi_1^*T^\ver_{\Z_1}\otimes
T^*_\Sigma), & \theta_{1a_1}&\in \Gamma(\phi^*\pi_1^*
T^{\ver *}_{\Z_1}),\\
\eta_2^{\ba_2}&\in\Gamma(\phi^*\pi_2^*{\bar T}^\ver_{\Z_2}), &
\rho_2^{a_2}&\in \Gamma(\phi^*\pi_2^*T^\ver_{\Z_2}\otimes
T^*_\Sigma), & \theta_{2a_2}&\in \Gamma(\phi^*\pi_2^* T^{\ver
*}_{\Z_2}).
\end{align}
The latter give rise to the 0-form
$$
\eta^{\bar i}\in \Gamma(\phi^*{\bar T}_Y),
$$
the 1-form
$$
\rho^i\in\Gamma(\phi^*T_Y\otimes T^*_\Sigma)
$$
which is the constant Fourier mode of the component of $\chi^i$
along $\Sigma$, and the 0-form
$$
\theta_i=\int \Omega_{ij'}\chi_3^{j'} dx^3 \in \Gamma(\phi^*T^*_Y).
$$
Their BRST transformations are
\begin{align*}
\delta_{Q,W}
\phi_1^{\ba_1}&=\eta_1^{\ba_1}+V^{\ba_1}_{1\bi}\eta^\bi,
&\delta_{Q,W} \phi_1^{a_1}&=0,\\
\delta_{Q,W} \eta_1^{\ba_1}&=-\left(\p_{\bb_1} V_{1\bi}^{\ba_1}
\right)\eta_1^{\bb_1} \eta^{\bi} +\half \cR^{\ba_1}_{1\bi\bj}
\eta^{\bar i}\eta^{\bar j},
&\delta_{Q,W}\theta_{1a_1}&=-\partial_{a_1}W_1,\\
\delta_{Q,W} \rho_1^{a_1}&=\nabla\phi_1^{a_1}-\cR^{a_1}_{1i \bar
j}\eta^{\bar j}\rho^i- \left(\p_{\bb_1} V_{1i}^{a_1}
\right)\eta_1^{\bb_1}\rho^i , & &\\
\delta_{Q,W}
\phi_2^{\ba_2}&=\eta_2^{\ba_2}+V^{\ba_2}_{2\bi}\eta^\bi,
&\delta_{Q,W} \phi_2^{a_2}&=0,\\
\delta_{Q,W}\eta_2^{\ba_2}&=-\left(\p_{\bb_2} V_{2\bi}^{\ba_2}
\right)\eta_2^{\bb_2} \eta^{\bi} +\half \cR^{\ba_2}_{2\bi\bj}
\eta^{\bar i}\eta^{\bar j},
&\delta_{Q,W} \theta_{2a_2}&=\partial_{a_2}W_2,\\
\delta_{Q,W}\rho_2^{a_2}&=\nabla\phi_2^{a_2}-\cR^{a_2}_{2i \bar
j}\eta^{\bar j}\rho^i- \left(\p_{\bb_2} V_{2i}^{a_2}
\right)\eta_2^{\bb_2}\rho^i, & &\\
\delta_{Q,W}\rho^i &=d\phi^i, &\delta_Q\eta^\bi &=0,
\end{align*}
\begin{multline}\label{BRSTtransftheta}\delta_{Q,W}\theta_i=\left(\p_{\bar
b_1}V^{a_1}_{1i}\eta_1^{\bar b_1}+\cR^{a_1}_{1i \bar j}\eta^{\bar
j}\right)\theta_{1a_1} +\left(\p_{\bar b_2}V^{a_2}_{2i}\eta_2^{\bar
b_2}+\cR^{a_2}_{2i \bar j}\eta^{\bar
j}\right)\theta_{2a_2}\\
-\left(\partial_iW_1+V_{1i}^{a_1}\partial_{a_1}W_1\right)+\left(\partial_i
W_2+V_{2i}^{a_2}\partial_{a_2}W_2\right).
\end{multline}

These transformations look complicated, but this is an artefact of
using a nonholomorphic trivialization of the tangent bundle of
$\Z_1\times_Y\Z_2$. If we assemble the fields
$\eta_1^{\ba_1},\eta_2^{\ba_2},\eta^\bi$ into a single field
$\eta^{\bar A}\in \Gamma(\phi^*{\bar T}_{\Z_1\times_Y \Z_2})$, the
fields $\rho_1^{a_1},\rho_2^{a_2},\rho^i$ into a single field
$\rho^A\in\Gamma(\phi^* T_{\Z_1\times_Y\Z_2}\otimes T^*_\Sigma)$ and
the fields $\theta_{1a_1},\theta_{2a_2},\theta_i$ into a single
field $\theta_A\in\Gamma(\phi^*T^*_{\Z_1\times_Y \Z_2})$, and use
the holomorphic coordinate trivialization, we find
\begin{align*}
\delta_{Q,W}\phi^{\bA}&=\eta^{\bA}, &\delta_{Q,W}\phi^A&=0,\\
\delta_{Q,W}\eta^{\bA}&=0, &\delta_{Q,W}\theta_A&=-\partial_A (W_1-W_2),\\
\delta_{Q,W}\rho^A&=d\phi^A. & &
\end{align*}
These are the BRST transformations of the curved B-model with target
$\Z_1\times_Y\Z_2$ and the curving $\pi_1^*W_1-\pi_2^* W_2$.
Presumably, quantum corrections further deform this result.

\subsubsection{Different submanifolds, trivial fibrations}

Now let us allow $Y_1$ and $Y_2$ to be different, but assume that
there are no boundary degrees of freedom, i.e. $\Z_1=Y_1$ and
$\Z_2=Y_2$. The classical vacua of the RW theory on $S^1\times
\Sigma$ are constant field configurations satisfying the boundary
conditions. Since the left and right boundaries must be mapped to
$Y_1$ and $Y_2$, respectively, the set of classical vacua is
$Y_{12}=Y_1\bigcap Y_2$. An obvious guess for a 2d field theory with
this set of classical vacua is the B-model with target $Y_{12}$.
However, if $Y_{12}$ is not an embedded submanifold, it is not clear
what this means. Even if $Y_{12}$ is an embedded submanifold (e.g.
when $Y_{12}$ consists of several isolated points), the naive guess
is not always correct. This happens because in general the
directions transverse to $Y_{12}$ do not correspond to massive
degrees of freedom.

To make a more intelligent guess, let $\Sigma=S^1\times\RR$. Then
one can determine the quantum space of states by first reducing the
RW theory on $S^1$. As discussed above, reduction on $S^1$ yields a
B-model with target $X$ on $I\times \RR$. The boundary conditions
correspond to Lagrangian submanifolds $\iota_1: Y_1\hookrightarrow
X$ and $\iota_2:Y_2\hookrightarrow X$. The space of states of this
B-model is
\begin{equation}\label{exts}
\oplus_p \Ext^p_X\left(\iota_{1 *}
\cO_{Y_1},\iota_{2*}\cO_{Y_2}\right).
\end{equation}
More precisely, this is the space of boundary-changing operators.
Let us compute this space when $Y_{12}=Y_1\bigcap Y_2$ consists of a
single point $r\in X$. In the neighborhood of $r$ we can choose
complex Darboux coordinates $p_i,q^i$ so that $\smpf=dq_i dp^i$,
$Y_1$ is given by the equations
$$
p_i=\frac{\partial F_1(q)}{\partial q^i},
$$
and $Y_2$ is given by the equations
$$
p_i=\frac{\partial F_2(q)}{\partial q^i},
$$
for some holomorphic functions $F_1(q), F_2(q)$ on an open set $\cU$
in a vector space $V\simeq\CC^n$. The functions $F_1, F_2$ are
called generating functions of the Lagrangians $Y_1$ and $Y_2$. The
intersection of $Y_1$ and $Y_2$ is the critical set of $F_1-F_2$; by
assumption $r\in\cU$ is the only critical point of $F_1-F_2$ in
$\cU$. Given a choice of Darboux coordinates, we may identify an
open neighborhood of $r$ with an open neighborhood of the zero
section in $T^*_\cU$ and regard the intersections of $Y_1$ and $Y_2$
with this neighborhood as Lagrangian submanifolds of $T^*_\cU$.

It is obvious physically and can be proved mathematically that the
$\Ext$ groups depend only on the behavior of $Y_1$ and $Y_2$ in an
arbitrarily small open neighborhood of $r$. Therefore to compute the
$\Ext$ groups we can use a local Koszul resolution for a submanifold
$Y_1$ of $T^*_\cU$:
$$
\begin{CD}
\Lambda^n V\otimes \cO @>\alpha
>>\Lambda^{n-1}V\otimes\cO @>\alpha
>>\cdots @>\alpha >> V\otimes\cO @>\alpha>> \cO,
\end{CD}
$$
where $\cO=\cO_{T^*_\cU}$, and $\alpha$ is a contraction with a
section of $V^*\otimes \cO$ given by $(p_i-\partial_i F_1) dq^i$.
Applying $\Hom(-,\iota_{2*}\cO_{Y_2})$ to the Koszul resolution, we
get a complex
$$
\begin{CD}
\Lambda^n V^*\otimes \Gamma(\cU) @<\beta <<\cdots @<\beta <<
V^*\otimes \Gamma(\cU) @<\beta <<\Gamma(\cU),
\end{CD}
$$
where $\beta$ is exterior product with an element of
$V^*\otimes\Gamma(\cU)$ given by $(\partial_i F_2-\partial_i F_1)
dq^i$, and $\Gamma(\cU)=H^0(\cU,\cO_\cU)$. The cohomology of this
complex is nonvanishing only in the left-most term and is isomorphic
to the vector space
$$
\Lambda^nV\otimes\left(\Gamma(\cU)/A\right)
$$
where $A$ is the ideal generated by partial derivatives of
$F_2-F_1$. The factor $\Lambda^n V$ is unimportant, since the
Lagrangian submanifolds $Y_1$ and $Y_2$ are equipped with
holomorphic volume forms, which give a natural isomorphism
$\Lambda^n V\simeq\CC$.

The vector space $\Gamma(\cU)/A$ is the Jacobi ring of $F_1-F_2$ and
is most naturally obtained in the Landau-Ginzburg model with target
$\cU$ and superpotential $W=F_1-F_2$. This suggests that reduction
on the interval produces such a Landau-Ginzburg model.

If $n$ is odd, the vector space (\ref{exts}) is purely odd. The same
is also true for the space of states of the Landau-Ginzburg model
with an odd-dimensional target. In fact, the choice of absolute
grading in both cases is ambiguous: if instead of $\Ext$ groups one
considers ${\rm Tor}$ groups, and instead of the space of states for
the Landau-Ginzburg model one considers the space of local
operators, one gets the Jacobi ring of $F_1-F_2$ with even grading
for all $n$. This ambiguity is discussed in more detail in section
\ref{sec:ambig}.

We will now argue that when $Y_1$ and $Y_2$ intersect at an isolated
point, the RW theory on $I\times\Sigma$ reduces to the
Landau-Ginzburg model on $\cU$ with the superpotential $F_1-F_2$. We
have seen above that this is true in the case $Y_1=Y_2$. In general,
we observed in subsection \ref{sec:defs} that a deformation of the
Lagrangian submanifold by means of a generating function $F$ is
equivalent, to first order in $F$, to adding a boundary
superpotential $F$ and modifying the BRST transformation of
$\chi_3^{k'}$ by boundary terms:
$$
\delta_Q\chi_3^{k'}=\partial_3\phi^{k'}
-\delta(x_3)\smpf^{k'j}\partial_j
F_1+\delta(x_3-1)\smpf^{k'j}\partial_j F_2.
$$
Hence to first order in $F_1,F_2$ the field $\theta_i$ defined by
(\ref{theta}) transforms as
$$
\delta_Q\theta_i=-\partial_i (F_1-F_2).
$$
This is exactly the right transformation law for the Landau-Ginzburg
model with superpotential $F_1-F_2$. After dropping the term
involving $\partial_3\phi^{i'}$, the boundary action (\ref{e2.14})
gives the following contribution to the action of the 2d field
theory on $\Sigma$:
$$
\int_\Sigma \frac12\hat\nabla_i\partial_j (F_1-F_2) \rho^i\rho^j.
$$
This is the Landau-Ginzburg deformation of the B-model action with
the superpotential $W=F_1-F_2$ (see section \ref{app:curved} for a
brief review of the Landau-Ginzburg model). This proves our
statement to leading order in $F_1,F_2$.

To complete the argument, we use the ghost number symmetry. We note
that $F_1$ and $F_2$ both have weight $2$ under this symmetry, just
like $\smpf$. The fields $\eta$ and $\theta$ have weights $+1$,
while the fields $\rho$ have weight $-1$. Terms of higher order in
$F_1,F_2$ must be at least quartic in $\rho$. But since $\rho$ is a
1-form and the action must be a 2-form on $\Sigma$, no such terms
are possible.\footnote{The topological character of the theory
implies that the part of the action which is not BRST-exact cannot
depend on the metric on $\Sigma$.}

Next we generalize this result to the case when $Y_{12}=Y_1\bigcap
Y_2$ is arbitrary. For simplicity we will neglect the quantum
corrections arising from the nontrivial external geometry of
$Y_{12}$. Without loss of generality we may assume that $Y_{12}$ is
connected. We choose an open neighborhood $K$ of $Y_{12}$ and a
Lagrangian submanifold $\cU$ of $K$ containing $Y_{12}$. We require
that in the neighborhood of any $r\in Y_{12}$ there exist Darboux
coordinates $p_i,q^i,i=1,\ldots,n$ so that $\cU$ is given by the
equations $p_i=0, i=1,\ldots,n$, and Lagrangian submanifolds
$Y_1\subset K$ and $Y_2\subset K$ are described by generating
functions $F_1$ and $F_2$. Obviously, the generating functions
$F_1,F_2$ are constant on $Y_{12}$.

As before, locally the reduction of the RW model gives the
Landau-Ginzburg model with target $\cU$ and the superpotential
$W=F_1-F_2$.\footnote{Quantum corrections could deform this result.}
Unlike in the previous case, the critical points of this
superpotential are not necessarily isolated; rather, the critical
set is $Y_{12}$. It remains to check that $F_1-F_2$ is a
globally-defined holomorphic function on $\cU$. We note that $F_1$
and $F_2$ are defined up to an additive constant, so $d(F_1-F_2)$ is
a well-defined closed holomorphic 1-form on $\cU$. We also know that
this form vanishes on $Y_{12}\subset\cU$ and therefore its class in
$H^1(Y_{12})$ vanishes as well. By taking a sufficiently small
$\cU$, we can always ensure that the restriction map $H^1(\cU)\raa
H^1(Y_{12})$ is an isomorphism; then $d(F_1-F_2)$ can be integrated
to a holomorphic function on $\cU$ defined up to an additive
constant.

This answer simplifies in the case when $Y_{12}$ is a submanifold of
$X$ and the intersection of $Y_1$ and $Y_2$ is clean. This means
that for any $r\in Y_{12}$ we have
$$
T_rY_1\bigcap T_rY_2=T_rY_{12}.
$$
This implies that $F_1-F_2$ are Morse-Bott functions on $\cU$, with
$Y_{12}$ being the critical submanifold. As explained in section
\ref{app:curved}, in such a situation the Landau-Ginzburg model with
target $\cU$  and superpotential $F_1-F_2$ is equivalent to the
B-model with target $Y_{12}$. More precisely, this is true if
$n-\dim Y_{12}$ is even; if $n-\dim Y_{12}$ is odd, the
Landau-Ginzburg model with target $\cU$ and superpotential $F_1-F_2$
is equivalent to the Landau-Ginzburg model with target $Y_{12}\times
\CC$ and superpotential $x^2$, where $x$ is an affine coordinate on
$\CC$. (While the Landau-Ginzburg model with target $\CC$ and
superpotential $x^2$ is trivial on the closed worldsheet $\Sigma$,
this is not true when $\Sigma$ has boundaries, see section
\ref{app:curvedbranes}).

\subsubsection{Different submanifolds, nontrivial fibrations}

The most general case is a combination of all of the above. We have
two fibrations $\Z_1$ and $\Z_2$ over complex Lagrangian
submanifolds $Y_1$ and $Y_2$ whose intersection is a
not-necessarily-smooth $Y_{12}$. For simplicity when discussing
reduction we will neglect quantum corrections coming from the
nontrivial external geometry of $Y$. We also need to choose an open
neighborhood $K\supset Y_{12}$ and a Lagrangian submanifold $\cU$
containing $T$ and contained in $K$ such that locally there exist
Darboux coordinates such that $\cU$ is given by the equation
$p_i=0$, while $Y_1\bigcap K$ and $Y_2\bigcap K$ are given by the
equations $p_i=\partial_i F_1$ and $p_i=\partial_i F_2$. The
differential $d(F_1-F_2)$ is a globally-defined closed holomorphic
1-form on $\cU$ which can be integrated to a holomorphic function on
$\cU$ defined up to an additive constant.

As before, if we neglect quantum corrections, one can replace $K$
with an infinitesimal neighborhood of the zero section of $T^*_\cU$
which we continue to call $K$. We have the projection
$\kappa:T^*_\cU\raa \cU$ which identifies $Y_1\bigcap K$ and
$Y_2\bigcap K$ with $\cU$. We can use $\kappa$ to push the
fibrations $\Z_1$ and $\Z_2$ to $\cU$; we will call the resulting
fibrations $\U_1$ and $\U_2$. The curvings $W_1$ and $W_2$ are push
forward to curvings $W_{\U_1}$ and $W_{\U_2}$ on $\U_1$ and $\U_2$,
respectively.

We can view $Y_s$ equipped with a fibration $\Z_s$ as a deformation
of $\cU$ equipped with $\U_s$. The deformations are described by
generating functions $F_s$, $s=1,2$. If we neglect the deformations
altogether, the reduced theory will be the curved B-model with
target $\U_1\times_\cU \U_2$ and curving $\pi_1^*W_{\U_1}-\pi_2^*
W_{\U_2}$, where $\pi_s,$ $s=1,2$ denotes the projection from
$\U_1\times_\cU \U_2$ to $\U_s$. To first order in $F_1, F_2$, we
get an extra piece in the transformation law for $\theta_i$:
$$
\tdelta_{Q,W}\theta_i=\delta_{Q,W}\theta_i-\partial_i (F_1-F_2),
$$
where $\delta_{Q,W}$ denotes the BRST-variation
(\ref{BRSTtransftheta}). Then in the holomorphic coordinate
trivialization we get
$$
\tdelta_{Q,W}\theta_A=-\partial_A(W_1+F_1-W_2-F_2).
$$
This is the BRST transformation for the curved B-model with the
target $\U_1\times_\cU \U_2$ and the curving
$$
\pi_1^*W_{\U_1}-\pi_2^* W_{\U_2}+p_{12}^*(F_1-F_2),
$$
where $p_{12}$ is the projection $\U_1\times_\cU \U_2\raa \cU$. The
modification of the boundary action due to $F_1,F_2$ gives the
following correction to the action of the reduced theory:
$$
\int_\Sigma \frac12\hat\nabla_i\partial_j(F_1-F_2)\rho^i\rho^j.
$$
This is the Landau-Ginzburg deformation corresponding to the
superpotential $p_{12}^*(F_1-F_2)$.

When considering reduction on $S^1\times\Sigma$, we found that for a
curved worldsheet $\Sigma$ the action of the reduced theory contains
a dilatonic coupling not usually present in the B-model. Such terms
could also arise in the case of reduction on the interval. We leave
the analysis of such terms to future work.

\subsubsection{The grading ambiguity}\label{sec:ambig}

The above discussion of dimensional reduction brings out a subtlety
in the quantization of the RW model with boundaries: the absolute
$\ZZ_2$-grading on the space of states is not well-defined if the
complex dimension of $X$ is not divisible by four, i.e. if $n$ is
odd. The underlying reason is that for odd $n$ there is no
completely canonical choice of a BRST-invariant measure on the space
of $\chi$ zero modes. For simplicity, let us assume that the
boundary condition on $\partial M$ is given by a complex Lagrangian
submanifold $Y$; then the zero modes of $\chi$ normal to $Y$ take
values in $H^1(M,\partial M)$, and the zero modes tangent to $Y$
take values in $H^1(M)$. The BRST-invariant measure involves volume
forms on $H^1(M)$ and $H^1(M,\partial M)$ raised to the $n^{\rm th}$
power. If $n$ is odd, changing the orientation of $H^1(M)$ or
$H^1(M,\partial M)$ results in a change of sign of the
BRST-invariant measure.

One can explain this ambiguity in a different way. Let $\Sigma$ be
an oriented 2-manifold with a boundary. Given a complex Lagrangian
submanifold $Y$, the RW model attaches to $\Sigma$ a vector space
with a well-defined relative $\ZZ_2$ grading. To define an absolute
grading, it is sufficient to specify a distinguished vector whose
grading is defined to be even. One can attempt to define such a
distinguished vector by picking an oriented 3-manifold $M$ whose
boundary is split by a 1-manifold into $\Sigma$ and $\Sigma'$, where
$\Sigma'$ is some oriented 2-manifold with the same boundary as
$\Sigma$. Given such $M$, one can consider the path-integral for the
RW model on $M$ with the boundary condition corresponding to $Y$
specified on $\Sigma'$. This gives a state in the vector space
associated to $\Sigma$. This procedure works if different choices of
$M$ produce vectors which have the same relative grading. For odd
$n$ it turns out that this is not true, and consequently there is no
canonical choice of absolute grading.

As an example, consider $\Sigma=S^1\times I$, where the boundary
conditions on the two components of the boundary are identical and
given by a Lagrangian submanifold $Y$ in $X$. If we reduce the
theory on a circle, we obtain a B-model with target $X$ on an
interval, with boundary conditions corresponding to a B-brane $Y$.
Its space of states is $H^\bul(\Lambda^\bul N_Y)$. It is a Frobenius
algebra; in particular, it has a trace function which is
nonvanishing only on the component $H^n(\Lambda^n N_Y)$, where
$n=\dim_\CC Y$. It is natural to regard this component as even; this
determines an absolute $\ZZ_2$ grading. From the three-dimensional
viewpoint, this absolute grading corresponds to taking the
3-manifold $M$ to be a solid torus whose boundary $S^1\times S^1$ is
glued from two annuli $\Sigma$ and $\Sigma'$.

On the other hand, if we reduce the theory on an interval, we obtain
a B-model with target $Y$ on a circle. Its space of states is
$H^\bul(\Lambda^\bul T_Y)$. It is again a Frobenius algebra, with a
trace function which is nonvanishing only on the component
$H^n(\Lambda^n T_Y)$. It is therefore natural to regard this
component as even. From the three-dimensional viewpoint this
definition of absolute grading corresponds to $M$ being a solid
cylinder whose boundary is glued from an annulus $\Sigma$ and two
disks.

Now, since $Y$ is Lagrangian, we have $N_Y\simeq T_Y^*$, and since
$Y$ is a Calabi-Yau manifold, we have $\Lambda^p N_Y\simeq
\Lambda^{n-p} T_Y$. Thus the two vector spaces are naturally
isomorphic as ungraded vector spaces, but the ``natural'' $\ZZ_2$
gradings agree only for even $n$.

Note that when $\Sigma$ has no boundary, the grading ambiguity does
not arise \cite{RW}. While there is more than one way to ``fill''
$\Sigma$ with a 3-manifold $M$ such that $\partial M=\Sigma$, all
these ways correspond to the same grading on the vector space
attached to $\Sigma$.

\section{Topological defects}\label{sec:defects}

\subsection{Surface operators}

A surface operator in a 3d TFT is a topological defect of
codimension $1$. One can reduce to a large extent the study of
surface operators to a study of boundary conditions using the
``folding trick'': if the location of the surface operator is given
by $x^3=0$, one can identify the region $x^3<0$ with $x^3>0$ by
means of the parity-reversing involution $x^3\mapsto -x^3$ and
regard the surface operator as a boundary condition for a TFT which
is a product of the original TFT and the parity-reversed TFT.

In the case of the RW model, parity-reversal is equivalent to
replacing $\smpf$ with $-\smpf$. If $X$ is a complex symplectic
manifold with a holomorphic symplectic form $\smpf$, let $X^*$
denote the same complex manifold with the holomorphic symplectic
form $-\smpf$. Thus a surface operator for the RW model with target
$X$ is the same as a boundary condition for the RW model with target
$X^*\times X$.

Among all surface operators, there is a special one, corresponding
to the diagonal $\Delta\subset X^*\times X$. It is obviously a
Lagrangian submanifold, thanks to the relative minus sign between
the symplectic forms of the two factors. The boundary conditions for
this surface operator say that all fields and the first derivatives
of bosonic fields vary continuously across $x^3=0$. Such a surface
operator is equivalent to no surface operator at all, so we will
refer to it as the invisible surface operator.

One can generalize the notion of a surface operator by considering
the situation when the TFTs for $x^3>0$ and $x^3<0$ are not
necessarily isomorphic. For example, they could be RW models with
different target spaces $X_1$ and $X_2$. Such a surface operator is
equivalent to a boundary condition for the RW model with target
$X_1^*\times X_2$. The usual boundary condition is a special case of
such a generalized surface operator corresponding to $X_1=\{pt\},$
$X_2=X$.

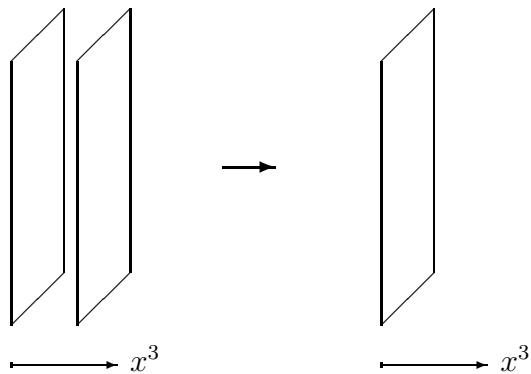
\begin{figure}\label{fig:monoidalsurface}

\begin{picture}(300,160)(0,0)
\put(20,20){\line(0,1){100}}\put(40,40){\line(0,1){100}}\put(20,20){\line(1,1){20}}
\put(20,120){\line(1,1){20}}

\put(25,0){\put(20,20){\line(0,1){100}}\put(40,40){\line(0,1){100}}\put(20,20){\line(1,1){20}}
\put(20,120){\line(1,1){20}}}

\put(20,5){\vector(1,0){40}}\put(20,4){\line(0,1){2}}\put(65,3){$x^3$}

\put(100,80){\thicklines\vector(1,0){20}}

\put(140,0){\put(20,20){\line(0,1){100}}\put(40,40){\line(0,1){100}}\put(20,20){\line(1,1){20}}
\put(20,120){\line(1,1){20}}}

\put(140,0){\put(20,5){\vector(1,0){40}}\put(20,4){\line(0,1){2}}\put(65,3){$x^3$}}

\end{picture}
\caption{Fusion of surface operators.}
\end{figure}

One important difference between boundary conditions and surface
operators is that the set of surface operators has a monoidal
structure (i.e. an associative multiplication with a unit object).
Indeed, if we consider two surface operators located at $x^3=0$ and
$x^3=\eps$, by taking the limit $\eps\raa 0+$ one gets another
surface operator (Fig. \ref{fig:monoidalsurface}). In the
topological theory it is not really necessary to take the limit,
since changing $\eps$ can be effected by a diffeomorphism of $M$.
The invisible surface operator is the identity object for this
monoidal structure.

One can also compose generalized surface operators, if we regard
them as morphisms in a category whose objects are holomorphic
symplectic manifolds, so we only consider composition when the
``target'' of the first surface operator is the same as the
``source'' of the second one. In fact, as explained below, the set
of generalized surface operators between a fixed pair of holomorphic
symplectic manifolds itself has the structure of a 2-category. Thus
holomorphic symplectic manifolds are objects of a 3-category. This
is discussed in more detail in \cite{KRS2}.

In symplectic geometry one sometimes considers the ``symplectic
category'' whose objects are symplectic manifolds and morphisms are
Lagrangian correspondences, i.e. a morphism between $X_1$ and $X_2$
is a Lagrangian submanifold in $X_1^*\times X_2$
\cite{Weinstein,WB}. This is not a true category because the
composition of morphisms is defined only when Lagrangian
submanifolds have a clean intersection \cite{Gui,Weinstein1}. The
category of surface operators on the other hand is well-defined by
construction, i.e. compositions of morphisms are always defined.
However, it is not easy to compute this composition in general.
Neither is it obvious that the set of surface operators that we
constructed is closed with respect to composition, although we
believe this to be the case.

To compare the ``symplectic category'' and the category of surface
operators, note that objects of the former are also objects of the
latter. Now consider two Lagrangian correspondences $Y_{12}\subset
X_1^*\times X_2$ and $Y_{23}\subset X_2^*\times X_3$. Let
$\Delta_{X_2}$ denote the diagonal in $X_2\times X_2^*$. The
composition $Y_{12}\circ Y_{23}$ in the ``symplectic category'' is
defined as the image of
\begin{equation}\label{intersection}
\left(Y_{12}\times Y_{23}\bigcap X_1^*\times\Delta_{X_2}\times
X_3\right)\subset X_1^*\times X_2\times X_2^*\times X_3
\end{equation}
under the projection to $X_1^*\times X_3$. When the intersection
(\ref{intersection}) is clean, its image is a Lagrangian submanifold
in $X_1^*\times X_3$ \cite{Gui,Weinstein1}. In general, the fiber of
the projection over a point in the image may have dimension greater
than zero. It is zero-dimensional precisely when the intersection is
transverse.

On the other hand, we may consider the composition of $Y_{12}$ and
$Y_{23}$ as generalized surface operators. The intersection
(\ref{intersection}) is precisely the space of classical vacua on
$\Sigma=\RR^2_{x_1,x_2}$ with the insertion of surface operators at
$x_1=0$ and $x_1=\eps$. If the intersection is clean, the projection
to $X_1^*\times X_3$ defines a fibration over its image whose fiber
describes degrees of freedom living on the composite surface
operator with the bulk fields fixed. Thus if the intersection is
clean  but not transverse, the composition of $Y_{12}$ and $Y_{23}$
as generalized surface operators is described not by the Lagrangian
submanifold $Y_{12}\circ Y_{23}$, but by a fibration over it. Only
in the case of transverse intersection do the two compositions
agree.

We conclude this section by considering the composition of
generalized surface operators in the case when the corresponding
Lagrangian correspondences need not intersect cleanly. For
simplicity, let $X_1={\pt}$, $X_2=\CC^{2n}$ with coordinates
$q^1,\ldots,q^n,p_1,\ldots,p_n$, $\smpf=dq^i dp_i$ and
$X_3=\CC^{2n}$ with coordinates
$\tq^1,\ldots,\tq^n,\tp_1,\ldots,\tp_n$, $\tsmpf=d\tq^i d\tp_i.$
Suppose further that $Y_{12}\subset X_2$ is given by the generating
function $f(q)$ and $Y_{23}\subset X_2^*\times X_3$ is given by the
generating function $F(q,\tq)$. Since $X_1={\pt}$, the generalized
surface operator corresponding to $Y_{12}$ is simply a boundary
condition for the RW model with target $X_2$. The fusion of the
surface operator $Y_{23}$ with this boundary condition produces a
new boundary condition for the RW model with target $X_3$.

Imagine now that the boundary is located at $x^3=0$, and the surface
operator $Y_{23}$ is located at $x^3=\eps>0$. The degrees of freedom
of the RW model with target $X_2$ on the interval $[0,\eps]$ should
be thought of as boundary degrees of freedom for the new boundary
condition. As for fields of the RW model with target $X_3$ on the
half-line $x^3>\eps$, they should be kept fixed. Now we can appeal
to the results of section \ref{sec:intervalred} and conclude that
the boundary degrees of freedom for the new boundary condition are
described by a Landau-Ginzburg model with target $\CC^n_q$ and the
superpotential $W=f(q)-F(q,\tq)$. This superpotential depends on the
fields $\tq$, so in the end we get a trivial fibration over the
Lagrangian submanifold $\CC^n_{\tq}$ with fiber $\CC^n_q$ and the
superpotential $f(q)-F(q,\tq)$.

\subsection{Line operators}
\subsubsection{Generalities}
A line operator in a 3d TFT is a topological defect of codimension
two. The corresponding one-dimensional submanifold can be in the
interior of $M$, can split the boundary of $M$, or can split a
two-dimensional submanifold on which a surface operator is inserted.
The last case is the most general one, since one can regard a
boundary line operator as a line operator on a surface operator
between $X_1=\{pt\}$ and $X_2=X$, and one can regard the ``bulk''
line operator as a line operator on the invisible surface operator
in $M$.

\begin{figure}\label{fig:lineops}

\begin{picture}(100,120)(0,0)

\put(0,0){\line(0,1){100}}\put(60,20){\line(0,1){100}}\put(0,0){\line(3,1){60}}\put(0,100){\line(3,1){60}}
\put(30,10){\thicklines\line(0,1){100}}\put(5,90){$B_1$}\put(45,102){$B_2$}\put(32,60){$C$}

\put(80,60){$C\in\Hom(B_1,B_2)$}

\end{picture}
\caption{Line operator $C$ is a morphism between surface operators
$B_1$ and $B_2$.}

\end{figure}
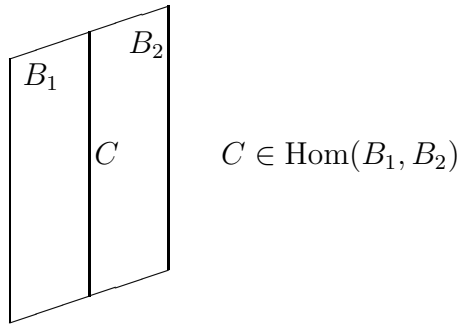

Note that the submanifold corresponding to a line operator may
separate the ``worldsheet'' of the surface operator into two
disconnected pieces, so that one may have different surface
operators on the two sides of the line operator (Fig.
\ref{fig:lineops}). One may regard line operators as morphisms in a
category whose objects are surface operators. Since surface
operators themselves form a category, one can think of line
operators as 2-morphisms in a 2-category whose objects are complex
symplectic manifolds and whose 1-morphisms are surface operators. In
fact, the set of line operators separating two fixed surface
operators has the structure of a $\CC$-linear category, with
morphisms corresponding to point operators separating different line
operators. Thus this 2-category is really a 3-category. Its objects
are holomorphic symplectic manifolds, its 1-morphisms are
generalized surface operators, its 2-morphisms are line operators on
generalized surface operators, and its 3-morphisms are point
operators.

\subsubsection{Line operators on the boundary}

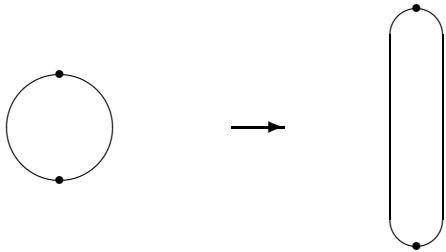
\begin{figure}\label{fig:reduction}
\begin{picture}(300,90)(0,0)
\put(25,47){\circle{40}}\put(25,27){\circle*{3}}\put(25,67){\circle*{3}}
\put(90,47){\thicklines\vector(1,0){20}}

\put(150,0){\put(10,47){\oval(20,90)}\put(10,2){\circle*{3}}\put(10,92){\circle*{3}}}

\end{picture}

\caption{Boundary line operators in 3d TFT correspond to branes in
the effective 2d TFT obtained by reducing 3d TFT on an interval.}
\end{figure}

The study of line operators can be reduced to a large extent to the
study of branes in 2d TFT. Without loss of generality, we may
consider line operators on the boundary of $M$, since line operators
on a surface separating RW models with targets $X_1$ and $X_2$ are
equivalent to line operators on the boundary of the RW model with
target $X_1^*\times X_2$. Let us further suppose that the line
operator is a morphisms between boundary conditions corresponding to
curved fibrations $(\Z_1,W_1)$ and $(\Z_2,W_2)$ over Lagrangian
submanifolds $Y_1$ and $Y_2$ respectively. Let $M$ be
$D^2\times\RR$, where the unit disc $D^2$ is regarded as space and
$\RR$ as time. We will further assume that there are two marked
points $r_0,r_1$ on $\partial D^2$, and two line operators located
at $\{r_0\}\times\RR$ and $\{r_1\}\times\RR$. In the limit when
$D^2$ becomes an infinitely narrow oval, the RW model on such $M$
reduces to a 2d TFT on $I\times \RR$, and the line operators become
boundary conditions for this 2d TFT (see Fig. \ref{fig:reduction}).
Point operators located on line operators become boundary-changing
operators in the 2d TFT, so the category of line operators is
equivalent to the category of branes in the 2d TFT.

In the case of the RW model, the effective 2d TFT was determined in
subsection \ref{sec:intervalred}. It is simple to describe it in the
classical approximation in the case when $Y_1=Y_2=Y$. On the
classical level the reduced theory is a curved B-model with target
$\Z_1\times_Y \Z_2$ and the curving $\pi_1^* W_1-\pi_2^* W_2$, where
$\pi_s$ is the projection from $\Z_1\times_Y\Z_2$ to $\Z_s$. In what
follows we will not show the curving explicitly and will write
$\Z_s$ instead of $(\Z_s,W_s)$. We will also denote by $\Z_s^*$ the
pair $(\Z_s,W_s)$. Thus the category of branes in the effective 2d
TFT is the curved derived category of $\Z_1\times_Y \Z_2^*$. Here it
was assumed that the orientation of the part of the boundary
corresponding to $\Z_1$ agrees with the orientation of the line
operator, so we should regard this category as the category of
morphisms from the 1-object $\Z_2$ to the 1-object $\Z_1$. The
category of morphisms from $\Z_1$ to $\Z_2$ is the curved derived
category of $\Z_1^*\times_Y\Z_2$. If the class $[\beta]\in
H^1(T_Y\otimes N_Y^*)$ is nonvanishing, quantum corrections may
deform the target of the reduced theory; for example, as explained
in section \ref{sec:intervalred}, without fibrations the leading
deformation is described by the Beltrami differential
$\mu=\frac{\beta}{2}(\partial W_1+\partial W_2)$. We also showed in
section \ref{sec:intervalred} that if $\Z_1=\Z_2=Y,$ and
$W_1=W_2=0$, the classical answer is exact. That is, in this special
case the category of boundary line operators is exactly the
$\ZZ_2$-graded derived category of $Y$.

Another simple case is when fibrations are absent and $Y_1$ and
$Y_2$ intersect over a finite set $\{p_1,\ldots,p_\ell\}\subset X$.
In this case the reduced 2d TFT is a Landau-Ginzburg model whose
target is a union of open sets $\cU_k\subset \CC^n$,
$k=1,\dots,\ell$, with superpotentials $W_1,\ldots,W_\ell$. The
superpotential $W_k$ on $\cU_k$ is determined by the local geometry
of $Y_1$ and $Y_2$ in a neighborhood of $p_k$: given a holomorphic
symplectomorphism from this neighborhood to $T\cU_k^*$, $W_k$ is
given by $F_1-F_2$, where $F_1$ and $F_2$ are the holomorphic
generating functions on $\cU_k$ corresponding to $Y_1$ and $Y_2$,
respectively.

Instead of utilizing the known results about branes in curved
B-models, one can analyze the category of line operators on the
boundary more directly. For definiteness, let us consider the case
$Y_1=Y_2=Y$; as explained in subsection \ref{sec:intervalred}, this
entails no real loss of generality, since the general case can be
reduced to this one. Let us show explicitly how to construct a
boundary line operator given an object of the curved derived
category of $\Z_1\times_Y\Z_2^*$. For simplicity, we will do on the
classical level, i.e. neglecting boundary corrections to BRST
transformations originating from the second fundamental form of $Y$.

We consider a connected component $\Sigma$ of $\partial M$ and a
one-dimensional submanifold $\gamma\subset\Sigma$ which splits
$\Sigma$ into $\Sigma_1$ and $\Sigma_2$. We will assume that the
orientation of $\gamma$ agrees with the orientation of $\Sigma_1$
and disagrees with the orientation of $\Sigma_2$. The boundary
condition on $\Sigma_s$ is described by the fibration $\Z_s$,
equipped with the curving $W_s\in H^{0,{\hat 0}}(\Z_s)$, $s=1,2$.
The BRST variation of the part of the boundary action coming from
$\Sigma_s$ is
$$
\delta_Q S^{bry}_s=\int_{\Sigma_s} \left(\smpf^{k'i} A^K_{k'}
\left(\cF^a_{s i\bar A} \theta_a H_s^{\bar A} {\delta S^{bulk}\over
\delta \chi_3^K}+\nabla_i W_s\frac{\delta S^{bulk}}{\delta
\chi_3^K}\right) + dW_s^{(1)}\right),
$$
where $W_s^{(1)}$ is given by (\ref{W1}) with $W$ replaced with
$W_s$. The BRST transformation of $\chi_3^K$ involves boundary terms
as in (\ref{e3.22}); thanks to these boundary terms the BRST
variation of the total action is
\begin{equation}\label{delta12tot}
\delta_{Q,W}(S_1^{bry}+S_2^{bry}+S^{bulk})=\int_\gamma
\left(W_1^{(1)}-W_2^{(1)}\right).
\end{equation}
Here $W_1-W_2$ is regarded as a $\bpartial$-closed form of type
$(0,{\hat 0})$ on $\Z_{12}=\Z_1\times_Y\Z_2$. This BRST variation
must be canceled by the BRST variation of the line operator on
$\gamma$.

Now consider an object of the curved derived category of $\Z_{12}$.
It is a pair $(E,D)$, where $E$ is a smooth $\ZZ_2$-graded vector
bundle on $\Z_{12}$ and $D$ is an odd differential operator on
$\obul(E)$ satisfying (\ref{Dcond}). The restrictions of the fields
$\phi^A, H^{\bar A}$ and $P^A$ to $\gamma$ define a map
$\Phi:\gamma\raa \Pi{\bar T}^*_{\Z_{12}}$ and a fermionic 1-form
$\rho^A$ on $\gamma$ with values in the pull-back of $T_{\Z_{12}}$.
The BRST operator $\delta_Q$ acts on the space of pairs
$(\Phi,\rho^A)$. As explained in section \ref{app:curvedbranes},
given an object $(E,D)$ of the curved derived category of $\Z_{12}$
one can construct a function $\exp(-S^\gamma)$ on the space of pairs
$(\Phi,\rho^A)$ whose BRST variation is the right-hand-side of
(\ref{delta12tot}). This function is the supertrace of the holonomy
of a certain connection on $\Phi^*(E)$ constructed from $D$ and the
fields $\Phi$ and $\rho^A$. We take $\exp(-S^\gamma)$ as the line
operator corresponding to $(E,D)$.

The function $\exp(-S^\gamma)$ greatly simplifies in the case when
$\Z_1=\Z_2=Y$, and $(E,D)$ is a holomorphic vector bundle on $Y$.
Then the line operator takes the form
$$
\Tr\, \Hol_\gamma\left( A_i d\phi^i+A_\bi
d\phi^\bi+F_{i\bj}\chi^i\eta^\bj\right),
$$
where $(A_i,A_\bi)$ are components of a connection 1-form on $E$,
and $F_{i\bj}$ is the curvature of this connection.

\subsubsection{Fusion of line operators on the boundary}

In the case $\Z_1=\Z_2=\Z$, the category of line operators is the
category of endomorphisms of the object $\Z$ in the 2-category of
boundary conditions. Therefore it has a monoidal structure realized
physically by fusing the line operators on the boundary. The
invisible line operator is the unit object with respect to this
monoidal structure. Mathematically, it is represented by the
``fibered diagonal'' $\Delta_Y$ in the fibered product
$\Z\times_Y\Z$. Note that the curving vanishes on $\Delta_Y$, so
presumably this is a valid object of the curved derived category
(see \cite{KapRoz} where this is explained in the special case of
the Landau-Ginzburg model).

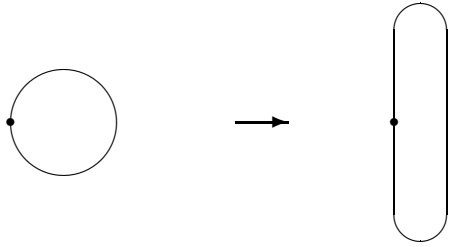
\begin{figure}\label{fig:linemono}
\begin{picture}(300,90)(0,0)
\put(25,47){\circle{40}}\put(5,47){\circle*{3}}
\put(90,47){\thicklines\vector(1,0){20}}

\put(150,0){\put(10,47){\oval(20,90)}\put(0,47){\circle*{3}}}

\end{picture}

\caption{Boundary line operators in 3d TFT also correspond to bulk
line operators in the effective 2d TFT obtained by reducing 3d TFT
on an interval.}
\end{figure}

As explained above, to any line operator on the boundary described
by the fibration $\Z$ one can associate a boundary condition in the
curved B-model with target $\Z^*\times_Y \Z$ (classically) or its
deformation (quantum-mechanically). From this viewpoint, the
monoidal structure on the category of line operators is not natural.
Nevertheless, there is a purely two-dimensional interpretation of
this monoidal structure. Consider a a 3-manifold $M=\RR\times
\Sigma$ where $\Sigma$ is a disc. On the boundary of $M$ we specify
a boundary condition described by $\Z$ and insert a line operator
which is located at a point on the boundary of $\Sigma$. Consider
squashing $\Sigma$ into an interval (see Fig. \ref{fig:linemono}).
According to section \ref{sec:intervalred}, the RW model reduces to
the curved B-model on $I\times\RR$ with target $\Z^*\times_Y\Z$
(more precisely, its deformation), and the boundary line operator
becomes a bulk line operator in the B-model located at a point in
the interior of $I$. Fusing boundary line operators in the RW model
obviously reduces to fusing bulk line operators in the B-model.

The category of bulk line operators in a curved B-model with target
$Z$ is equivalent to the curved derived category of $Z^*\times Z$
where $Z^*$ denotes $Z$ with the opposite curving. The monoidal
structure is given by the convolution product. Equivalently, an
object of the curved derived category of $Z^*\times Z$ can be
thought of as a functor from the curved derived category of $Z$ to
itself, and the monoidal structure is given by the composition of
such functors. This monoidal structure is not symmetric (i.e. it is
noncommutative).

Bulk line operators of the reduced theory which originate from
boundary line operators in the parent RW theory on an interval form
a monoidal subcategory in the curved derived category of $Z^*\times
Z$, where $Z$ is a deformation of $\Z^*\times_Y \Z$. In general,
there is no reason to expect this monoidal subcategory to be
symmetric.

The curved derived category of $\Z^*\times_Y \Z$ has an obvious
monoidal structure given by the convolution product. Equivalently,
we can think of an object of the curved derived category of
$\Z^*\times_Y\Z$ as a functor from the curved derived category of
$\Z$ to itself, and the obvious monoidal structure is given by the
composition of such functors. It is easy to see that on the
classical level this is the right monoidal structure. We will now
argue that it receives quantum corrections which depend not only on
the geometry of $\Z$, but also on the way $Y$ is embedded into $X$.

To simplify the discussion, we will consider in detail the case
$\Z=Y$ and vanishing curving. In this case the category of boundary
line operators is simply the $\ZZ_2$-graded derived category of $Y$,
and the obvious monoidal structure is given by the tensor product of
vector bundles on $Y$. This monoidal structure is symmetric and the
unit object is the trivial rank-one bundle on $Y$. It is obvious
that fusion of line operators on the classical level corresponds to
tensoring the corresponding vector bundles. We would like to
determine the leading quantum correction to this result.

The most straightforward way is to evaluate the product of boundary
line operators in perturbation theory; in section \ref{app:fusion}
we do it to leading order in $\hbar$. In this section we will take
an indirect route which is less computationally demanding. Let
$(E,D)$ be an object of the $\ZZ_2$-graded derived category of $Y$,
i.e. $E$ is a smooth $\ZZ_2$-graded vector bundle on $Y$ and $\cK$
is an odd differential operator on $\obul(E)$ satisfying
(\ref{Dcond}) with $W=0$. We can represent the operation of
tensoring with $(E,D)$ by an object of the $\ZZ_2$-graded derived
category of $Y\times Y$, namely the push-forward of $(E,D)$ from the
diagonal in $Y\times Y$. An infinitesimal deformation of this
functor corresponds to an infinitesimal deformation of the
corresponding object on $Y\times Y$. The latter are classified by
odd elements of the endomorphism vector space of $\delta_*(E,D)$,
where $\delta$ is the diagonal embedding of $Y$ into $Y\times Y$.
This space is isomorphic to the cohomology of $D$ on the space
$\obul(\End(E))\otimes\Lambda^\bul T_Y$ (this is proved in the same
way as in the usual derived category).

We can determine the leading-order deformation without any
computations as follows. The boundary line operator corresponding to
a pair $(E,D)$ is the holonomy of a certain superconnection $\cN$
defined by (\ref{cNdef}). This superconnection depends on the
restrictions of various fields to the 1-dimensional submanifold
$\gamma\in\partial M$. Quantum corrections to the composition of
line operators are obtained by expanding the holonomy of $\cN$ in a
power series in the fields and evaluating the contractions of
fermionic and bosonic fields. The contractions arise from
interactions and therefore must involve the curvature of $Y$, the
second fundamental form of $Y\subset X$, and their derivatives. From
the algebro-geometric viewpoint the curvature represents an element
of $H^1(\Omega^1_Y\otimes\End T_Y)$ (the Atiyah class of the tangent
bundle of $Y$), while the second fundamental form $\beta$ represents
an element of $H^1(T_Y\otimes N_Y^*)$ determined by the exact
sequence (\ref{exactsequenceY}). For a Lagrangian submanifold one
has $N_Y^*\simeq T_Y$, and $[\beta]\in H^1(\Sym^2 T_Y)$. The leading
term in the expansion of the holonomy of $\cN$ is proportional to
the curvature of $D$, i.e.
$$
F=\nabla D+D\nabla\in \Omega^1_Y \otimes \obul(\End(E)).
$$
The form $F$ is $D$-closed and can be regarded as a representative
of the Atiyah class of the object $(E,D)$ in the $\ZZ_2$-graded
derived category.

These considerations uniquely determine the leading deformation to
be
$$
\beta \corn F\in T_Y\otimes \obul(\End(E)),
$$
where $\corn $ denotes a contraction between a holomorphic bi-vector
field and a holomorphic 1-form. This class is odd, as required, and
$D$-closed (because $\beta$ is $\bpartial$-closed and $F$ is
$D$-closed). Thus it defines a leading deformation of the
pushforward of $(E,D)$ from $Y$ to $Y\times Y$.

It remains to check whether the coefficient of the leading
deformation is really nonzero (the normalization of the coefficient
is unimportant). This is done in section \ref{app:fusion} by
evaluating the OPE of the Wilson loops to leading order in $\beta$.
As a preparation, let us make the deformation of the monoidal
structure a bit more explicit in the case when $(E,D)$ is a
holomorphic vector bundle on $Y$. Then $F$ is the ordinary curvature
of $E$, and the deformation corresponds to an element
$$
[\beta \corn F]\in H^2(T_Y\otimes \End\, E).
$$
Thus if $E_1$ and $E_2$ are two holomorphic vector bundles on $Y$,
the composition of the corresponding line operators will be the
smooth vector bundle $E_{12}=E_1\otimes E_2$ equipped with a
differential operator
$$
D_{12}=\bpartial+\beta \corn (F_1\wedge F_2)
$$
on $\obul(E_1\otimes E_2)$. Here $\bpartial$ is the ordinary
$\bpartial$-operator on $E_1\otimes E_2$ and $F_1$ and $F_2$ are
curvature forms of $E_1$ and $E_2$. Note that the quantum correction
to $\bpartial$ changes sign under the exchange of $E_1$ and $E_2$
which means that the deformed monoidal structure is not symmetric.
Note also that the correction is a $(0,3)$ form on $Y$, so the
composition of line operators corresponding to holomorphic line
bundles is no longer a holomorphic line bundle but a more general
object of the $\ZZ_2$-graded derived category of $Y$.

If the class $[\beta]$ vanishes, then to leading order in the Planck
constant the monoidal structure of the category of boundary line
operators is not deformed. However, it is likely that at higher
orders in the Planck constant the monoidal structure will be
affected by the geometry of higher-order infinitesimal neighborhoods
of $Y$ in $X$. For example, at quadratic order in the Planck
constant we have a class $[\beta']\in H^1(\Sym^3 T_Y)$ which likely
deforms the associativity morphism in the category of boundary line
operators. We will see an example of this below.

To conclude this section we will show that the invisible line
operator on $Y$ (i.e. the unit object in the category of boundary
line operators on $Y$) does not receive quantum corrections. On the
classical level, the unit object is the structure sheaf $\cO_Y$. A
quantum correction would correspond to an element $\mu\in H^{\hat
1}(\cO_Y)$ which has ghost number $1$. This element must be
constructed from the curvature tensor of $Y$ and the classes
$\beta^{(\ell)}\in H^1(\Sym^\ell T_Y)$. Suppose $\mu$ contains $m$
curvature tensors as well as classes $\beta^{(\ell_i)},$
$i=1,\ldots,N$. Ghost number symmetry implies
$$
m+N+\sum_i \left(2-2\ell_i\right)=1.
$$
Hence $N=m+1\, {\rm mod}\, 2$. This also ensures that the
cohomological degree of $\mu$ is odd, as required.

Now we note that the action of the RW theory has a parity symmetry
if we require all fields to be parity-even and $\smpf$ to be
parity-odd. The boundary condition corresponding to $Y$ is
parity-invariant, and so is the invisible line operator on the
boundary. Thus the class $\mu$ must be parity-even. By definition,
$\beta^{(\ell)}$ contains $\smpf^{-1}$ raised to the power $\ell-1$
and so its parity is $\ell-1\, {\rm mod}\, 2$. This implies
$$\sum_i\ell_i=N\, {\rm mod}\, 2$$
and therefore
$$
\sum_i\ell_i=m+1\, {\rm mod}\, 2.
$$
But then it is impossible to contract all holomorphic indices to get
an element of $H^{\hat 1}(\cO_Y)$. Thus the invisible line operator
is undeformed.

\subsubsection{Line operators in the bulk}

The monoidal 2-category of surface operators in the RW model has a
unit object: the invisible surface operator. Its category of
endomorphisms is the Hochschild cohomology of the 2-category of
surface operators. From the physical viewpoint, it should be thought
of as the category of line operators in the bulk.

Since the invisible surface operator is represented by the diagonal
$\Delta\hookrightarrow X\times X$ with a trivial curving, the
category of line operators in the bulk is the $\ZZ_2$-graded derived
category of $X$. An object of this category can be represented by a
$\ZZ_2$-graded smooth vector bundle $E$ on $X$ equipped with a
differential operator $D$ on $\obul(E)$ satisfying $D^2=0$ and the
Leibniz rule
$$
D(\omega\wedge\sigma)=\bpartial\omega\wedge
\sigma+(-1)^{\deg_{\ZZ_2}\omega}\,\omega\wedge D\sigma,\\
\forall \omega\in\obul(X), \forall \sigma\in\obul(E).
$$
The operator $D$ is called a $\bpartial$-superconnection. One can
get such an object from a complex of holomorphic vector bundles by
letting $D=\bpartial+\cK$, where $\cK$ is the differential in the
complex. This generalizes the observation of \cite{RW} that one can
define a BRST line operator for any holomorphic vector bundle on
$X$.

If we take $E$ to be trivial rank-one vector bundle, then the Wilson
line operator is simply the identity operator, i.e. it is invisible.

The category of bulk line operators has a monoidal structure
obtained by fusing the operators together. The invisible line
operator is the unit object; it corresponds to the structure sheaf
$\cO_X$. The second fundamental form for the diagonal in $X^*\times
X$ vanishes, so to leading order in the Planck constant the monoidal
structure is undeformed. But already at second order in the Planck
constant we expect corrections, since the the second-order
infinitesimal neighborhood of the diagonal is not isomorphic to the
second-order infinitesimal neighborhood of the zero section in
$T^*_X$ \cite{Kapranov}. The deviation is parameterized by the
Atiyah class of $T_X$, which can be regarded as an element of
$H^1(\Sym^3 T_X)$. Such a correction cannot affect the tensor
product itself (i.e. cannot affect the isomorphism class of
$E_1\otimes E_2$), but it can affect the associativity morphism
which tells us how to identify $(E_1\otimes E_2)\otimes E_3$ and
$E_1\otimes (E_2\otimes E_3)$. Thus at quadratic order in the Planck
constant we expect a correction to the associativity morphism
proportional to the contraction of this class with the Atiyah
classes of the three objects involved.

The deformation of the associativity morphism is related to the fact
that the category of bulk line operators is a braided monoidal one,
with a nontrivial braiding \cite{RW,RobertsWillerton}. The braided
structure comes about because diffeomorphisms of a disc with marked
points (at which line operators are inserted) act functorially on
the category which TQFT associates to such a disk, and
diffeomorphisms isotopic to identity act trivially. On the classical
level, the braiding is trivial, but at linear order in the Planck
constant a nontrivial braiding appears proportional to the Atiyah
classes of the two objects involved \cite{RW,RobertsWillerton}.

\subsubsection{The categorified boundary-bulk map}

In the 2d TFT, there is a boundary-bulk map which sends an object of
the category of branes to a local operator. This maps factors
through the K-theory of the category of branes and can be regarded
as a homomorphism from the K-theory to the additive group of the
space of local operators. Pictorially, it is represented by a disc
1-correlator with an insertion of a bulk operator, or equivalently
by an annulus whose inner boundary component is a cut boundary, and
the outer boundary component is the brane boundary.

In 3d TFT the same geometric structure corresponds to a map which
sends an object of the 2-category of boundary conditions to a bulk
line operator. One can interpret it as a categorified version of the
Chern character. Let us determine it for boundary conditions in the
RW model.

Consider first the case when the boundary condition is described by
a complex Lagrangian submanifold $Y$. Let $(E,D)$ be an object of
the $\ZZ_2$-graded derived category of $X$. We need to compute the
vector space associated to a disc with an insertion of the bulk line
operator $(E,D)$ and the boundary condition $Y$. The bosonic zero
modes parameterize $Y$, while the fermionic zero modes are
components of $\eta$ tangent to $Y$. The Hilbert space of the zero
modes is therefore the space of smooth sections of
$\obul(E\vert_Y)$, and the vector space associated to the punctured
disc is the cohomology of $\obul(E\vert_Y)$ with respect to $D$.
This means that the line operator corresponding to $Y$ is simply the
structure sheaf of $Y$ pushed forward to $X$.

To determine the answer for a fibration $\Z$ over $Y$, note that one
can deform the disc so that it looks like the surface of a cigar.
One can then imagine the circumference of the cigar shrink to zero
size. In this limit the boundary condition corresponding to $Y$
reduces to a boundary condition in the B-model with target $X$, and
therefore the boundary-bulk map is simply reduction on a circle.
Reduction on a circle for boundary conditions in the RW model has
been described in section \ref{sec:redcircle}. Namely, given a
fibration $\Z$ with compact fibers we take the flat vector bundle on
$Y$ given by the fiberwise de Rham cohomology and push it forward to
$X$.

\subsubsection{Line operators corresponding to submanifolds in the
target space}

In the case of the B-model one can define a boundary condition
corresponding to an arbitrary complex submanifold $Y$ in the target
space $X$. It is believed that the resulting brane corresponds to an
object of the derived category of $X$ which is the pushforward of
$\cO_Y$ to $X$. Such an object has a resolution by locally free
sheaves, and therefore is isomorphic to a certain complex of
holomorphic vector bundles on $X$. Thus as far as the B-model is
concerned, it is sufficient to consider branes corresponding to
complexes of holomorphic vector bundles on $X$.

If $X$ is a complex symplectic manifold, we can use the resolution
of $\iota_*\cO_Y$ to construct a line operator in the RW model with
target $X$. Different resolutions will correspond to isomorphic line
operators, but this is far from obvious from our construction of
line operators. It is interesting to ask whether in the RW model
there is an alternative description of the same line operator which
is manifestly independent of the choice of resolution. By analogy
with the B-model, one expects that such a line operator will be
defined by the condition that $\phi$ maps the support $\gamma$ of
the line operator to $Y$.

In this section we will describe a candidate for such a line
operator in the RW model. Since it appears unnatural to impose
conditions on $\phi$ on a codimension-two submanifold, we will
regularize the problem by excising a tubular neighborhood $\cM$ of
$\gamma$ and will define the line operator by imposing suitable
boundary conditions on fields on the boundary of $M\backslash\cM$.

We will use cylindrical coordinates in the neighborhood of $\gamma$:
$(r,\varphi,t)$, $r\in [0,\infty)$, $\varphi\in [0,2\pi),$ $t\in
\RR$ so that $\cM$ is given by $r < r_0$. In a topological field
theory, one can assume $r_0$ to be arbitrarily small, since the
correlators do not depend on it. Keeping this in mind, we will
impose conditions at $r=r_0$ which are nonlocal in the
$\varphi$-direction but local in the $t$-direction. This nonlocality
will not matter when we take $r_0$ to be very small.

Near $r=r_0$ we can expand each field into a Fourier series, for
example $\phi^I=\sum_{n\in \ZZ} \phi^I_n(r,t) e^{in\varphi},$ etc.
To ensure that the limit $r_0\raa 0$ is non-singular we will require
\be \label{e6.1} \phi_n^{I}=0,\quad \phi_n^{\bar I}=0,\quad
\eta_n^{\bar I}=0,\quad
\Lambda\chi_{t,n}^{I}-\chi_{\varphi,n}^I=0,\quad n\ne 0,\quad
I,{\bar I}=1,\ldots,2n\ee where the parameter $\Lambda^{-1}$ has
mass dimension one. These conditions do not depend on the
submanifold $Y$.

To write down the conditions for the constant Fourier-modes, suppose
that $Y$ is locally defined by the equations $\phi^{i'}=0$,
$i'=k+1,\ldots,2n$. We may use $\phi^i, i=1,\ldots,k,$ as local
coordinates on $Y$. We will require
\begin{equation} \label{e6.2}
\phi_0^{i'}=0,\quad \phi_0^{\bar i'}=0,\quad \eta_0^{\bar
i'}=0,\quad \Lambda \chi_{t,0}^{i'}-\chi_{\varphi,0}^{i'}=0,\quad
i',{\bar i'}=k+1,\ldots,2n,\end{equation}
\begin{equation}
\label{e6.3} \pi_{i,0}=0,\quad \pi_{\bi,0}=0,\quad g_{J\bar
i}\chi_{r,0}^J=0,\quad g_{J\bar i}\chi_{\varphi,0}^J=0,\quad i,{\bar
i}=1,\ldots,k.
\end{equation}
Here $\pi_i$ and $\pi_\bi$ are canonical momenta for $\phi^i$ and
$\phi^\bi$, if regard $r$ as the time coordinate. The system of
constraints (\ref{e6.1}-\ref{e6.3}) is BRST invariant and ensures
that $\partial\cM$ is mapped to the $k$-dimensional submanifold $Y$.

In particular, if $Y=X$, the limit $r_0\raa 0$ of these conditions
simply says that all fields are smooth at $r=0$. This means that
this line operators is trivial, or invisible.

\subsection{Point operators}

Let us make a few remarks about point operators, i.e. operators
localized at points. A more standard name for them is local
operators. The most general case is when such an operator is
inserted at a joining point of two line operators. The space of such
operators is the space of morphisms in the category of line
operators. As explained above, the categories of line operators
arising in the RW model are all equivalent to curved derived
categories of certain complex manifolds, so in this way we get a
complete description of point operators and their algebra. Let us
give a few examples.

Consider first a boundary condition corresponding to a fibration
$\Z$ over a Lagrangian submanifold $Y$. A point operator on such a
boundary can be thought of as sitting on an invisible line operator,
which is represented by the fibered diagonal in $\Z\times_Y \Z$. The
space of such line operators is the endomorphism algebra of the
structure sheaf of the fibered diagonal $\Delta_Y$ which on the
classical level is isomorphic to
\begin{equation}\label{bdrylocalfibr}
\oplus_r \Ext^r(\cO_{\Delta_Y},\cO_{\Delta_Y})\simeq \oplus_{p,q}
H^p(\Lambda^q T^\ver_\Z).
\end{equation}
As discussed above, there may be nontrivial quantum corrections to
this result due to nontrivial external geometry of $Y$; thus
(\ref{bdrylocalfibr}) is merely the first term in a spectral
sequence which converges to the quantum algebra of boundary
observables.

In the special case $\Z=Y$, (\ref{bdrylocalfibr}) reduces to
\begin{equation}\label{bdrylocalnofibr}
\oplus_p H^p(\cO_Y),
\end{equation}
which agrees with section \ref{sec:defs}. In this case no quantum
corrections are possible. Indeed, we have shown above that in this
case neither the category of boundary line operators nor the unit
object receive quantum corrections. Hence the endomorphism algebra
of the unit object is also uncorrected.

Our last  example is a point operator in the bulk, which can be
thought of as sitting on an invisible line operator in the bulk. The
category of bulk line operators is equivalent to the $\ZZ_2$-graded
derived category of $X$. The invisible line operator corresponds to
the structure sheaf of $X$, so the space of point operators is
$$
\oplus_p \Ext^p(\cO_X,\cO_X)=\oplus_p H^p(\cO_X).
$$
This agrees with \cite{RW}. Again, no quantum corrections are
possible.

The algebra of local observables in any $N$-dimensional
$\ZZ_2$-graded TFT is a supercommutative algebra with a Lie bracket
of degree $N-1 {\rm mod}\, 2$. These two algebraic structures are
compatible in an obvious sense. The Lie bracket arises as follows:
to any local observable one can associate a tower of descendants.
The $(N-1)^{\rm th}$ descendant is a $N-1$ form of ghost number $1-N
{\mod 2}$ which is closed up to BRST-exact terms. Thus one can think
of it as a conserved current which generates an infinitesimal
symmetry of the TFT. All such symmetries form a Lie superalgebra
with a bracket of degree $0$. One can show that the symmetries
arising from descendants of local observables form a subalgebra of
this Lie superalgebra, and therefore the algebra local observables
inherits a Lie bracket of degree $N-1~~{\rm mod~~ 2}$.

For example, in the curved B-model the algebra of local observables
carries an odd bracket. A supercommutative algebra with a compatible
odd Lie bracket bracket is called a $\ZZ_2$-graded Gerstenhaber
algebra. It is well-known that the Hochschild cohomology of any
$\ZZ_2$-graded DG-category carries such a structure; in the case of
the curved B-model the relevant category is the category of branes.

In the RW model, we expect that the algebra of local observables
carries an even Lie bracket; the resulting structure is simply a
Poisson algebra. Indeed, since $X$ is a complex symplectic manifold,
the sheaf $\cO_X$ is obviously a sheaf of Poisson algebras, and
therefore its cohomology is a Poisson algebra. In the compact
K\"ahler case the Poisson bracket is trivial; analogously, in the
B-model with a compact K\"ahler target the Gerstenhaber bracket is
trivial.

\section{Relationship with categorified algebraic
geometry}\label{sec:concl}

In this paper we have studied topological boundary conditions and
topological defects of various codimensions for the RW model. We
have seen that boundary conditions form a 2-category: morphisms
between two boundary conditions are boundary line operators
separating two parts of the boundary, and 2-morphisms between line
operators are local operator insertions on line operators. We have
described a large class of boundary conditions for the RW model with
target $X$, namely complex fibrations over complex Lagrangian
submanifolds in $X$. The structure of this 2-category is rather
intricate, and it would be desirable to have a more algebraic
description for it. A good analogy is the B-model: the structure of
the category of B-branes is greatly clarified by relating it to the
category of coherent sheaves, i.e. a category of modules over the
sheaf of holomorphic functions.

While we do not know such an algebraic description in general, we
can propose a candidate description for the full sub-2-category
whose objects are fibrations over a fixed complex Lagrangian
submanifold $Y$. Let ${\mathsf A}_0$ denote the boundary condition
corresponding to $\Z=Y$. Let $\mathsf A$ be any other boundary
condition supported on $Y$. The category of boundary-changing line
operators between $\mA_0$ and $\mA$ (i.e. the category ${\rm
Hom}(\mA_0,\mA)$ in the 2-category of boundary conditions) is a
module over the monoidal category of boundary operators on $\mA_0$
(i.e. the category $\Hom(\mA_0,\mA_0)$).\footnote{Monoidal category
is a categorification of an algebra. A module over a monoidal
category is defined by categorifying the usual definition of a
module over an algebra. Modules over a monoidal category form a
2-category.} If $\mA$ is supported on $Y$, one might conjecture that
the category ${\rm Hom}(\mA_0,\mA)$ determines $\mA$; then one could
hope to identify the 2-category of boundary conditions with the
2-category of modules over the monoidal category ${\rm
Hom}(\mA_0,\mA_0)$.

By analogy with the B-model, we expect that this conjecture is true
if $Y$ is a (smooth) affine variety; otherwise we need to
``sheafify'' the categories ${\rm Hom}(\mA_0,\mA)$ and ${\rm
Hom}(\mA_0,\mA_0)$, i.e. consider a sheaf of categories over $Y$
which is a module over a sheaf of monoidal categories.

On the classical level, we can make all this rather explicit. The
monoidal category ${\rm Hom}(\mA_0,\mA_0)$ is the $\ZZ_2$-graded
derived category of coherent sheaves on $Y$, or better the
DG-category of $\ZZ_2$-graded complexes of $\cO_Y$-modules, with the
obvious monoidal structure. Any fibration $\Z\raa Y$ with a curving
$W$ is naturally a module over this category: the module structure
is given by tensoring an object on $\Z$ by a pull-back of an object
on $Y$.

Quantum corrections deform the monoidal category
$\Hom(\mA_0,\mA_0)$, so for the boundary condition $\mA$ to make
sense on the quantum level, the module $\Hom(\mA_0,\mA)$ must admit
a deformation as well. In general, as in the case of modules over an
algebra, there are obstructions for finding such a deformation. This
is what we found in section \ref{sec:fibrations}; the leading
deformation turned out to be quadratic in the Atiyah class of the
fibration $\Z$ and the curving $W$ on $\Z$ and linear in the class
$[\beta]$. Presumably, all obstructions can be determined by
studying the deformation theory of modules over the monoidal
category $\Hom(\mA_0,\mA_0)$.

As discussed above, deformations of $Y$ in $X$ are equivalent to
turning on a nontrivial curving $W$ on $Y$. This suggests that the
2-category of modules over $\Hom(\mA_0,\mA_0)$ also includes
boundary conditions which are supported on a formal neighborhood
$Y$. In general, this neighborhood is a deformations of $T^*_Y$, and
the monoidal structure on the category $\Hom(\mA_0,\mA_0)$ receives
corrections from the classes $\beta^{(\ell)}\in H^1(\Sym^\ell T_Y)$
describing this deformation. Thus the RW model ``solves'' the
problem of deformation quantization of the monoidal category
$D_{\ZZ_2}(\Coh(Y))$.

A particularly simple and instructive case is when all classes
$\beta^{(\ell)}$ vanish. In this case we can simply take $X=T^*_Y$.
We conjecture that the 2-category of boundary conditions for the RW
model with target $T^*_Y$ is equivalent to the 2-category of modules
over the $\ZZ_2$-graded derived category of $Y$, or better the
category of modules over the DG-category of complexes of $\cO_Y$
modules, regarded as a monoidal DG-category. We can view an object
of this 2-category as a sheaf of DG-categories over $Y$.

A very similar 2-category, called the 2-category of derived
categorical sheaves over $Y$, was introduced in \cite{To-Ve} (see
also \cite{BFN}). The aim of these papers was to develop a kind of
categorification of algebraic geometry, with sheaves of
DG-categories taking place of complexes of coherent sheaves. One
apparent difference is that complex fibrations appear to be rather
special sheaves of DG-categories, analogous in some sense to
holomorphic vector bundles in the B-model. One may wonder if more
general sheaves of categories, e.g. skyscraper sheaves, play a role
in the RW theory. In analogy with the B-model, one could try to
build more complicated boundary conditions by considering
``complexes'' of fibrations. However, it is not clear what this
could mean. Instead, we will argue in \cite{KRS2} that skyscraper
sheaves of DG-categories and other more exotic categorical sheaves
appear in the RW theory automatically if we allow for nontrivial
curving. In a sense, we will see that turning on a curving on a
fibration is analogous to deforming a graded holomorphic vector
bundle into a complex of holomorphic vector bundles.

Another apparent difference compared to \cite{To-Ve,BFN} is that the
authors of \cite{To-Ve,BFN} work with $\ZZ$-graded complexes of
$\cO_Y$ modules, rather than with $\ZZ_2$-graded ones. We can
account for this difference by noting that in the case $X=T^*_Y$ the
$\ZZ_2$ ghost number symmetry of the RW model can be promoted to a
$U(1)$ symmetry. Indeed, $T^*_Y$ has a $U(1)$ action with respect to
which the holomorphic fiber coordinates have weight $2$. It
preserves the metric and therefore the term $\cL_1$ in the
Lagrangian. On the other hand, the holomorphic symplectic form has
weight $2$ under this action, so this geometric $U(1)$ symmetry does
not preserve $\cL_2$. However, the term $\cL_2$ is invariant under
the combined action of the geometric $U(1)$ and the naive ghost
number symmetry. We conclude that the RW model with target $T^*_Y$
has a $U(1)$ ghost number symmetry with respect to which the fiber
coordinates on $T^*_Y$ have weight two. We can require all boundary
conditions and line operators to be equivariant with respect to this
symmetry.

The relationship between categorified algebraic geometry and 3d TFT
has been already pointed out in \cite{BFN}, from a somewhat
different viewpoint. First of all, the definition of a 3d TFT used
in \cite{BFN} is different from ours. Instead of considering the
3-category of decorated structures of dimension $n\leq 3$, the
authors consider the $\infty$-category of two-dimensional
cobordisms. The TFT is defined, roughly, as a functor from this
$\infty$-category to the 2-category of DG-categories. The authors of
\cite{BFN} construct a TFT in this sense for any complex manifold
$Y$ (and more generally for any stack satisfying certain
properties).

To compare with the discussion in \cite{BFN}, note that the TFT in
the sense of \cite{BFN} assigns a DG 2-category to a point, a
braided monoidal DG-category to a circle and a DG-vector space to a
Riemann surface. For a complex manifold $Y$, the 2-category assigned
to a point is the 2-category of categorical sheaves over $Y$. As
discussed above, this agrees with the RW theory. (On the quantum
level, one has to require $Y$ to be a Calabi-Yau manifold. This
condition does not arise in \cite{BFN}, because the authors do not
try to extend their theory to three-dimensional cobordisms.)

Further, according to \cite{BFN} one assigns to a circle the
DG-category of complexes of quasicoherent sheaves on the
supermanifold $T_Y[-1]$ (i.e. the odd tangent bundle where the fiber
coordinates have degree $-1$). By Koszul duality, it is equivalent
to the category of complexes of quasicoherent sheaves on the graded
manifold $T^*_Y[2]$ (i.e. $T^*_Y$ with a $\CC^*$-action such that
the fiber coordinates have weight $2$). This is the same as the
category of bulk line operators in the $\ZZ$-graded version of the
RW model with target $T^*_Y$. In principle, one also needs to
compare the braided monoidal structures on the two categories; we
conjecture that they agree.

The unit object in the monoidal category attached to a circle is the
structure sheaf of $T^*_Y$ (with the obvious $\CC^*$-action). From
the physical viewpoint, it represents the invisible line operator in
the bulk. Its Dolbeault complex is the vector space attached to a
2-sphere, or in physical terms the space of local operators in the
bulk. The cohomology of this vector space can be identified with
$$
H^\bul(\Sym^\bul T_Y[2]).
$$
It is a supercommutative algebra with a compatible Poisson bracket
of degree $-2$; as explained above, the Poisson bracket comes from
the Poisson bracket on forms on $T^*_Y$. This graded Poisson algebra
is the space of infinitesimal deformations of the RW model with
target $T^*_Y$; from the mathematical viewpoint it describes
infinitesimal deformations of the 2-category of modules over the
monoidal DG-category of complexes of $\cO_Y$ modules.

\appendix

\section{The B-model and its category of branes}\label{app:notcurved}

\subsection{BRST transformations, action, and
observables}\label{app:Bmodel}

Let $X$ be a complex manifold with local complex coordinates
$\phi^i$, $i=1,\ldots,n$. The fields of the B-model with target $X$
consist of a map $\phi:\Sigma\raa X$, fermionic 0-forms
$\eta\in\Gamma(\phi^*{\bar T}_X)$ and
$\theta\in\Gamma(\phi^*T^*_X)$, and a fermionic 1-form
$\rho\in\Gamma\left(\phi^*T_X\otimes T^*\Sigma\right)$. The BRST
transformations are
\begin{align}
\delta_Q \phi^i&=0, & \delta_Q\phi^{\bar i}&=\eta^{\bar i},\\
\delta_Q \eta^{\bar i}&=0, & \delta_Q\theta_i&=0,\\
\delta_Q \rho^i&=d\phi^i. &  &
\end{align}
The action is
$$
S_0=\int_\Sigma \left(\delta_Q \left( g_{i\bar j}\rho^i \wedge
*d\phi^{\bar j}\right)-\theta_i
\nabla\rho^i-R^i_{jk\bar l}\theta_i\rho^j\rho^k\eta^{\bar l}\right).
$$
Here $g_{i\bar j}$ is a K\"ahler metric on $X$, $\nabla$ is the
covariant differential with respect to the pull-back of the
Levi-Civita connection, and $R^i_{jk\bar l}$ is the curvature tensor
of the Levi-Civita connection.

This action defines a topological field theory, because it depends
on the worldsheet metric $h$ only through BRST-exact terms. It is
called the B-model with target $X$ \cite{Wittenmir}. Its algebra of
topological observables (i.e. the cohomology of $\delta_Q$ on the
space of local functions of fields) is isomorphic to
$$
\cA=\oplus_{p,q} H^p\left(\Lambda^qT_X\right).
$$
The isomorphism acts as follows:
$$
\frac{1}{p!q!}\omega_{\bi_1\ldots\bi_p}^{i_1\ldots i_q}
d\phi^{\bi_1}\ldots d\phi^{\bi_p}
\partial_{i_1}\ldots\partial_{i_q}\mapsto \frac{1}{p!q!}
\omega_{\bi_1\ldots\bi_p}^{i_1\ldots i_q} \eta^{\bi_1}\ldots
\eta^{\bi_p} \theta_{i_1}\ldots\theta_{i_q}.
$$
Correlators of topological observables depend only on the complex
structure on $X$, because the variation of $S_0$ with respect to the
metric $g_{i\bj}$ is BRST-exact \cite{Wittenmir}.

On the quantum level, one usually requires $X$ to be a Calabi-Yau
manifold, i.e. one requires the existence of a holomorphic volume
form $\Psi$ on $X$. The volume form is used to write down a
BRST-invariant measure on the space of bosonic and fermionic zero
modes. Given $\Psi$, one defines a trace function on $\cA$ as
follows: it is nonvanishing only on $H^n(\Lambda^n T_X)$ and is
given by
$$
\omega\mapsto \int_X \Psi\wedge \omega\corn \Psi.
$$
This trace function makes $\cA$ into a Frobenius algebra.

The formula for the trace function makes sense in the case when the
canonical line bundle $K_X$ of $X$ is not necessarily trivial, but
its square is. We will call such $X$ a half-CY manifold. The closed
sector of the B-model is well-defined in this slightly more general
case.

The B-model has a $U(1)$ symmetry known as the ghost number
symmetry. It acts trivially on the bosonic fields; the fermionic
0-forms have weight $1$; the fermionic 1-form $\rho$ has weight
$-1$. The BRST charge itself has weight $1$. One may choose to keep
track only of the $\ZZ_2$ subgroup of the ghost number symmetry; the
resulting theory is a $\ZZ_2$-graded version of the B-model.

\subsection{Topological branes in the B-model}\label{app:branes}

To any topological field theory one can associate a category of
topological boundary conditions usually called the category of
branes. In the case of the B-model with a Calabi-Yau target $X$,
this is the bounded derived category of coherent sheaves (or more
precisely, the bounded derived category of complexes of
$\cO_X$-modules with coherent cohomology). In the $\ZZ_2$ graded
version of the theory, this presumably becomes the derived category
of 2-periodic complexes of $\cO_X$-modules with coherent cohomology.

Let us describe objects and morphisms in the category of B-branes.
For definiteness, we will work with the $\ZZ_2$-graded version of
the B-model. An object of the category of B-branes is a
$\ZZ_2$-graded smooth vector bundle $E=E^+\oplus E^-$ equipped with
a flat $\bpartial$-superconnection $D$. A flat
$\bpartial$-superconnection is an odd element of the $\ZZ_2$-graded
vector space
$$
\End_\CC(\obul(E))
$$
satisfying
\begin{multline}\label{Dcond}
D^2=0,\quad D(\omega\wedge\sigma)=\bpartial\omega\wedge
\sigma+(-1)^{\deg_{\ZZ_2}\omega}\,\omega\wedge D\sigma,\\
\forall \omega\in\obul(X), \forall \sigma\in\obul(E).
\end{multline}
One can write $D$ as a sum of a $\bpartial$-connection
$\bpartial_E:\Omega^{0,p}(E)\raa\Omega^{0,p+1}(E)$ preserving the
decomposition $E=E^+\oplus E^-$ and an odd section of the
$\ZZ_2$-graded vector bundle $\obul(\End(E))$. Note that
$\bpartial_E$ does not square to zero by itself, in general, so $E$
does not have a natural holomorphic structure. This description of
B-branes goes back to Lazaroiu \cite{Laz2} and Diaconescu
\cite{Dia}; more recently it was discussed in \cite{HHP,Bergman}.

Given two such pairs $(E_1,D_1)$ and $(E_2,D_2)$ we consider the
$\ZZ_2$-graded vector space
$$
\obul\left(\Hom(E_1,E_2)\right)
$$
and its odd endomorphism $D_{12}$ defined by
$$
D_{12}f=D_2\,f - (-1)^{\deg_{\ZZ_2}f}\,f D_1,\quad f\in
\obul(\Hom(E_1,E_2)).
$$
It is easy to see that $D_{12}^2=0$. We define the space of
morphisms from $(E_1,D_1)$ to $(E_2,D_2)$ to be the cohomology of
$D_{12}$.

Note that given a $\ZZ_2$-graded complex of holomorphic vector
bundles $E$ on $X$ one gets an object of the category of B-branes by
letting $D=\bpartial+\cK$, where $\cK$ is the differential in the
complex . The space of morphisms between two such objects is the
hypercohomology vector spaces of the $\Hom$-complexes. This shows
that for smooth algebraic $X$ the derived category of 2-periodic
complexes of coherent sheaves on $X$ is equivalent to a full
subcategory of the category of B-branes. We conjecture that in
general the $\ZZ_2$-graded version of the category of B-branes is
equivalent to the derived category of 2-periodic complexes of
$\cO_X$-modules with coherent cohomology. For a detailed discussion
of the $\ZZ$-graded case see \cite{Block,Bergman,BondalRosly}.

To see how a pair $(E,D)$ defines a B-brane, we need to choose
$\partial$-connections on $E^+$ and $E^-$ which in a local
trivialization we write as
$$
\partial^\pm=\partial+\cA^\pm,\quad
\cA^\pm\in\Omega^{1,0}(U,\End(E^\pm)),\ U\subset X.
$$
Further, we write
$$
D=\bpartial\cdot \xId+\cK=\bpartial\cdot\xId+\begin{pmatrix}\cB^+ & \cT\\
\cS & \cB^- \end{pmatrix},
$$
where
\begin{multline}
\cB^\pm \in\Omega^{0,\hat 1}(U,\End(E^\pm)),\ \cT\in
\Omega^{0,\hat 0}(U,\Hom(E^-,E^+)),\\
\cS \in \Omega^{0,\hat 0}(U,\Hom(E^+,E^-)),\ U\subset X.
\end{multline}
Let us introduce the supermanifold $\Pi{\bar T}_X$ with odd
coordinates $\eta^{\bi}$. There is an obvious map $\pi:\Pi{\bar
T}_X\raa X$, and we may regard $\cK$ as a locally-defined odd
section of the $\ZZ_2$-graded bundle $\pi^*\End(E)$. Similarly,
$\bpartial$ can be reinterpreted as an odd vector field
$\eta^{\bi}\partial_{\bi}$ on $\Pi{\bar T}_X$, and $D$ is a
first-order differential operator on $\pi^*E$.

Now let $\gamma$ be a connected component of $\partial\Sigma$, i.e.
a circle. Let $t\in [0,1)$ be a coordinate on $\gamma$. Given a map
$\Phi=(\phi,\eta):\gamma\raa \Pi{\bar T}_X$ and a section $\rho^i_t
dt$ of $\phi^* T_X\otimes T^*_\gamma$, we consider a connection
1-form on the pull-back of $\Phi^*(E)$:
\begin{equation}\label{cNdef}
\cN=\begin{pmatrix} \cA^+_i \dt\phi^i+\cB^+_\bi \dt\phi^\bi
+F^+_{i\bj}\rho^i_t\eta^\bj &
\rho^i_t\nabla_i\cT+\dt\phi^\bi \frac{\partial\cT}{\partial\eta^\bi}\\
\rho^i_t \nabla_i\cS+\dt\phi^\bi \frac{\partial
\cS}{\partial\eta^\bi} & \cA^-_i \dt\phi^i+\cB^-_\bi \dt\phi^\bi
+F^-_{i\bj}\rho^i_t\eta^\bj
\end{pmatrix}
\end{equation}
Here $\dt=d/dt$, $\nabla_i$ denotes the covariant derivatives with
respect to the $\partial$-connections on $\Hom(E^\pm,E^\mp)$, and
$F^\pm_{i\bj}$ is defined by
$$
F^\pm_{i\bj}\eta^{\bj}=\partial_i \cB^\pm-\partial_\bj
\cA^\pm_i\eta^{\bj}+[\cA^\pm_i,\cB^\pm].
$$

One can check that this connection on $\Phi^*(E)$ satisfies
$$
\delta_Q(\dt+\cN)=[\dt+\cN,\Phi^*\cK],
$$
where the brackets denote the graded commutator. We can use this
identity to construct a suitable boundary action as follows. Let
$\U(0,t)$ be the parallel transport operator of the connection
$\cN$, i.e. the unique solution of the first order differential
equation
$$
(\dt+\cN)\U(0,t)=0
$$
satisfying $\U(0,0)=1$. We define
$$
\exp(-S^{bry})=\STr\ \U(0,1),
$$
where $\STr$ denotes the supertrace on $\End_\CC(E_{\Phi(0)})$. Then
it is easy to check that
$$
\delta_Q S^{bry}=0.
$$

Now suppose the circle $\gamma$ is split into two segments $(0,t_0)$
and $(t_0,1)$ by the points $t=0$ and $t=t_0$. Suppose further than
on the segment $(0,t_0)$ the boundary conditions is described by
$(E_1,D_1)$, while on the segment $(t_0,1)$ it is described by
$(E_2,D_2)$. Given a morphism $f_{12}$ from $(E_1,D_1)$ to
$(E_2,D_2)$ and a morphism $f_{21}$ from $(E_2,D_2)$ to $(E_1,D_1)$
consider the following expression to be inserted in the
path-integral:
$$
\STr\ f_{21}(\Phi(1))\U_2(1,t_0)f_{12}(\Phi(t_0))\U_1(0,t_0).
$$
Here $\U_1$ and $\U_2$ are constructed from $(E_1,D_1)$ and
$(E_2,D_2)$ as above, and we regard the bundle-valued forms $f_{12}$
and $f_{21}$ as sections of $\pi^*\Hom(E_1,E_2)$ and
$\pi^*\Hom(E_2,E_1)$. It is easy to check that this expression is
BRST-invariant. Therefore we may regard morphisms $f_{12}$ and
$f_{21}$ as local boundary topological observables to be inserted at
the joining point of boundary conditions $(E_1,D_1)$ and
$(E_2,D_2)$.

\section{The curved B-model and its category of branes}\label{app:curved}

\subsection{BRST transformations, action, and
observables}\label{app:curvedBmodel}

The curved B-model can be thought of as a generalization of the
Landau-Ginzburg deformation of the B-model, so we begin by briefly
reviewing the latter.

Let $W$ be a holomorphic function on a manifold $X$ such that the
line bundle $K_X^2$ is trivial. Then one can deform the B-model with
target $X$ by modifying the transformation law for $\theta_i$:
$$
\delta_Q \theta_i=-\partial_i W
$$
and adding to the action a term
$$
S_W=\frac12 \int\nabla_i\partial_j W\rho^i\rho^j.
$$
(Strictly speaking, to write down this deformation one needs only a
closed holomorphic 1-form $dW$, not $W$ itself. However, if the
cohomology class of $dW$ is nontrivial, the model does not admit
interesting boundary conditions.)

We will call this model the Landau-Ginzburg model with target $X$
and superpotential $W$. Note that the postulated transformation law
for $\theta$ and the action are compatible only with the $\ZZ_2$
subgroup of the ghost number symmetry. Thus the Landau-Ginzburg
model is $\ZZ_2$-graded.

The usual action of the Landau-Ginzburg model differs from ours by a
BRST-exact term
\begin{align}
-\int_\Sigma vol_\Sigma \delta_Q\left(g^{i\bar j}\theta_i
\partial_{\bar j}{\bar W}\right)=\int_\Sigma vol_\Sigma \left(g^{i\bar
j}\partial_i W\partial_{\bar j} {\bar W}-g^{i\bar j}\nabla_{\bar
k}\partial_{\bar j}{\bar W}\theta_i\eta^{\bar k}\right),
\end{align}
where $vol_\Sigma$ is the volume form with respect to a Riemannian
metric on $\Sigma$.

The algebra of topological observables for the Landau-Ginzburg model
is the hypercohomology of the complex
$$
\Lambda^n T_X\raa \Lambda^{n-1} T_X\raa\hdots\raa T_X\raa \cO_X,
$$
where the differential is contraction with the holomorphic 1-form
$-\partial W$. In the case when $X$ is a contractible open subset of
$\CC^n$ and the critical points of $W$ are isolated, the
hypercohomology of this complex is the Jacobi algebra
$$
H^0(X,\cO_X)/A_{\partial W}
$$
where $A_{\partial W}$ is the ideal generated by partial derivatives
of $W$. Note that this depends only on the behavior of $W$ in the
neighborhood of the critical set of $W$. In particular, if $W$ has
an isolated non-degenerate critical point, then the algebra is
isomorphic to $\CC$, and the Landau-Ginzburg model becomes trivial
(such a model is called massive).

More generally, if the critical set of $W$ is a closed submanifold
$T$ and $W$ is a Morse-Bott function (i.e. the Hessian of $W$ is
non-degenerate in the directions normal to $T$), then the directions
normal to $T$ are ``massive'' and can be integrated out. Thus the
model becomes equivalent to the $\ZZ_2$-graded version of the
B-model with target $T$. More precisely, this is true in the case
when $\Sigma$ has empty boundary; if $\Sigma$ has boundaries and the
number of transverse dimensions is odd, the Landau-Ginzburg model
differs from the B-model with target $T$ in a subtle way (see
below).

One can generalize the above construction by replacing $W$ with an
inhomogeneous even $\bpartial$-closed form on $X$. We will call this
generalization the curved B-model. Consider a local observable
$W(\phi,\eta)$ representing an even element of
$$
\oplus_p H^{p}(\cO_X).
$$
Its descendants are
\begin{align}
W^{(1)}&=\rho^i \partial_i W +d\phi^\bi\frac{\partial
W}{\partial\eta^\bi},\\
W^{(2)}&=\frac12 \rho^i\rho^j \nabla_i \partial_j W+\rho^i d\phi^\bj
\partial_i \frac{\partial W}{\partial\eta^\bj}+\frac12 d\phi^\bi
d\phi^\bj \frac{\partial^2 W}{\partial\eta^\bi\partial\eta^\bj}
\end{align}
The descendants satisfy
$$
\delta_Q W^{(1)}=dW,\quad \delta_Q W^{(2)}= d W^{(1)}+\partial_i
W\frac{\delta S_0}{\delta \theta_i}.
$$
We can deform the action of the B-model by a term
$$
S_W=\int_{\partial M} W^{(2)},
$$
and simultaneously modify the BRST transformation for $\theta_i$:
$$
\delta_Q\theta_i=-\partial_i W.
$$
Then the total action is BRST-invariant up to a total derivative:
\begin{equation}\label{brststotW}
\delta_Q (S_0+S_W)=\int_\Sigma d W^{(1)}.
\end{equation}
The resulting theory is the curved B-model. It is $\ZZ_2$-graded,
just like the Landau-Ginzburg model.

The algebra of observables in the curved B-model is computed in
essentially the same way as in the Landau-Ginzburg model. Namely,
the algebra of observables is the cohomology of a certain
differential $\delta_Q$ in the space of $(0,p)$ forms with values in
polyvector fields of type $(q,0)$. This differential is given by
$$
\delta_Q=\bpartial-\partial W\corn ,
$$
where $\corn$ denotes contraction on the holomorphic indices and
exterior product on the antiholomorphic ones. If $X$ is compact and
K\"ahler, then one can always find a form in the cohomology class of
$W$ which is $\partial$-closed. For such $X$ the differential
$\delta_Q$ reduces to $\bpartial$, and the algebra of topological
observables is the same as in the ordinary B-model.

\subsection{Topological branes in the curved B-model}\label{app:curvedbranes}

The category of branes associated to a curved B-model can be
regarded as a deformation of the $\ZZ_2$-graded derived category; we
will call it the curved derived category of $X$. There are several
equivalent ways to define such a deformation; the most natural one
from the physical viewpoint goes as follows. Let $W$ be an even
$\bpartial$-closed element of $\obul$. An object of the curved
derived category is a $\ZZ_2$-graded smooth vector bundle
$E=E^+\oplus E^-$ equipped with a curved differential $D$. A curved
differential is an odd element of the $\ZZ_2$-graded vector space
$$
\End_\CC(\obul(E))
$$
satisfying
\begin{multline}\label{Dcond}
D^2=\xId_E\otimes W,\quad
D(\omega\wedge\sigma)=\bpartial\omega\wedge
\sigma+(-1)^{\deg_{\ZZ_2}\omega}\,\omega\wedge D\sigma,\\
\forall \omega\in\obul(X), \forall \sigma\in\obul(E).
\end{multline}
One can write $D$ as a sum of a $\bpartial$-connection $\bpartial_E$
preserving the decomposition $E=E^+\oplus E^-$ and an odd section of
the $\ZZ_2$-graded vector bundle $\obul(\End(E))$. Note that
$\bpartial_E$ does not square to zero, in general, so $E$ does not
have a natural holomorphic structure.

Given two such pairs $(E_1,D_1)$ and $(E_2,D_2)$ we consider the
$\ZZ_2$-graded vector space
$$
\obul\left(\Hom(E_1,E_2)\right)
$$
and its odd endomorphism $D_{12}$ defined by
$$
D_{12}f=D_2\,f - (-1)^{\deg_{\ZZ_2}f}\,f D_1,\quad f\in
\obul(\Hom(E_1,E_2)).
$$
It is easy to see that $D_{12}^2=0$. We define the space of
morphisms from $(E_1,D_1)$ to $(E_2,D_2)$ to be the cohomology of
$D_{12}$.

To see how an object of the curved derived category defines a brane
for the curved B-model, we need to choose $\partial$-connections on
$E^+$ and $E^-$ which in a local trivialization we write as
$$
\partial^\pm=\partial+\cA^\pm,\quad
\cA^\pm\in\Omega^{1,0}(U,\End(E^\pm)),\ U\subset X.
$$
Further, we write
$$
D=\bpartial\cdot \xId+\cK=\bpartial\cdot\xId+\begin{pmatrix}\cB^+ & \cT\\
\cS & \cB^- \end{pmatrix},
$$
where
\begin{multline}
\cB^\pm \in\Omega^{0,\hat 1}(U,\End(E^\pm)),\ \cT\in
\Omega^{0,\hat 0}(U,\Hom(E^-,E^+)),\\
\cS \in \Omega^{0,\hat 0}(U,\Hom(E^+,E^-)),\ U\subset X.
\end{multline}
Let us introduce the supermanifold $\Pi{\bar T}_X$ with odd
coordinates $\eta^{\bi}$. There is an obvious map $\pi:\Pi{\bar
T}_X\raa X$, and we may regard $\cK$ as a locally-defined odd
section of the $\ZZ_2$-graded bundle $\pi^*\End(E)$. Similarly,
$\bpartial$ can be reinterpreted as an odd vector field
$\eta^{\bi}\partial_{\bi}$ on $\Pi{\bar T}_X$, and $D$ is a
first-order differential operator on $\pi^*E$.

Now let $\gamma$ be a connected component of $\partial\Sigma$, i.e.
a circle. Let $t\in [0,1)$ be a coordinate on $\gamma$. Given a map
$\Phi=(\phi,\eta):\gamma\raa \Pi{\bar T}_X$ and a section $\rho^i_t
dt$ of $\phi^* T_X\otimes T^*_\gamma$, we consider a connection
1-form on the pull-back of $\Phi^*(E)$:
\begin{equation}
\cN=\begin{pmatrix} \cA^+_i \dt\phi^i+\cB^+_\bi \dt\phi^\bi
+F^+_{i\bj}\rho^i_t\eta^\bj &
\rho^i_t\nabla_i\cT+\dt\phi^\bi \frac{\partial\cT}{\partial\eta^\bi}\\
\rho^i_t \nabla_i\cS+\dt\phi^\bi \frac{\partial
\cS}{\partial\eta^\bi} & \cA^-_i \dt\phi^i+\cB^-_\bi \dt\phi^\bi
+F^-_{i\bj}\rho^i_t\eta^\bj
\end{pmatrix}
\end{equation}
Here $\dt=d/dt$, $\nabla_i$ denotes the covariant derivatives with
respect to the $\partial$-connections on $\Hom(E^\pm,E^\mp)$, and
$F^\pm_{i\bj}$ is defined by
$$
F^\pm_{i\bj}\eta^{\bj}=\partial_i \cB^\pm-\partial_\bj
\cA^\pm_i\eta^{\bj}+[\cA^\pm_i,\cB^\pm].
$$

One can check that this connection on $\Phi^*(E)$ satisfies
$$
\delta_Q(\dt+\cN)=[\dt+\cN,\Phi^*\cK]-W^{(1)},
$$
where the brackets denote the graded commutator. We can use this
identity to construct a suitable boundary action as follows. Let
$\U(0,t)$ be the parallel transport operator of the connection
$\cN$, i.e. the unique solution of the first order differential
equation
$$
(\dt+\cN)\U(0,t)=0
$$
satisfying $\U(0,0)=1$. We define
$$
\exp(-S^{bry})=\STr\ \U(0,1),
$$
where $\STr$ denotes the supertrace on $\End_\CC(E_{\Phi(0)})$. Then
it is easy to check that
$$
\delta_Q S^{bry}=-\int_\gamma W^{(1)}.
$$
This precisely cancels the BRST-variation of the bulk action
(\ref{brststotW}).

Now suppose the circle $\gamma$ is split into two segments $(0,t_0)$
and $(t_0,1)$ by the points $t=0$ and $t=t_0$. Suppose further than
on the segment $(0,t_0)$ the boundary conditions is described by
$(E_1,D_1)$, while on the segment $(t_0,1)$ it is described by
$(E_2,D_2)$. Given a morphism $f_{12}$ from $(E_1,D_1)$ to
$(E_2,D_2)$ and a morphism $f_{21}$ from $(E_2,D_2)$ to $(E_1,D_1)$
consider the following expression to be inserted in the
path-integral:
$$
\STr\ f_{21}(\Phi(1))\U_2(1,t_0)f_{12}(\Phi(t_0))\U_1(0,t_0).
$$
Here $\U_1$ and $\U_2$ are constructed from $(E_1,D_1)$ and
$(E_2,D_2)$ as above, and we regard the bundle-valued forms $f_{12}$
and $f_{21}$ as sections of $\pi^*\Hom(E_1,E_2)$ and
$\pi^*\Hom(E_2,E_1)$. It is easy to check that the BRST-variation of
this expression cancels the BRST variation of $\exp(-S_0-S_W)$.
Therefore we may regard morphisms $f_{12}$ and $f_{21}$ as local
boundary topological observables to be inserted at the joining point
of boundary conditions $(E_1,D_1)$ and $(E_2,D_2)$.

In the special case when $W$ is a holomorphic function on $X$, the
curved B-model reduces to the Landau-Ginzburg model with
superpotential $W$ and the curved derived category of $X$ reduces to
the category of Landau-Ginzburg branes. In particular, if $X$ is a
Stein manifold, the latter is equivalent to the category of matrix
factorizations of $W$ \cite{KLione,Brunneretal,Laz}. An alternative
algebro-geometric definition of this category was given by D.~Orlov
\cite{Orlov:sing}.

The category of Landau-Ginzburg branes has a remarkable property
known as the Kn\"orrer periodicity: its equivalence class does not
change when one replaces $X$ with $X\times\CC^2$ and $W$ with
$W+xy$, where $x,y$ are coordinates on $\CC^2$ \cite{Orlov:Knorrer}.
In particular, if $X=\CC^{2n}$ and $W$ is a non-degenerate quadratic
form, Kn\"orrer periodicity implies that the category of
Landau-Ginzburg branes is equivalent \cite{KLione,Orlov:Knorrer} to
the category of $\ZZ_2$-graded vector spaces (which is the category
of Landau-Ginzburg branes when $X$ is a point). On the other hand,
if $X=\CC^{2n+1}$ and $W$ is a non-degenerate quadratic form, the
category of Landau-Ginzburg branes is equivalent to the category of
modules over the Clifford algebra with one generator \cite{KLione}.
It is easy to see that Kn\"orrer periodicity remains true for the
more general case of the curved derived category.

\section{Perturbative computation of the fusion of Wilson lines on the
boundary}\label{app:fusion}

In section 4.2.3 we discussed that on the classical level fusion of
the two line operators on the boundary corresponds to tensoring
vector bundles and argued that the first quantum correction  is the
deformation of the resulting connection by a $(0,3)$-form
\begin{equation}
\label{defapp} C_0 \beta_{\bar m}^{ij} F_{i {\bar k}}\tF_{j {\bar
p}} d\phi^\bm d\phi^\bk d\phi^\bp
\end{equation}
Here $C_0$ is a complex number, $F$ and $\tF$ are curvatures of
connections on the two holomorphic vector bundles on $Y$, and
$\beta^{ij}_{\bar m}$ is given by
$$
\beta^{ij}_{\bar m}=\smpf^{k'i}\partial_{\bar m} A^i_{k'},
$$
where $A^K_{k'}$ is defined by (\ref{Adef}) and defines an embedding
of $N_Y$ into $T_X\vert_Y$. Essentially, $\beta^{ij}_{\bar m}$ is
the second fundamental form of $Y$. In Darboux coordinates,
$\beta^{ij}_{\bar m}$ is symmetric in the upper indices.

In this section we verify the presence of the correction
(\ref{defapp}) by computing the OPE of Wilson line operators on the
boundary to leading order in $\beta$. To simplify the computation,
note first that the BRST-invariant connection 1-form entering the
definition of the Wilson line
$$
\cO^{(1)}(A)=A_i d\phi^i+A_\bi d\phi^\bi +F_{i\bj}\chi^i\eta^\bj
$$
is a descendant of a BRST-invariant 0-form observable
$$
\cO(A)=A_\bi \eta^\bi.
$$
That is, one has
$$
d\cO(A)=\delta\cO^{(1)}(A).
$$
The observable $\cO(A)$ does not determine $\cO^{(1)}$ uniquely, but
it does define it up to BRST-exact terms. Instead of computing the
OPE of two Wilson lines, we may therefore compute the OPE of
$\cO(A)$ and $\cO^{(1)}(\tA)$, where $A$ and $\tA$ are connections
on the two holomorphic vector bundles.

One can further simplify the computation by noting that the expected
correction is antisymmetric with respect to the exchange of $A$ and
$\tA$. The antisymmetric part of the quantum correction to the OPE
is given by the contour integral over the insertion point of
$\cO^{(1)}(\tA)$:
\begin{equation}\label{descans}
\oint_{\gamma} \cO^{(1)}(\tA)\cO(A)(0).
\end{equation}
Here $\gamma$ is a circle at the boundary of $M$ centered at $0$. We
are going to show that to leading order in $\beta$ this contour
integral is
$$
C_0 \beta_{\bar m}^{ij} F_{i {\bar k}}\tF_{j {\bar p}} \eta^\bm
\eta^\bk \eta^\bp .
$$

The computation is performed by expanding around a constant bosonic
background $\phi^I=\phi_0^I$, where $\phi_0^I$ belongs to $Y$. It is
convenient to use a complex frame adapted to the boundary
conditions. That is, we will work in the basis where the metric
$g_{I\bJ}(\phi_0)$ does not have mixed components (with both primed
and unprimed indices), while the holomorphic symplectic form
$\Omega_{IJ}(\phi_0)$ has only mixed components. In this coordinate
system, $\beta^{ij}_{\bar m}$ is given by
$$
\beta^{ij}_{\bar m}=-\smpf^{k'i}g^{\bk j}\partial_{\bar m}
g_{k'\bk}=\smpf^{ik'}g^{i\bar P}g_{k'\bar J}\Gamma^{\bar J}_{\bar
m\bar P}.
$$
We will denote by $x$ a local complex coordinate on $\partial M$,
while the coordinate in the normal direction will be denoted $x_3$.
Thus we can expand the 1-form field $\chi^I$ as follows:
$$
\chi^I(x,x_3)=\rho^I(x,x_3) dx+\bar\rho^I(x,x_3)
d\bx+\chi_3^I(x,x_3) dx^3.
$$
We also use the notation that unprimed lower-case indices
$i,j,k,\ldots,$ run from $1$ to $n$ and label coordinates on $Y$,
while primed lower-case indices $i',j',k',\ldots,$ run from $n+1$ to
$2n$ and label coordinates in the directions normal to $Y$.

The key ingredients in the computation are the following OPEs of
local fields on $\partial M$:
\begin{align}\label{locope}
\rho^i(x,0)\delta\phi^j(0,0) &\sim \frac{C}{x}\beta^{ij}_\bk
\eta^\bk(0,0)+\ldots,\\
\bar\rho^i(x,0)\delta\phi^j(0,0) &\sim \frac{-C}{\bx}\beta^{ij}_\bk
\eta^\bk(0,0)+\ldots.
\end{align}
Here $\delta\phi^i=\phi^i-\phi_0^i$, $C$ is a complex number, and
dots denote terms which are either nonsingular or contain
derivatives of $\beta$.

Inspecting the action, we see that to leading order in perturbation
theory we need to compute a free-field OPE
\begin{equation}\label{opeapp1}
\rho^i(x,0)\delta\phi^j(0,0)\int d^3y \sqrt{h} g_{K{\bar
J}}h^{\mu\nu}\chi_{\mu}^K(y,y_3)
 \partial_{\nu}\phi^{\bar N}(y,y_3)
\Gamma^{\bar J}_{{\bar N} {\bar M}} \eta^{\bar M}(y,y_3).
\end{equation}
The terms we are interested in are obtained by contracting $\rho^i$
with $\chi$ and $\delta\phi^j$ with $\phi^{\bar N}$. In the presence
of the boundary the nonvanishing free-field contractions we need are
given by
\begin{align}
\delta\phi^i(x,x_3)\delta\phi^\bj(y,y_3) &\sim \frac{g^{i\bar
j}}{4\pi|\bfx-\bfy|}+\frac{g^{i\bar j}}{4\pi|\bfx-\Pi\bfy|},
\\
\rho^i(x,x_3)\brho^{k'}(y,y_3)&\sim
-\frac{i\Omega^{ik'}}{8\pi}\left(\frac{x_3-y_3}{|\bfx-\bfy|^3}-\frac{x_3+y_3}{|\bfx-\Pi\bfy|^3}\right),\\
\rho^i(x,x_3)\chi_3^{k'}(y,y_3) &\sim
\frac{i\Omega^{ik'}}{8\pi}\left(\frac{\bx-\by}{|\bfx-\bfy|^3}+\frac{\bx-\by}{|\bfx-\Pi\bfy|^3}\right),\\
\brho^i(x,x_3)\chi_3^{k'}(y,y_3) &\sim -
\frac{i\Omega^{ik'}}{8\pi}\left(\frac{x-y}{|\bfx-\bfy|^3}+\frac{x-y}{|\bfx-\Pi\bfy|^3}\right),\\
\rho^i(x,x_3)\eta^{\bar k}(y,y_3)&\sim -\frac{g^{i{\bar
k}}}{8\pi}\left(\frac{{\bar x}-{\bar y}}{|\bfx-\bfy|^3}+
\frac{{\bar x}-{\bar y}}{|\bfx-\Pi\bfy|^3}\right),\\
\brho^i(x,x_3)\eta^{\bar k}(y,y_3)&\sim -\frac{g^{i{\bar
k}}}{8\pi}\left(\frac{x-y}{|\bfx-\bfy|^3}+
\frac{x-y}{|\bfx-\Pi\bfy|^3}\right),
\end{align}
where $\bfx,\bfy\in \CC\times \RR=\RR^3$ have components
$(x,x_3),(y,y_3)$, respectively, and $\Pi \bfy \in\CC\times\RR$ has
components $(y,-y_3)$.

To detect the terms in the OPE we are interested in, it is
convenient to integrate the expression (\ref{opeapp1}) with respect
to $x$ along a circle $\gamma_\eps$ of radius $\eps$ centered at
$x=0$ and then take the limit $\eps\raa 0$. The result is
$$
\beta^{ij}_\bk \lim_{\eps\raa 0} \int d^3 y\, \eta^\bk(y,y_3)
F_\eps(y,y_3),
$$
where
$$
F_\eps(y,y_3)=\frac{-1}{16\pi^3} \oint_{\gamma_\eps} dx \,\frac{\bx
y_3}{|\bfy|^3 |\bfx-\bfy|^3}.
$$
A straightforward computation which we omit shows that in the limit
$\eps\raa 0$ the function $F_\eps(y,y_3)$ tends to $C
\delta^3(y,y_3)$ for some constant $C$ (in this computation we
assume that $y^3\geq 0$, since the integration in (\ref{opeapp1})
extends only over nonnegative $y^3$). This proves the first line of
(\ref{locope}). The second line is proved similarly.

Now we can compute the contour integral (\ref{descans}). The result
is (up to BRST trivial terms)
$$
- 2\pi i C F_{i\bj} \tF_{k\bl}\beta^{ik}_\bm
\eta^\bj\eta^\bl\eta^\bm .
$$
This observable (divided by two) represents the leading quantum
correction to the fusion of Wilson lines on the boundary.

\end{document}